\documentclass{jfm}
\usepackage{graphicx}
\usepackage{tikz}
\graphicspath{{image/}}
\usepackage{amsmath, epstopdf, epsfig, multirow, float}
\usepackage{subcaption}
\usepackage[hidelinks]{hyperref}
\usepackage[noabbrev]{cleveref}
\hypersetup{
  colorlinks = true,   %Colours links instead of ugly boxes
  urlcolor   = blue,   %Colour for external hyperlinks
  linkcolor  = blue,   %Colour of internal links
  citecolor  = blue    %Colour of citations
}

\shorttitle{Threshold of Drop Fragmentation}
\shortauthor{Parik, Truscott, and Dutta}

\title{On the Threshold of Drop Fragmentation under Impulsive Acceleration}

\author{Aditya Parik\aff{1}
  \corresp{\email{aditya.parik@usu.edu}},
  T. Truscott\aff{2},
 \and S. Dutta\aff{1} \corresp{\email{som.dutta@usu.edu}}}

\affiliation{\aff{1}Department of Mechanical and Aerospace Engineering,
                 \\ Utah State University,  UT 84321, USA 
                 \aff{2}Department of Mechanical Engineering,
                 \\ King Abdullah University of Science and Technology, Saudi Arabia}

\renewcommand{\Re}{\mathit{Re}}
\newcommand{\We}{\mathit{We}}
\newcommand{\Oho}{\mathit{Oh}_o}
\newcommand{\delP}{{\Delta P}_{drive}}
\newcommand{\Ohd}{\mathit{Oh}_d}
\newcommand{\Wecr}{\mathit{We}_{cr}}
\newcommand{\Cbr}{C_\mathit{breakup}}

\newcommand{\Vrel}{V_\mathit{rel}}
\newcounter{para}

\begin{document}

\maketitle

\begin{abstract}
Secondary fragmentation of an impulsively accelerated drop depends on fluid properties and velocity of the ambient. The critical Weber number $(\Wecr)$, the minimum Weber number at which a drop undergoes non-vibrational breakup, depends on density ratio $(\rho)$, the drop $(\Ohd)$, and the ambient $(\Oho)$ Ohnesorge numbers. The current study uses VoF based interface-tracking multiphase flow simulations to quantify the effect of different non-dimensional groups on the threshold at which secondary fragmentation occur. For $\Ohd \leq 0.1$, a decrease in $\Ohd$ was found to significantly influence the breakup morphology, plume formation, and $\Wecr$. The balance between the pressure difference between the poles and the periphery, and the shear stresses on the upstream surface, was found to be controlled by $\rho$ and $\Oho$. These forces induce flow inside the initially spherical drop, resulting in deformation into pancakes and eventually the breakup morphology of forward/backward bag. The evolution pathways of the drop morphology based on their non-dimensional groups have been charted. With inclusion of the data from the expanded parameter-space, the traditional $\Wecr-\Ohd$ diagram used to illustrate the dependence of critical Weber number on $\Ohd$, was found to be inadequate in predicting the minimum initial $\We$ required to undergo fragmentation. A new non-dimensional parameter $C_{breakup}$ is derived based on the competition between the forces driving the drop deformation and the forces resisting the drop deformation. Tested using available experimental data and current simulations, $C_{breakup}$ is found to be a robust predictor for the threshold of drop fragmentation. 
\end{abstract}


\section{Introduction} \label{section:introduction}

Drop fragmentation, also known as secondary fragmentation, is the process of through which a drop breaks up under the action of external aerodynamic forces induced by the ambient flow. These forces originate due to a velocity deficit between the drop and the ambient medium. There are two fundamental ways a drop might experience a velocity deficit: a uniform ambient flow impacts a stationary drop in a gravity-free environment, called ``impulsive acceleration'' \citep{Han2001}; or an initially stationary drop accelerates under the action of a constant body force, while experiencing aerodynamic forces, called ``free-fall'' \citep{Jalaal2012}. For both cases, a liquid drop experiences aerodynamic forces that cause the drop to deform, which may lead to its fragmentation at a Weber number $\We_0$ higher than the critical value $\Wecr$ \citep{Hinze1949, Hinze1955}.
During free-fall, the drop starts with zero aerodynamic forces that gradually increase to a maximum, at either its terminal or its breakup velocity. On the other hand, an impulsively accelerated drop starts its deformation process with the largest velocity deficit, and the corresponding large aerodynamic stresses acting on it. These stresses gradually reduce as the drop decelerates with respect to the ambient flow. It should be noted that as the drop decelerates, it also simultaneously deforms causing an increase in its frontal area, which can in turn increase surface shear stresses, given the velocity deficit is still substantial.

\begin{figure}
  \centering \includegraphics[width=1\textwidth]{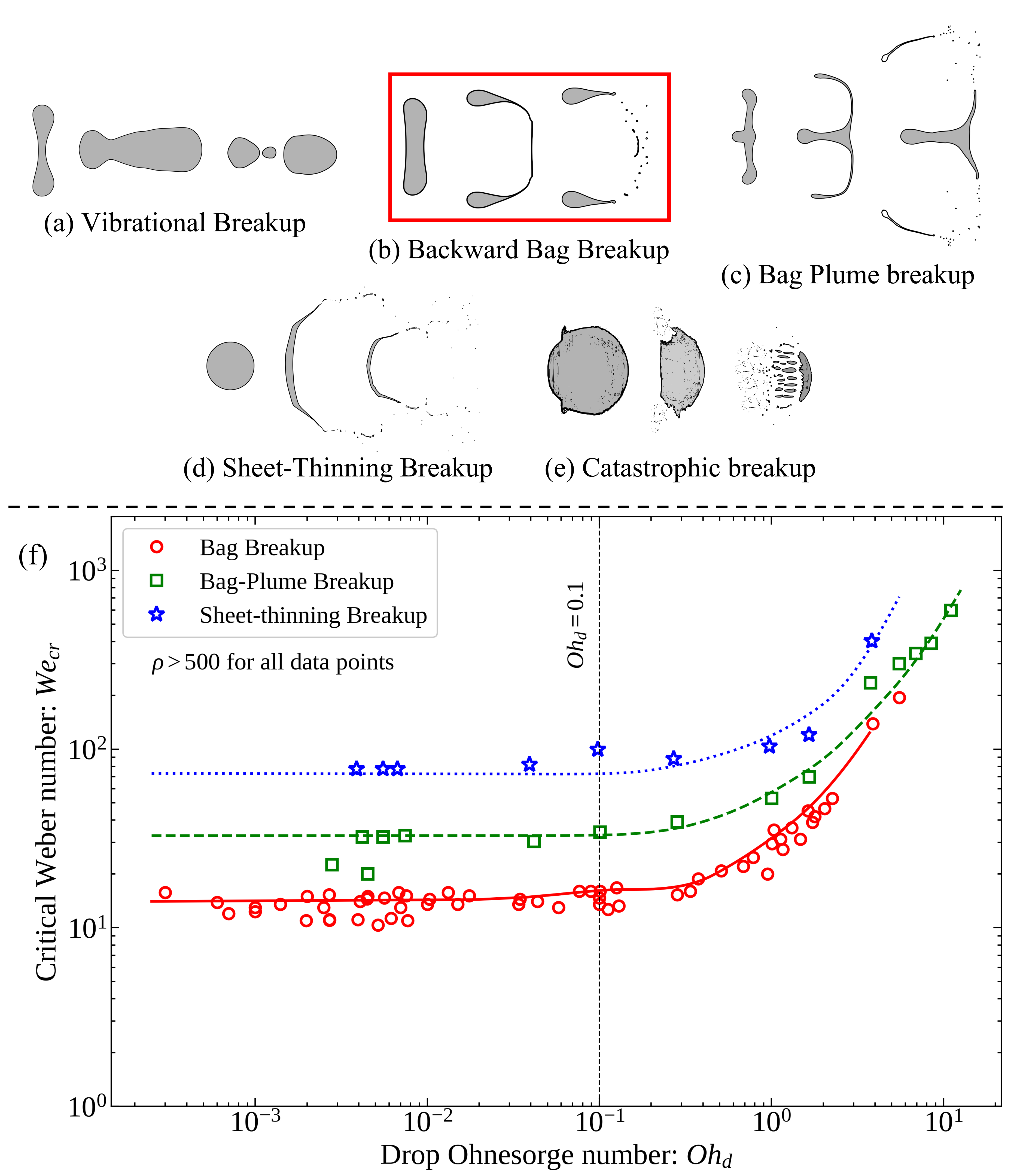}
  \caption{  Types of drop breakup morphologies \citep{guildenbecher_secondary_2009,theofanous_aerobreakup_2011} observed in experiments in order of increasing threshold Weber numbers are illustrated through figures (a) to (e).
  Figure (f) plots all the experimental data on threshold Weber numbers required to produce different fragmentation morphologies, based on a similar plot in \cite{Hsiang1995} and data from \cite{Krzeczkowski1980,Pilch1987,wierzba_deformation_1990,dai_temporal_2001,Kulkarni2014,jackiw_aerodynamic_2021,Han2001,jain_secondary_2019}.
  Backward bag breakup, as shown is figure (b) in red box and the red curve in figure (f), has been the predominantly observed critical non-vibrational fragmentation morphology.}
  \label{fig:lit_available_data}
\end{figure}

Most applications such as the internal combustion engine, spray painting, etc., involve a secondary fragmentation due to impulsive acceleration. Among impulsive acceleration cases, there can be different experimental systems such as a drop introduced into a uniform cross-flow, or a drop exposed to a shockwave in a wind tunnel. The timescales of an impulsive drop breakup process is relatively small, resulting in both the aforementioned experimental setup predicting similar critical Weber number for secondary fragmentation \citep{hsiang_near-limit_1992,Hsiang1995}.
Most of the experimental studies conducted on secondary fragmentation \citep{Pruppacher1970,Krzeczkowski1980,wierzba_deformation_1990,hsiang_near-limit_1992,Gelfand1996,theofanous_aerobreakup_2004,Szakall2009,jain_secondary_2015,Kulkarni2014} have focused on impulsive acceleration, and those results are summarized in Figure \ref{fig:lit_available_data}.

After starting from an initial spherical shape, drops start the deformation process with flattening of the downstream face under the influence of pressure forces \citep{Villermaux2009,jain_secondary_2019,jackiw_aerodynamic_2021}. This is followed by the formation of a pancake of one of the following two types: \textbf{(a)} a flat disk-like structure with both upstream and downstream faces showing an increase in radius of curvatures (henceforth called ``flat pancake''); or \textbf{(b)} a pancake with a concave-shaped downstream surface, corresponding with minimal change in curvature of the upstream surface (henceforth called ``forward pancake'') \citep{Han2001}. These differences in pancake shapes have been observed for differences in physical and flow parameters such as density-ratio $\rho$, and initial Reynolds number $\Re_0$ (or outside Ohnesorge number $\Oho$). However, the exact physical mechanism that causes this difference in pancake morphology has not yet been explored in the literature. Beyond the formation of a pancake, the pancake deforms further and starts to form a toroidal periphery (rim), which then leads to further deformation and possibly even breakup through different morphologies (discussed in the next paragraph). This stage, which marks the completion of pancake formation and the start of a visible peripheral rim, can be temporally indicated by a non-dimensional time $t^* = t/\tau \approx 1$. Here, $\tau = D\sqrt{\rho}/V_0$ is the drop deformation timescale \citep{ rimbert_spheroidal_2020}, where $V_0$ is the uniform initial velocity of the ambient medium relative to the drop, $D$ is the drop's volume-averaged diameter, and $t$ represents the elapsed dimensional time during the deformation process. This timescale is the same as the dimensionless time for Rayleigh-Taylor or Kelvin-Helmholtz instabilities specified by Pilch \& Erdman \citep{Pilch1987}. $\tau$ includes the effect of $\rho$, thus making $t^*$ a useful temporal scale when comparing cases with different density-ratios.

Following the formation of a pancake after $t^*>1$, the drop may further deform and ultimately breakup through one the the following morphologies (as illustrated in \cref{fig:lit_available_data}(a) to (e)): 
\textbf{(a)} Vibrational mode where the drop oscillates about a maximum deformation state, and does not show consistent breakup \citep{hsiang_near-limit_1992,rimbert_spheroidal_2020};
\textbf{(b)} Simple Bag Breakup which involves the formation of a toroidal rim and the inflation of a thin film (bag) in between, which ultimately ruptures due to Rayleigh-Plateau instabilities \citep{Kulkarni2014,jackiw_aerodynamic_2021};
\textbf{(c)} a bag breakup with morphological features in addition to a bag, such as stamen/plume \citep{Hsiang1995,jain_secondary_2015} or multiple bags \citep{Cao2007,jackiw_aerodynamic_2021};
\textbf{(d)} Sheet Thinning breakup where thin sheets and ligaments are removed from the periphery of the pancake, and are blown downstream relative to the drop core due to their low local inertia, ultimately breaking up due to instabilities \citep{khosla_detailed_2006,guildenbecher_secondary_2009}; and 
\textbf{(e)} catastrophic breakup where unstably growing surface waves pierce through the entire pancake and cause it to catastrophically disintegrate \citep{theofanous_aerobreakup_2011}.

For liquid drops in air under standard atmospheric conditions, $\rho>500$ and $0.0005<\Oho<0.005$. Consequently, in the context of density ratio, most existing experimental works are concentrated within $\rho>500$. For all such experimental works, the critical drop breakup morphology for $\Ohd<0.1$ has been observed to always be a simple bag breakup, with minimal dependence of $\Wecr$ on density ratio. Through the advent of petascale/exascale computing in the last two decades, both axisymmetric and 3D DNS of large $\rho$ cases has become possible, thus allowing computational exploration of the entire density-ratio space for drops, i.e., $\rho>1$. As a result, substantial research has been conducted in the high $\rho$ space \citep{theofanous_aerobreakup_2011, tavangar_cfd_2015, strotos_predicting_2016, yang_transitions_2017, guan_numerical_2018,  dorschner_formation_2020, lingDetailedNumericalInvestigation2023, tangBagFilmBreakup2023}, where threshold breakup morphologies similar to the experiments have been observed. On the other hand, computational works exploring low density ratio cases ($\rho<100$) tell a different story. \cite{Han2001} was one of the earliest computational works to explore the role of density ratio in the impulsive acceleration of a drop for various ambient and drop viscosities. The work focused on $\rho=1.15$ and $10$, and found $\rho$ to significantly influence the threshold Weber numbers and the corresponding fragmentation morphology (e.g., forward pancake and bag formation). \cite{jain_secondary_2019} explored the effect of $\rho$ on drop deformation for a specific $\Re_0$ and viscosity ratio for a range of $\We_0$ from $20$ to $100$, and observed the immense impact $\rho$ has on the orientation of bags and pancakes, the drop velocities, and the total observed deformations. Similar conclusions were reached by \cite{marcotte_density_2019}, where a higher threshold Weber number for both fragmentation and the transition from bursting to stripping was observed for the low density ratio cases.

By the 1990s, experimental and theoretical studies \citep{Karam1968,Krzeczkowski1980,Pilch1987} had established the important role of drop Ohnesorge number $\Ohd$ on $\Wecr$. Hsiang \& Faeth's review paper in 1995 \citep{Hsiang1995} significantly advanced this understanding. They aggregated all available experimental data from existing works, including their own experiments, into $\Wecr$ vs. $\Ohd$ plots. Their findings showed that the threshold $\We_0$ for the onset of all types of breakup morphologies (i.e., simple backward bag and other higher $\We_0$ breakup morphologies) follows a similar trend with respect to $\Ohd$ (see figure 1 of \cite{Hsiang1995} or \cref{fig:lit_available_data}(f)), with the threshold $\We_0$ almost independent with respect to $\Ohd$ for $\Ohd<0.1$, and then increasing rapidly for $\Ohd>0.1$. Furthermore, the critical breakup morphology (for the onset of breakup) for all the explored works were found to be simple bag breakups. It is crucial to emphasize that all the findings presented in \cite{Hsiang1995} were derived from experiments in the high density ratio regime, i.e., $\rho>500$. This focus on high density ratios is also reflected in the majority of computational studies investigating the effect of $\Ohd$ on drop breakup. For instance, \cite{yang_transitions_2017} observed an increase in the transition Weber number from squeezing to bag breakup for increasing $\Ohd$ for drops with $\rho=800$. Similarly, \cite{jain_secondary_2019} explored the effect of viscosity ratio (and hence $\Ohd$) on breakup morphologies through simulations of drops of $\rho=1000$ and two different viscosity ratios. They noted a decrease in $\We_0$ with decreasing $\Ohd$ and the emergence of a plume at the upstream pole for lower viscosity ratios in both 3D and equivalent axisymmetric simulations. \cite{tangBagFilmBreakup2023} examined the influence of a range of $\Ohd$ values less than $0.1$ on the time of the start of breakup of the bag, discovering an exponential increase with $\Ohd$. All the aforementioned computational studies operate within $\rho \approx 800$, and thus result in a weak dependence of the threshold to $Oh_d$, similar to experiments. This is apparent in Figure 8 in \cite{yang_transitions_2017}, where the $\Wecr$ is observed to be within $10<\Wecr<20$ for all $\Ohd<0.1$. However, the threshold $\We_0$ can exhibit substantial variation with $\Ohd$ for lower density ratios and ambient Ohnesorge numbers ($\Oho$). This is highlighted by some (although limited) computational studies, that have explored the effect of $\Ohd$ at lower density ratios. \cite{Han2001}, for a density ratio of $10$, observed a significant decrease in the amount of deformation for higher drop viscosities. This decrease in deformation was speculated to lead to an increase in $\Wecr$ values. \cite{Kekesi2014} linked the fragmentation morphology to the ratio of the characteristic times for shear and bag breakup, proportional to the ratio of viscosity ratio and density ratio, supported by 3D simulations. \cite{farsoiyaRoleViscosityTurbulent2023} discovered that $\Wecr$ (based on turbulence dissipation rate) low density ratio ($\rho=1$) drops in isotropic turbulence is significantly influenced by the viscosity ratio.

\cite{Villermaux2009} was the first to analytically describe the bag breakup process for an inviscid drop and derived a constant threshold value of 6 for $\Wecr$, an underestimation compared to experimentally seen threshold values. Their work was extended to include the viscosity of the drop fluid, first by \cite{Kulkarni2014} and most recently by \cite{jackiw_aerodynamic_2021}, resulting in a function of $\Ohd$ that describes $\Wecr$. This corrected for the underestimation and led to a great match with previous experimental results. The analytical works mentioned above do not take into account ambient fluid properties such as ambient viscosity ($\Oho$) and density ($\rho$) (which in turn dictates the relative velocity of the drop with the ambient) in influencing the resulting deformation characteristics. However, the density and viscosity contrasts between the ambient and drop fluids are generally very large for experimental systems, thus minimizing the relative significance of ambient fluid properties relative to that of the drop. This allows for a good match between the analytical solution and corresponding experimental $\Wecr$ values, even with the aforementioned assumptions. However (as will be explored in detail in this work), for systems where the contrast between the (physical properties of) ambient and drop fluids is not substantial, these factors must be taken into account for the correct estimation of threshold $\We_0$ values.

Initial Reynolds number $Re_0$ (or alternatively Ambient Ohnesorge Number $\Oho$) also remains to be exhaustively explored, especially in the context of critical drop breakup threshold. \cite{Han2001} did simulations for different $\Re_0$ values for some low density-ratio cases, and observed large reduction in drop deformations for low $\Re_0$ values. They speculated that this reduction in deformation might lead to an increase in $\Wecr$ values. Very few other works have explored or commented on the role of $\Oho$ on drop breakup \citep{guildenbecher_secondary_2009,jain_secondary_2019,marcotte_density_2019}. \cite{jain_secondary_2019} once again was one of the very few works to analyze the impact of $\Re_0$ on high density-ratio drops ($\rho=1000$) and observed higher incidences of plume formation in backward bag morphology for higher $\Re_0$ values.

Hence, there is the need for a single cohesive study analyzing the effect of all the relevant non-dimensional parameters, $\Oho$, $\Ohd$, and $\rho$ on drop deformation and breakup, and their impact on the threshold Weber number observed for critical breakup ($\Wecr$). In this study we quantify the effect of each of these parameters using high-fidelity multiphase flow simulations. It should be noted that a distinction between liquid-gas and liquid-liquid drop-ambient systems has been maintained in the existing literature. However, fundamentally, the only differentiating factor between the two systems is the density and viscosity ratios, as well as the surface tension of the fluid interface. It is expected that the need for this distinction should vanish for a study that covers a sufficiently large parameter sweep involving $\rho$, $\Oho$ and $\Ohd$. Thus, a wide range of values of $\rho$ and $\Oho$ have been quantified. $\Ohd$ values explored in this study are restricted to $\Ohd \leq 0.1$, to focus at a range rarely explored in existing studies.

The paper starts with a description of the relevant impulsive acceleration problem (\cref{subsec:problem_description}), the high-fidelity numerical model (\cref{subsec:model_description} and \ref{subsec:validation}). The parameter space to be numerically explored is described in detail in \cref{subsec:paramter_space}. The effects of $\Ohd$, $\Oho$, and $\rho$ on the drop deformation, given that other parameters are constant, are described in detail, illustrating the forces and internal flow observed in the drops (\cref{sec:results}). During the course of the parameter sweep, by simulating a range of Weber number values for every non-dimensional parameter set, the corresponding critical Weber number can be discovered. Based on the insights gained from the simulations, a novel non-dimensional parameter ($C_{breakup}$) has been derived, incorporating the effects of all the relevant non-dimensional numbers. $C_{breakup}$ is found to be more effective in predicting the threshold of drop fragmentation, over the currently used $We_{cr}$.

\section{Problem Description and formulation} \label{sec:problem_description_and_formulation}
  \subsection{Problem Description and Non-dimensionalisation}
\label{subsec:problem_description}
Let us consider a drop of diameter $D$ containing a fluid of density $\rho_d$ and dynamic viscosity $\mu_d$ (subscript $d$ implies properties associated with the drop). It is impulsively accelerated by a uniform flow of velocity $V_0$, of density $\rho_o$ and viscosity $\mu_o$ (subscript $o$ implies properties associated with the ambient medium, i.e., outside the drop), starting at $t=0$. The surface tension of the drop-ambient interface is $\sigma$.

Based on these initial conditions, we can define an initial Weber number $\We_0$ ($\rho_o V_0^2 D/\sigma$), which represents the competition between the dynamic pressure forces that drive the deformation of the drop and the capillary forces that resist this deformation. An increase in $\We_0$ corresponds to an increase in the maximum deformation achieved by the drop before it retracts to its equilibrium shape of a sphere, due to the action of surface tension. We then expect there to be a maximum $\We_0$ for which the surface tension forces barely prevent fragmentation in the drop. This threshold is called the critical Weber number $\Wecr$ ($\rho_o V_{cr}^2 D/\sigma$), where $V_{cr}$ corresponds to the critical (lowest possible) $V_0$ required to consistently realize a non-vibrational breakup. A Buckingham-Pi analysis (\cref{eq:nondimensionalterms}) for this system reveals that
\begin{equation}
  \Wecr = F \left( \rho, \Oho, \Ohd \right).
  \label{eq:nondimensionalterms}
\end{equation}

The density-ratio $\rho$ = ($\rho_d/\rho_o$) is a measure of the inertia of the drop relative to the ambient medium, reflecting its responsiveness to external forces. The dynamic pressure forces exerted by the ambient medium scale with $\rho_o$, i.e., its effectiveness in inducing accelerations in specific parts (such as the peripheral rim or the core) or in the entirety of the drop is inversely proportional to $\rho$. The drop Ohnesorge number $\Ohd =$ ($\mu_d/\sqrt{\rho_d\,\sigma\,D}$) is a ratio of capillary and momentum diffusion time scales and provides an estimate of how the energy supplied to a drop by external forcing is distributed among surface energy and viscous dissipation. The ambient Ohnesorge number $\Oho =$ ($\mu_o/\sqrt{\rho_o\,\sigma\,D}$) provides a non-dimensional, velocity-independent analogue for the initial Reynolds number $\Re_0$ (since $\Oho = \sqrt{\We_0}/\Re_0$), and represents ambient viscosity given that other parameters are the same. Finally, we define an instantaneous Reynolds number $\Re = $ ($\rho_0 V_\mathit{rel} D_\mathit{rel}/\mu_o$) in addition to $\Re_0$, in order to accomodate the generally significant change in center-of-mass accelerations and frontal area across the timescales over which a drop deforms and fragments. $\Re_0$ is based on the drop's instantaneous velocity deficit ($V_\mathit{rel}$) with the ambient medium and its frontal radius for the instantaneous deformed shape ($D_\mathit{rel}$).

If we choose $\rho_o$, $V_0$ and $D$ as the basis variables for non-dimensionalising the system, the dimensional variables can be non-dimensionalized as follows: $\tilde{\rho_d} = \rho$, $\tilde{\sigma} = 1/\mathit{We}_0$, $\tilde{\mu}_o = \Oho/ \sqrt{\mathit{We}_0}$, and $\tilde{\mu}_d = \Ohd \sqrt{\rho/\mathit{We}_0}$.

Hence, for a specific ambient-drop fluid combination (i.e. fixed physical properties), a specific drop diameter, and a specific inflow velocity, a set of $\{ \rho, \Oho, \Ohd, \We_0 \}$ can be obtained that fully characterizes the system. A drop can be simulated in Basilisk to assess whether it undergoes a non-vibrational breakup for the corresponding non-dimensional set. If the drop does not fragment, $\We_0$ is systematically increased (attributable to a decrease in $\sigma$ in non-dimensional space, or an increase in inflow velocity in dimensional space) and the impulsive acceleration simulation is rerun. These steps are repeated until the drop exhibits a non-vibrational breakup. The corresponding minimum $\We_0$ that marks the onset of non-vibrational breakup is the critical Weber number $\Wecr$ for that particular non-dimensional set $\{ \rho, \Oho, \Ohd, \We_0 \}$.
  \subsection{Model Description} \label{subsec:model_description}
The simulations have been performed using the open-source solver Basilisk, which is well-validated for two-phase flows on adaptive Cartesian meshes across a wide range of densities and viscosity ratios \citep{Popinet2003, Popinet2009, marcotte_density_2019}. For this work, we employ its two-phase Navier-Stokes solver. The numerical scheme is detailed in Appendix \ref{app:numerical_scheme}; this section outlines only the assumptions and parameters specific to our problem.

The extensive parametric sweep required by this study (Equation \ref{eq:nondimensionalterms}(c)) makes 3D fragmentation simulations computationally infeasible. We therefore utilize axisymmetric simulations, since the primary deformation process—from initial flattening to the onset of bag inflation—is predominantly axisymmetric. We note that asymmetries develop at later stages ($t^*>1$) due to more prominent instabilities in the interface and ambient flow, making a quantitative point-by-point accurate prediction using axisymmetric simulations difficult. However, the goal of this work is to characterize the drop's deformation pathway and ultimate fate by analyzing its qualitative temporal evolution, such as periods of growth/decay of the drop dimensions and the concavity of the corresponding temporal evolution. We hypothesize that such predictions do not require point-by-point quantitative accuracy. Therefore, axisymmetric simulations are deemed sufficient for this study. The validity of this assumption is assessed in \cref{subsec:validation}. 
\begin{figure}
  \centering \includegraphics[width=0.75\textwidth]{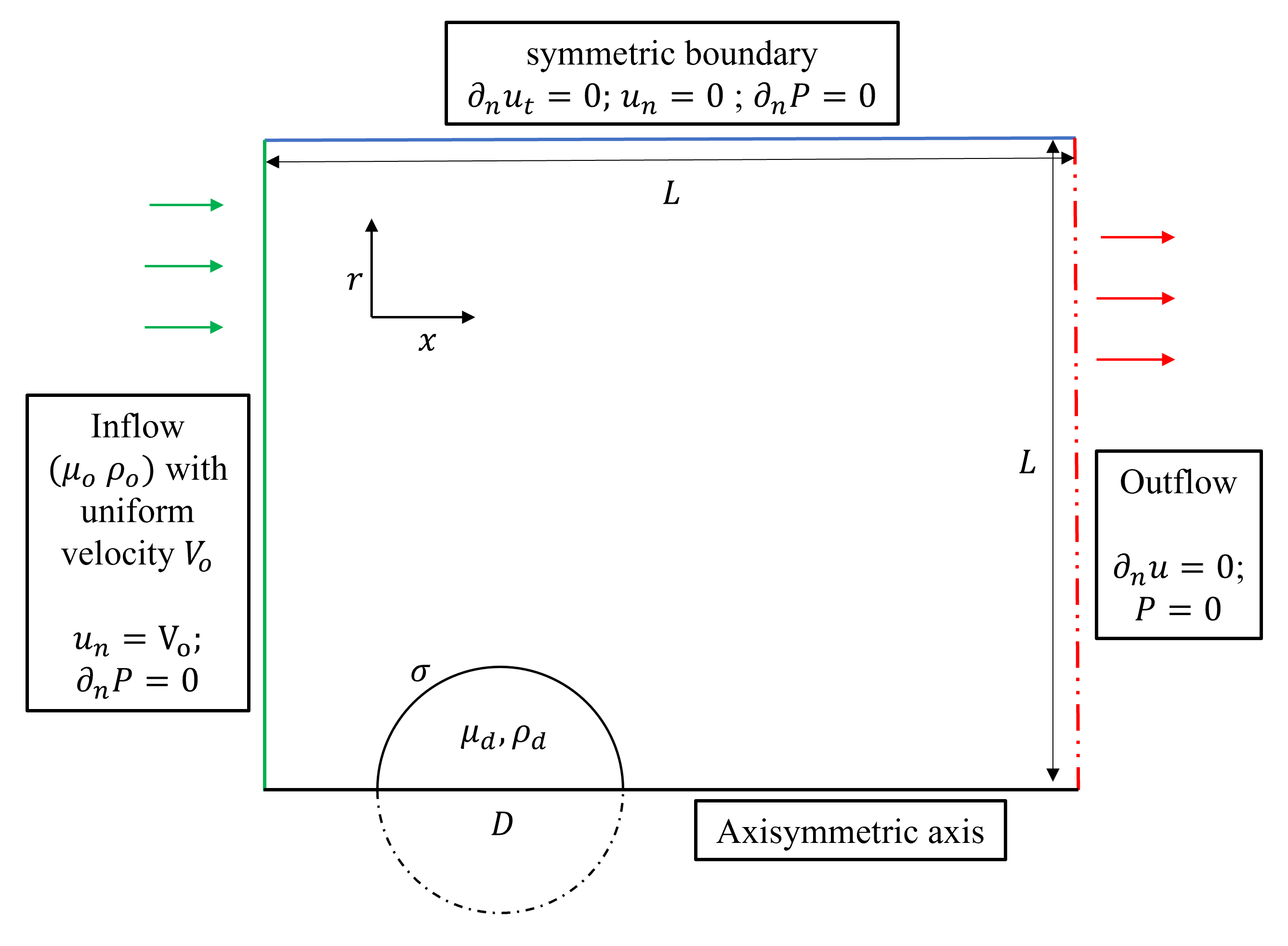}
  \caption{The axisymmetric domain used for all simulations in this work; At $t=0$, the simulation starts with a stationary axisymmetric spherical drop under impulsive acceleration.}
  \label{fig:axi_domain}
\end{figure}

The general simulation domain, used for the simulations in this study, is illustrated in \cref{fig:axi_domain}. It is a square (for compatibility with quadtree meshes) axisymmetric domain of size $L$ and its dimensions are carefully chosen to ensure that the drop always remains a sufficient distance from the boundaries. $L$ can be as small as $16$ for drops with high inertia ($\rho \ge 100$), and as large as $64$ for drops with low inertia ($\rho = 10$). The top boundary is symmetric ($\partial_nP=\partial_nu_t=u_n=0$) and the bottom boundary is the axisymmetry axis. The left boundary allows a uniform ambient fluid inflow into the domain ($u_n = V_0 = 1$), and the right boundary allows the flow to exit the domain freely ($\partial_nu_n=0$). A drop of diameter $D=1$ is initialized with its center on the axisymmetric axis and its initial velocity set to $0$.

To ensure sufficient solution accuracy, we enforce the maximum allowed wavelet errors to $\chi_u = 10^{-4}$ and $\chi_c = 10^{-6}$ for velocity and volume fraction fields, respectively. The maximum allowed residual for the Poisson solve, $\epsilon_p$, is set to $10^{-4}$. The minimum allowed cell size is set to $512$ cells per diameter, which corresponds to $2^{13}$ elements ($N=13$ levels of refinement) for $L=16$, except for cases with $\Oho=0.0001$ which correspond to the highest values of $\Re_0$, for which we use $1024$ cells per diameter ($N=14$ for $L=16$). Test simulations are run with different values for $\chi_c$, $\chi_u$, $\epsilon_e$, and $N$, detailed in \cref{app:convergence}, and the resulting volume fraction fields and streamwise and transverse lengths of the drops are compared. The results show that the chosen values of $\chi_c$, $\chi_u$, $\epsilon_e$, and $N$ are sufficient to ensure that the simulations result in drop shapes that are converged with respect to these parameters.

A significant fraction of the simulations in this study involve high density-ratios ($\rho > 500$). For such high density-ratios, a sharp interface can induce instabilities at the interface due to an unnatural spike in kinetic energy \citep{jain_secondary_2015}. The phenomenon has also been observed in all of our large $\rho$ simulations with low $\Ohd$ ($\Ohd \leq 0.001$), albeit not presented here. In these cases, the upstream face shows unnaturally large surface instabilities, which lead to the removal of micro-droplets from the main drop. To overcome these issues, the interface is smeared by vertex averaging the volume fraction field. This approach helps to reduce density gradients across the interface, preventing its premature breakup. The numerical scheme with this smearing will be validated for a high $\rho$ case in the next section.

At t=0, the ambient fluid is quiescent and the drop is initialized with zero initial velocity. Given the incompressible nature of the flow and an infinite propagation speed of any information across the domain, the end of the $1^{st}$ timestep sees the entire domain attain a flow consistent with the left inflow boundary conditions. This involves the establishment of an incompressible flow around the drop, requiring the velocity field to be solenoidal. In real life, this process occurs in a finite amount of time, dependent on the velocity of the acoustic wave velocity. However in a numerical system, this occurs in a single timestep and leads to a jump in drop center-of-mass velocity, without gaining any corresponding deformation. The magnitude of this velocity jump in its center-of-mass velocity $\delta V_{cm}$ is inversely proportional to $\rho$ \citep{marcotte_density_2019}. Hence, the effective relative velocity at undeformed state of the drop reduces to $V_{\mathit{eff}}=1-\delta V_{cm}$. Accounting for this effective velocity becomes imperative when calculating the associated $\We_0$ of the system. To address this, simulations have been conducted for each pertinent $\rho-\Oho$ pair, resulting in the following observed velocity jumps --- a substantial jump of approximately $0.14$ for $\rho=10$, a jump of $0.030$ for $\rho=50$, and a negligible velocity jump of $0.015$ $(\sim 1.5\%)$  for $\rho=100$. For all simulations with $\rho>100$, this $1^{st}$ timestep jump is deemed insignificant. All non-dimensional parameters in the current work have been corrected to incorporate this jump.
  \subsection{Verification of Axisymmetric drop Simulations} \label{subsec:validation}
To verify the capabilities of Basilisk in modeling high density-ratio drops subjected to impulsive acceleration, we reproduce the Bag breakup experiment as described in \cite{Flock2012}. An ethyl alcohol drop is released from a certain height above an approximately uniform jet of air. The drop undergoes a nearly quiescent free-fall for a height of $188$ mm before entering a jet of air of mean velocity of $10$ m/s and a peak velocity of $15$ m/s. The drop, having acquired some vertical velocity during its fall, has a shape that is close to but not perfectly spherical when it enters the air-jet. The drop then deforms as a result of aerodynamic forces exerted by the air-jet and finally breaks up according to a bag breakup morphology. As the drop enters the jet, it initially experiences aerodynamic forces applied by the boundary layer of the flow, and then moves into the main flow with peak flow velocities.

A simplified axisymmetric version of this experiment is simulated in Basilisk with non-dimensional parameters derived from the dimensional parameters specified in \cite{Flock2012}. The simulation parameters are $V=1$, $\Oho=2.3 \times 10^{-3}$, $\Ohd=5.9652 \times 10^{-3}$, choice of $\We_0$ depends on the choice of air-jet velocity between $10$ and $15$ m/s, $L=16$, and $D=1$. The simulation differs from the experiment in multiple, although minor ways. Firstly, the initial free fall of the drop is omitted since a gravity force perpendicular to the jet direction would render the system non-axisymmetric. Consequently, the slight deformation of the drop just before encountering the air-jet is be captured in the simulation. Secondly, unlike the experiments where the drop passes through a boundary layer of thickness approximately $3$ mm before experiencing the peak $15$ m/s jet velocity, the simulations instantaneously load the drop with the full velocity of the air-jet. Considering that the provided $\We_0=13$ is based on the mean jet velocity, it will be essential to find the $\We_0$ appropriate for our simulation conditions (instantaneous loading), corresponding to velocities between $10$ and $15$ m/s.
\begin{figure}
\centering
\begin{subfigure}{0.48\textwidth}
    \includegraphics[width=\textwidth]{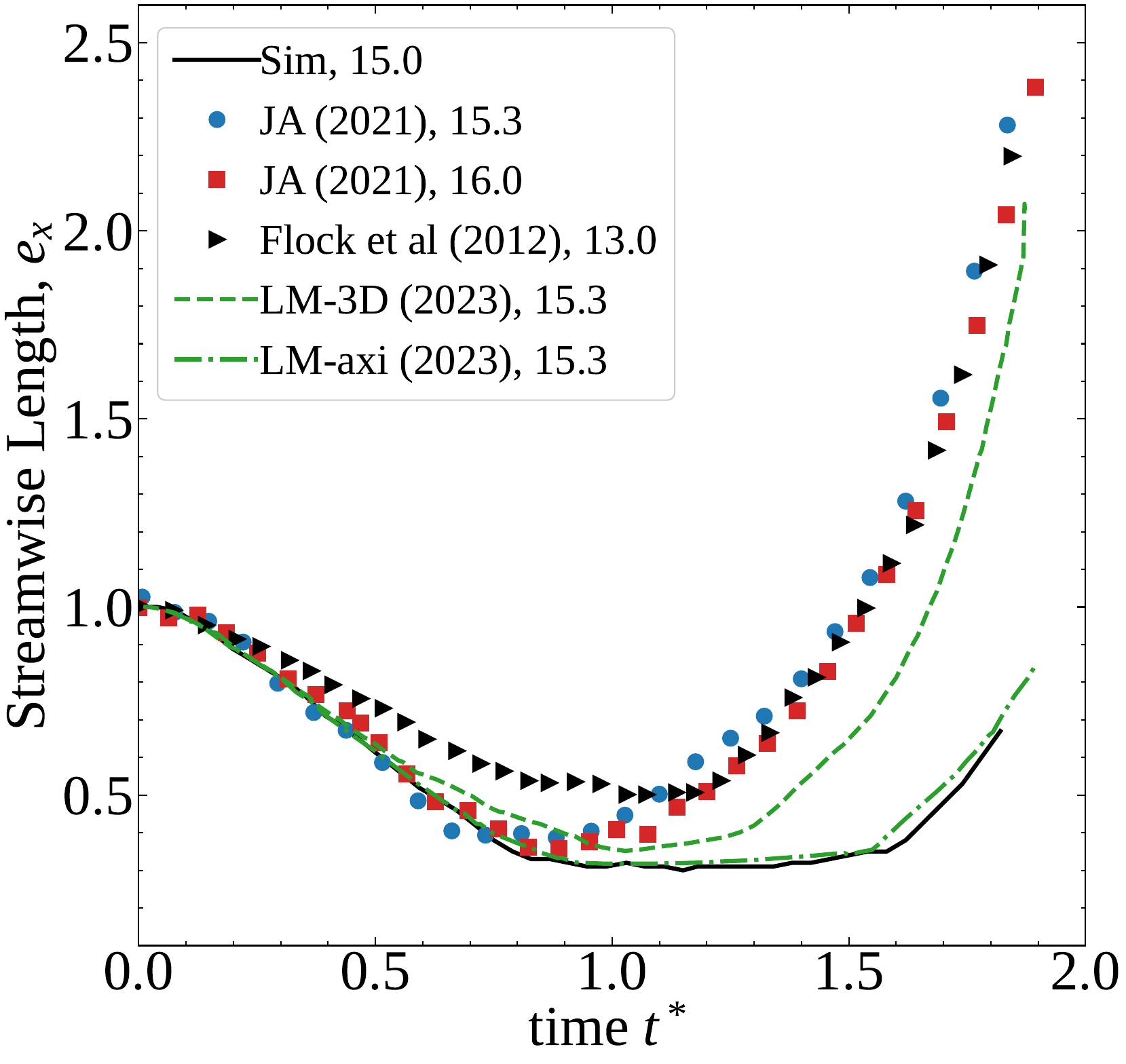}
    \caption{}
    \label{fig:exp_ex}
\end{subfigure}
\hfill
\begin{subfigure}{0.48\textwidth}
    \includegraphics[width=\textwidth]{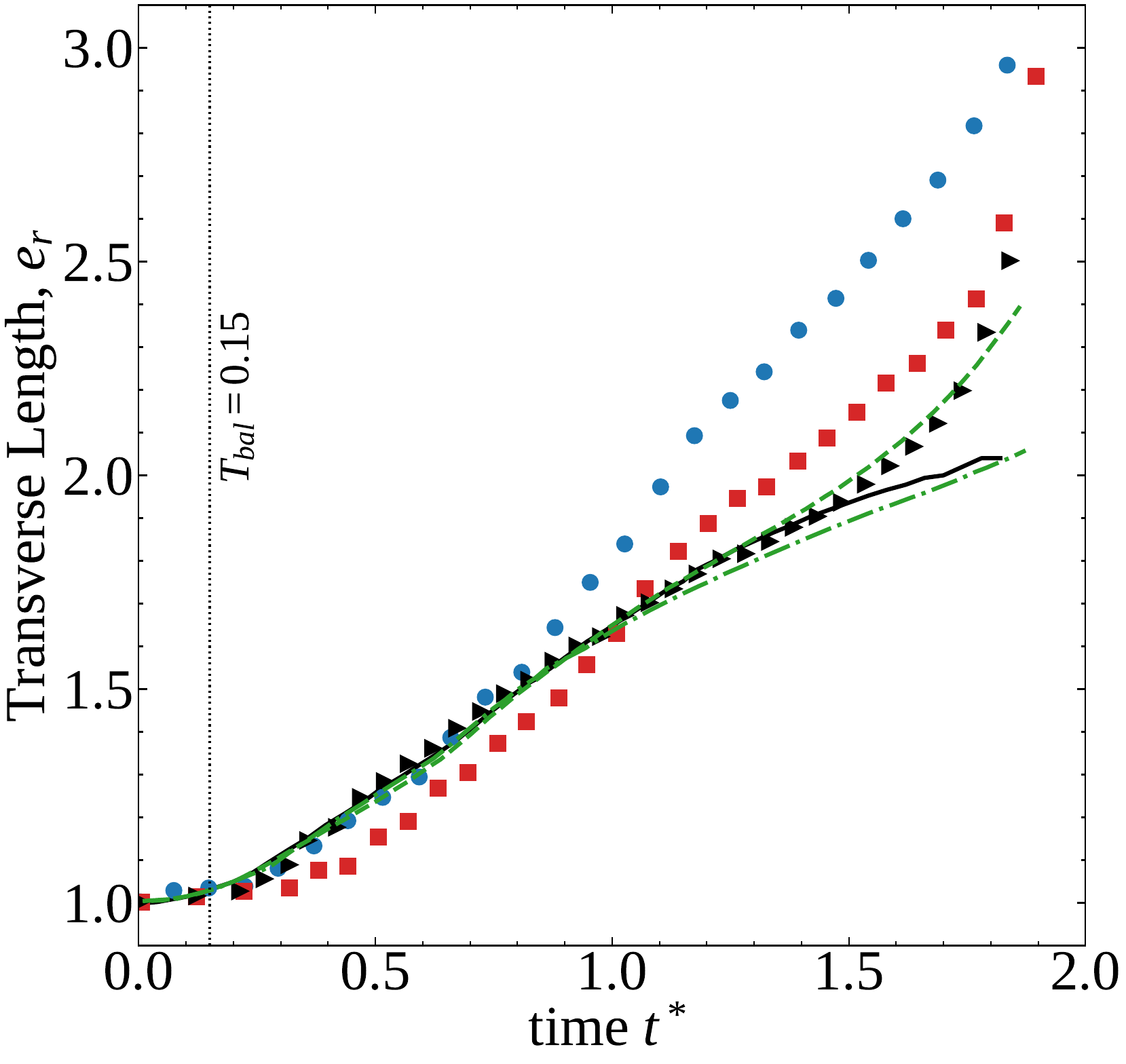}
    \caption{}
    \label{fig:exp_er}
\end{subfigure}
\caption{(a) and (b) compares the streamwise and transverse lengths (see \cref{fig:drop_params} for definitions of $e_x$ and $e_r$) obtained from the simulation to various experiments: JA (2021) \citep{jackiw_aerodynamic_2021}, \cite{Flock2012}, and axisymmetric and 3D simulation results from LM (2023) \citep{lingDetailedNumericalInvestigation2023}.}
\label{fig:exp_comp}
\end{figure}

\Cref{fig:exp_comp} plots the streamwise ($e_x$) and transverse ($e_r$) lengths of the drop as a function of time, obtained through an axisymmetric simulation in Basilisk, and compares them to the experimental and simulation results from some recent works. The streamwise and transverse lengths obtained from the experimental data in \cite{Flock2012} is plotted, along with the experiments performed by \cite{jackiw_aerodynamic_2021}. Lengths $e_x$ and $e_r$ corresponding to axisymmetric and 3D simulations performed by \cite{lingDetailedNumericalInvestigation2023} are also plotted for reference. The experiments performed by \cite{jackiw_aerodynamic_2021} had $\Oho$ and $\Ohd$ values close to \cite{Flock2012}, and hence are a good reference for comparison.

It is observed that the current axisymmetric simulations reasonably capture the streamwise length until $t^*\approx 1.2$ and transverse length until $t^*\approx 1.5$. The prediction of both the parameters worsens beyond this point.
This coincides with the period of rapid streamwise and transverse expansion that follows the initial inflation of the drop into a bag-like structure. This is evident in the temporal evolution of the transverse length, $e_r$, as shown in \cref{fig:exp_er}, which exhibits a rapid growth in the drop size after $t^*\approx 1.5$. The constraint of azimuthal symmetry, inherent to axisymmetric simulations, limits the accurate representation of the intrinsically three-dimensional phenomena associated with bag inflation and subsequent rupture. These phenomena are driven by surface instabilities \citep{lozano_instability_1998,bremond_bursting_2005,Villermaux2007,zhao_breakup_2011}. In addition, axisymmetric simulations tend to overestimate the stagnation pressures at the downstream pole of the drop, which artifically restricts the inflation of the bag \citep{lingDetailedNumericalInvestigation2023}. In contrast,3D simulations do not suffer from such limitations. The 3D simulations performed by \cite{lingDetailedNumericalInvestigation2023} show a substantially better agreement with experiments beyond $t^*\approx 1.2$. However, the current axisymmetric simulations exhibit excellent agreement with the axisymmetric results reported by \cite{lingDetailedNumericalInvestigation2023}, which employed the same boundary conditions and numerical methods. Therefore, the numerical setup employed in this work can be considered comparable to axisymmetric simulations in the recent literature.

It is worthwhile to note that the 3D simulations show a slight overestimation in streamwise deformation compared to the experiments. This manifests as a lower minimum $e_x$ in the 3D simulations compared to the experiments. This $e_x$ value corresponds to the drop's streamwise dimension at the onset of bag expansion, called the initiation time by \cite{jackiw_aerodynamic_2021}. This can be attributed to the difference in initial Weber number $\We_0$ between the experiments ($13$) and the simulations ($13$). This overestimation may also arise from the presence of gravity in the experimental setup which leads to a non-spherical initial shape of the drop when it enters the air-jet.

Even with the observed limitations of axisymmetric simulations in quantitatively predicting the point-by-point evolution of the drop dimensions, the simulations still capture the qualitative evolution of the drop morphology. The almost monotonic decrease in streamwise dimension until reaching an initiation time, and the subsequent growth resulting in a positive concavity, is captured by the axisymmetric simulations. In \cref{fig:exp_ex}, the axisymmetric simulations agree reasonably well with properties such as initiation time, and the constant radial expansion rate after balancing time $T_{bal}$, i.e., time when aerodynamic and surface tension forces balance \citep{jackiw_aerodynamic_2021}. The axisymmetric simulations also succeed in achieving a similar initial pancake shape of the drop, as well as the subsequent appearance of a bag for such high density low viscosity drops \citep{lingDetailedNumericalInvestigation2023}. For the purposes of this work, this is sufficient information to conduct a broad categorical analysis on the sensitivity of features to parameters such as $\Oho$, $\Ohd$ and $\rho$. Thus, the axisymmetric simulations are deemed sufficient for such water-air like systems.

However, in order to justify the use of axisymmetric simulations for the extensive parameter space relevant to this work, we would need to verify the ability of axisymmetric simulations in capturing the pancake shape and subsequent bag formation for a wider range of density ratios and ambient Ohnesorge numbers. This is done in \cref{app:3D_comparison}, where we compare 3D and axisymmetric simulations for a few benchmark cases, with a special focus on non-trivial cases that show forward pancake and forward bag formation. In short, we observe an good agreement between 3D and axisymmetric simulations for all cases except for $\Oho,\Ohd\le 0.001$, including the appearance of forward pancakes for high $\Oho$ cases and forward bags for low density ratio cases. The case with $\{\rho,\Oho,\Ohd\}=\{100,0.0001,0.001\}$ however shows a significant deviation in the post pancake deformation of the drop, with the axisymmetric simulations showing a forward bag formation while the 3D simulations show a backward bag formation. Thus, any conclusions drawn from axisymmetric simulations for such low viscosity systems should be treated with extreme caution. For all other systems, we can be confident that the axisymmetric simulations capture the qualitative evolution of the drop morphology, including the appearance of forward pancakes and forward bags.

Hence, all information pertinent to this work can be reliably obtained through axisymmetric simulations. The axisymmetric simulations are significantly less expensive than 3D simulations, and hence well-suited for the extensive parametric study proposed in this work.

  \subsection{Parameter space explored} \label{subsec:paramter_space}
\begin{table}
  \centering
  \begin{tabular}{ccc}
    Parameters      & Values                     & \multicolumn{1}{l}{Computational cells}                                                                 \\ \hline
    $\rho$          & $10, 50, 100, 500, 1000$   & \multirow{3}{*}{\begin{tabular}[c]{@{}c@{}}Min: $1.75 \times 10^5$\\ Max: $4 \times 10^6$\end{tabular}} \\
    $\mathit{Oh}_o$ & $0.1, 0.01, 0.001, 0.0001$ &                                                                                                         \\
    $\mathit{Oh}_d$ & $0.1, 0.01, 0.001$         &  \\
  \end{tabular}
  \caption{All values of $\rho$, $\Oho$, and $\Ohd$ which form the part of the parametric space to be explored through simulations are listed in this table. In total, 60 sets of $\{ \rho, \Oho, \Ohd \}$ are considered, each is simulated for multiple $\We_0$ values to obtain its $\Wecr$. The minimum and maximum number of cells in the computational domain associated with all simulations is listed in the third column. }
  \label{table:parameter_space}
\end{table}

The goal of this work is to systematically study the influence of the three non-dimensional parameters discussed, namely $\rho$, $\Oho$, and $\Ohd$, on drop deformation and breakup morphology. \Cref{table:parameter_space} summarizes the parameter space explored through simulations in this work, encompassing 60 sets of $\{\rho, \Oho, \Ohd\}$. Each set is simulated for different $\We_0$ values in order to identify its critical Weber number $\Wecr$ and the corresponding critical breakup morphology. This is achieved by simulating each $\{\rho, \Oho, \Ohd\}$ set with multiple $\We_0$ values to determine the lowest $\We_0$ value at which a non-vibrational breakup is observed, in short its $\Wecr$.

Given $Re_0 \propto 1/\Oho$, a range of $Re_0$ in the order of $(10, 10^4)$ corresponds to a $\Oho$ range of $(0.0001, 0.1)$. Any higher $\Re_0$ becomes computationally infeasible due to significant turbulent vortices in the domain leading to a requirement for higher mesh resolution, lower timestep sizes, and even 3D simulations. Hence, this justifies the parameter space specified for $\Oho$.

The influence of $\Ohd$ on $\Wecr$ has been investigated extensively in literature, primarily for systems with large $\rho$ and $\Oho$ values, conditions typical of most experimental setups. For such systems, $\Wecr$ has been observed to be relatively constant (typically within $10<\Wecr<20$) for such systems for $\Ohd<0.1$ \citep{Hsiang1995,guildenbecher_secondary_2009,yang_transitions_2017}. However, a similar comprehensive understanding of the influence of $\Ohd$ is lacking for low $\rho$ or $\Oho$ values. This study aims to address this gap by exploring the fragmentation threshold for the commonly seen values of $\Ohd$ across a parameter space encompassing variations in both $\rho$ and $\Oho$ beyond the typical experimental range. Naturally, to encompass the entire space of low and high density-ratio systems, $\rho$ values are varied from $10$ to $1000$.

The choice of such a comprehensive parameter exploration sets the stage for a detailed investigation into the nuanced interplay of these parameters on drop deformation and breakup characteristics.

\section{Results} \label{sec:results}
\begin{figure}
  \centering \includegraphics[width=1.0\textwidth]{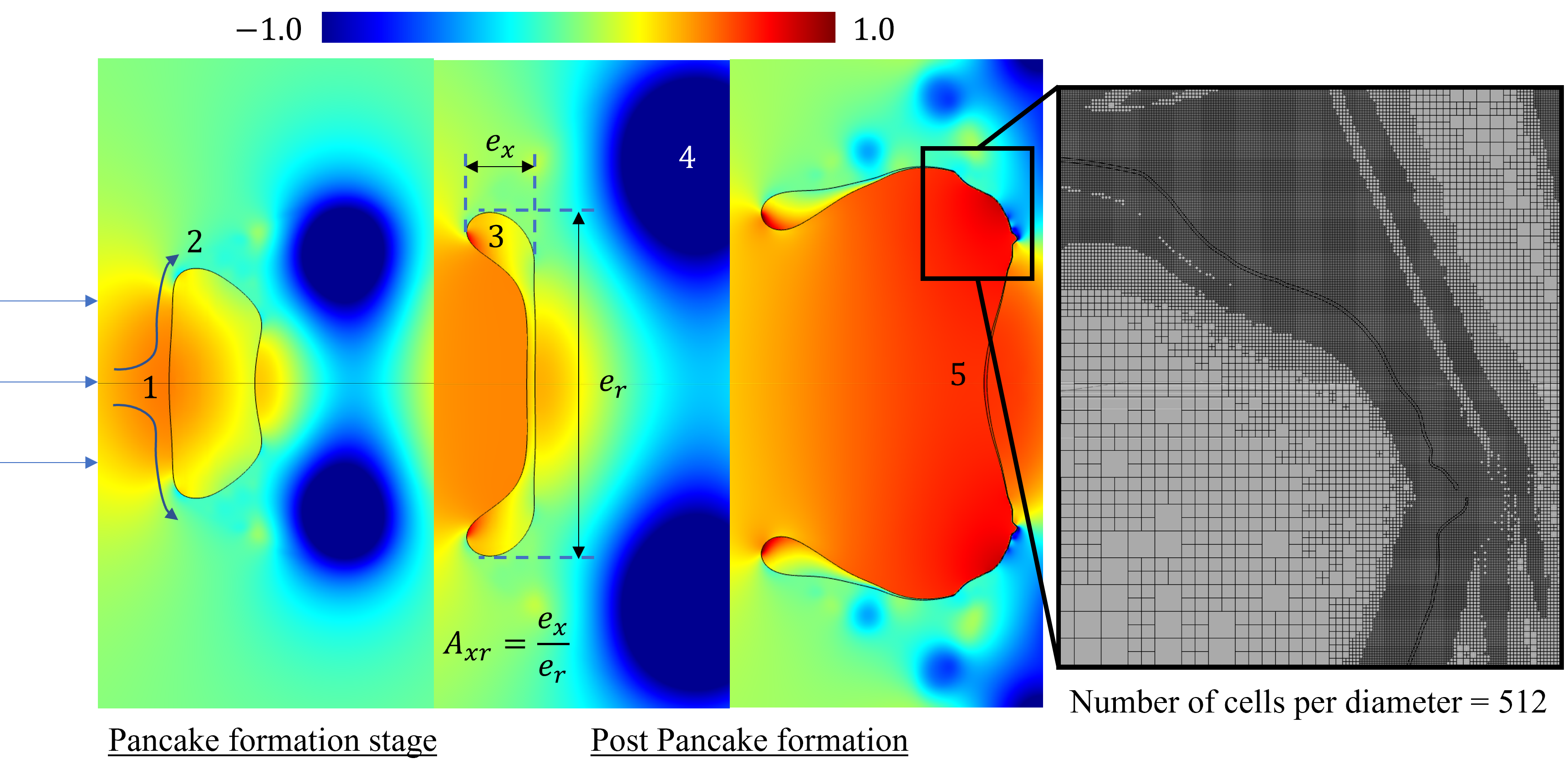}
  \caption{The figure shows the non-dimensional pressure field $P/(\rho_o V_0^2)$ renders for a drop undergoing backward bag fragmentation.
  The points of interest in a deforming drop marked in the figure are:
  \textbf{1.} Upstream pole / Center / Core of the drop;
  \textbf{2.} The periphery of the drop;
  \textbf{3.} The rim of the drop, which in general has a higher local inertia compared to its center;
  \textbf{4.} Downstream low pressure circulation zone, which can affect the motion of its rim if it is attached to the drop surface (unlike in the figure);
  \textbf{5.} The inflated bag, which inflates because of its low inertia and hence higher accelerations. }
  \label{fig:drop_params}
\end{figure}

Figure \ref{fig:drop_params} illustrates a typical drop deformation process, indicating the key features used to define several essential factors. Together, these factors provide an intuitive framework for understanding the deformation dynamics and will be used throughout this study.
\begin{enumerate}
\item The local variation in inertia across the drop, which determines the local accelerations of the constituent parts, e.g., the difference in local inertia between the center (1) and rim (3).
\item The pressure difference between the poles (1) and periphery (2), denoted by $\delP$. $\delP$ is directly proportional to the stagnation pressures observed at the upstream pole of the drop.
\item The surface stresses or viscous forces experienced by the upstream facing surface of the drop. These stresses are a function of the instantaneous Reynolds number $\Re$, which can in many cases approximated as equal to $\Re_0$ if the temporal evolution of center-of-mass velocities and drop deformation do not significantly alter $\Re$.
\item The drop Ohnesorge number $\Ohd$, which dictates the distribution of the total energy supplied by the ambient flow to the drop between surface energy and the fluid momentum gained by the drop.
\end{enumerate}

From the start of the deformation process until the formation of a distinct rim at $t^* \approx 1$, a drop does not exhibit any appreciable variation in local inertia along the lateral dimension. Hence, during this initial deformation phase, local inertia differences do not significantly influence the initial deformation; instead, the interplay between pressure and shear forces predominantly governs the process. Acceleration and hence the increase in velocity of the center-of-mass is inversely proportional to total inertia and directly affects instantaneous Reynolds number $\Re$ of the ambient flow past the drop. While $\Re$ dictates the shear stresses on the upstream surface, the relative velocity of the drop dictates the stagnation pressures at the upstream pole and hence $\delP$. The resulting pancake shape depends on the comparative strengths of the pressure difference and shear forces.

Once a drop develops local inertia variations across the lateral dimension as it deforms past the pancake stage, any subsequent deformation becomes strongly dependent on these variations in local accelerations. For the same external forces, regions of the drop with larger inertia experience much lower accelerations, and hence lag behind their lower inertia counterparts.

Reynolds number of the ambient flow past the drop dictates the strength, timescales, lengthscales, and location of the downstream vortices \citep{forouzi_feshalami_review_2022}. Thus, it is essential that we consider the interaction of these vortices with the rim for different Reynolds number values to understand the final shape. The sensitivity of the rim to these flow characteristics is decided almost solely by inertia relative to the ambient fluid, i.e., $\rho$. A large density-ratio drop is expected to exhibit very little sensitivity to downstream vortices, and vice verca.

If we consider the specific drop case shown in \cref{fig:drop_params}, the ratio of spatial extent of the drop along the axisymmetric axis to the spatial extent along the $r$-axis provides its aspect ratio $A_{xr}$. This parameter will be used in future sections to quantify the deformation shown by the drops for the parameter space. In the first image (from left to right), we observe a flat pancake, which occurs when $\delP$ predominantly drives the internal flow in the drop (over shear stresses). We also observe a clear toroidal rim (image 2) that has a large local inertia and is therefore expected to lag the lower inertia center of the drop. Due to the large inertia, the rim remains unaffected by the low-pressure zone created by the downstream vortex, which sheds a sufficient distance away from the rim and is not attached to the drop. Ultimately, the drop deforms into a backward bag morphology, where the center inflates into a bag under the action of pressure forces at the stagnation point.
  \subsection{Density-ratio} \label{subsec:rho}
\begin{figure}
	\centering
	\includegraphics[width=1\textwidth]{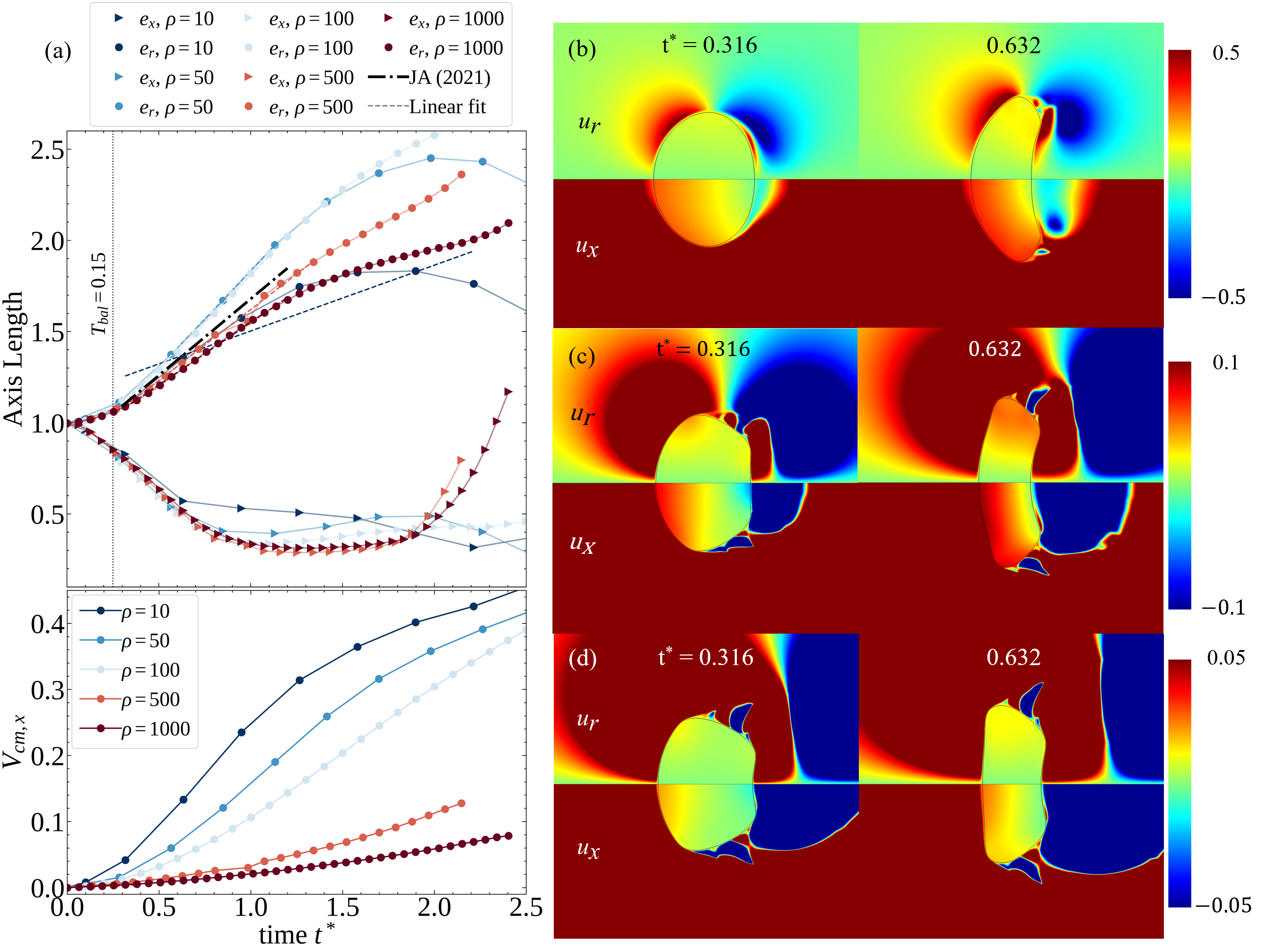}
	\caption{(a) shows the temporal variation of (streamwise $e_x$ and transverse $e_r$) axis lengths, and x-component of center-of-mass (cm) velocity of drops with different $\rho$ values.
	The analytical relationship for $\dot e_r$ as given by \cite{jackiw_aerodynamic_2021} is plotted for reference (see line labeled ``JA(2021)'').
	Dashed lines represent linear fit lines of $e_r$ from $t^*=0.3$ to $t^*=1.2$.
	Internal velocity fields for $\rho=10$, $\rho=100$, and $\rho=1000$ are plotted in (b), (c), and (d) respectively.
	The upper half shows r-velocities ($u_r$), whereas the lower half of each plot shows x-velocities ($u_x$).
	All drop systems presented have $\Oho=0.001$, $\Ohd=0.1$, $\We_0=20$.}
	\label{fig:rho_vel}
\end{figure}

This section illustrates the role of density-ratio in influencing drop deformation and breakup morphology. \Cref{fig:rho_vel} illustrates the variation of the streamwise ($e_x$) and transverse ($e_r$) lengths of the drop and x-component of center-of-mass (cm) velocities ($V_{cm,x}$), in the top and bottom of the plot, respectively. Density-ratios from $10$ to $1000$ are shown. All cases show a decrease in their transverse length until $e_x$ reaches a minima. We define this instant at which the drop achieves its minimum streamwise length as its pancake state. If the streamwise length of the pancake reaches a critical minimum, the drop unstably loses fluid from its core to its periphery, resulting in the formation of a thin fluid sheet near the core and a toroidal rim near the periphery, which eventually ruptures. The cases with high density-ratio, i.e., $\rho = 100,500,1000$, achieve their minimum $e_x$ at $t^* \approx 1$, which coincides with the initiation time as described by \cite{jackiw_aerodynamic_2021}, i.e., the time of initiation of inflation of bag. \cite{jackiw_aerodynamic_2021} suggested an initiation time of approximately $1$ based on their experiments of a water drop in an air-jet, which is consistent with the results presented here. All the high $\rho$ cases also achieve a minimum $e_x$ at $t^* \approx 1$. In addition, the minimum $e_x$, which can alternatively be interpreted as a maximum rim curvature, is nearly identical. Since all cases share the same surface tension coefficient (for the same Weber number), a likely hypothesis is that the drops deform streamwise until their surface tension forces, which are directly proportional to rim curvature, reach a similar magnitude. The rim curvature can be closely approximated by the reciprocal of the thickness of the flattened "pancake" shape \citep{Villermaux2009,Kulkarni2014}. Thus, for similar aerodynamic forces, the rim curvature is expected to be similar for two drops with the same surface tension coefficient.

In contrast to the high $\rho$ cases, the low $\rho$ cases ($\rho=10,50$) do not fragment for a initial Weber number of $\We_0 = 20$. This aligns with the fact that the drops for the two cases do not achieve a minimum $e_x$ as low as the high $\rho$ cases. Instead of unstably losing fluid from the core to the periphery, these drops begin retracting back towards a spherical shape after reaching their maximum deformation or pancake state. This hints at the lower aerodynamic forces experienced by the low $\rho$ cases, which dictates the magnitude of the corresponding surface tension forces required to balance the aerodynamic forces. This hypothesis is supported by examining $V_{cm,x}$ of the lowest $\rho$ drops, which rapidly lose their relative velocity with the ambient, and hence experience lower aerodynamic forces at later times.

It is also observed that the temporal development of transverse length $e_r$ very closely follows a linear trend after reaching the balancing time $T_{bal}$, which is consistent with the analytical relationship for $\dot e_r$ as given by \cite{jackiw_aerodynamic_2021}.
The analytical relationship for $\dot e_r$ is given by
\begin{equation}
  \label{eq:dot_er_JA2021}
  \dot e_r = \left( \dfrac{a}{2} \right)^2 \left( 1 - \dfrac{128}{a^2 \We} \right) T_{bal},
\end{equation}
where $a=6$ and $T_{bal} = 1/8$ as specified by \cite{jackiw_aerodynamic_2021}. The analytical relationship for $\dot e_r$ is plotted in \cref{fig:rho_vel}(a) as a dash-dotted line labeled ``JA(2021)'' for reference. The linear fit lines for $e_r$, from $t^*=0.3$ to the specific initiation time for each case, are plotted as dashed lines, and they show a good agreement with \ref{eq:dot_er_JA2021}.
\begin{figure}
	\centering
	\includegraphics[width=1\textwidth]{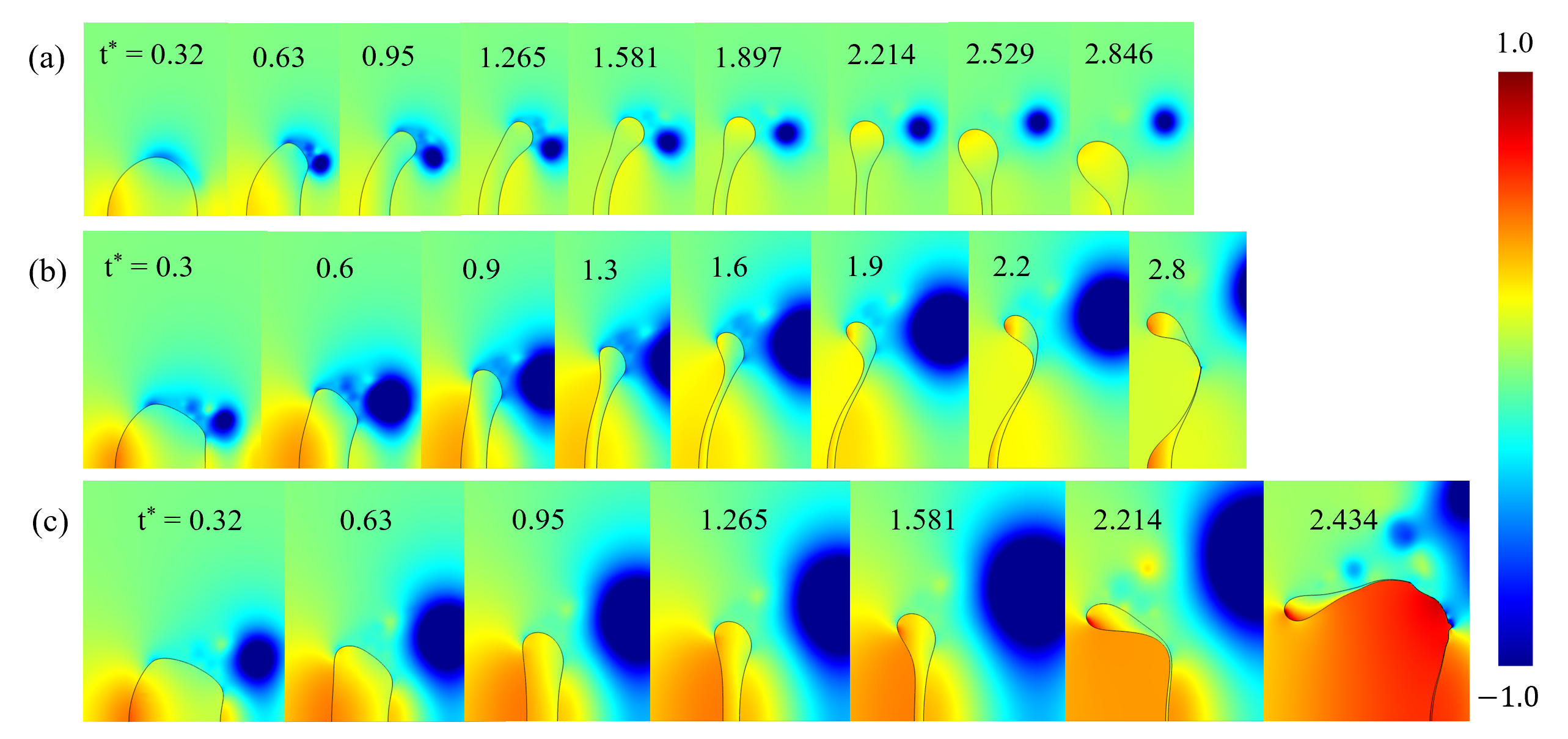}
	\caption{Pressure field plots for drops with three different density-ratios are plotted (a) $\rho=10$ (b) $\rho=100$ (c) $\rho=1000$ for a low $\Oho$ system. All drops referred here have the following common parameters: $\Oho=0.001$, $\Ohd=0.1$, $\We_0=20$.}
	\label{fig:rho_P_lowoho}
\end{figure}

\Cref{fig:rho_vel}(b), (c), (d) and \ref{fig:rho_P_lowoho}(a), (b), and (c), compares the internal velocity and pressure fields for drops of different density-ratios, for $\Oho = 0.001$ ($\Re_0 \approx 4472$), respectively. Such high $\Re_0$ results in low shear stresses acting on the drop's upstream surface, and almost the entirety of aerodynamic forcing is exerted through pressure forces. Until the formation of a distinct toroidal rim at $t\approx 1$, local inertia variations across the lateral dimension of the drop remain small and therefore do not significantly influence the corresponding local accelerations experienced by each region of the drop. Hence, during the pancake formation stage, the deformation is primarily dependent on the balance between the shear and pressure forces applied by the ambient medium.

Low $\rho$ drops, having lower total inertias, experience larger centroid accelerations. This leads to lower relative velocities, instantaneous Reynolds numbers ($\Re$), and stagnation pressures compared to high $\rho$ drops under identical conditions. Thus, low $\rho$ drops show a higher local and volume-averaged responsiveness to external forces (such as attached vortical structures). Consequently, the deformation process becomes highly sensitive to $\rho$.

For instance, pressure field for three different $\rho$ values is presented in \cref{fig:rho_P_lowoho}, and shows the lowest driving pressure forces ($\delP$) for $\rho=10$ case, allowing even the low shear stresses (due to high $\Re_0$) acting at the upstream surface of the drop to be significant contributor to the initial deformation and internal flow of the drop. As $\rho$ increases, the stagnation pressure (and hence $\delP$) grows, making the contribution of shear stresses increasingly irrelevant. Thus, for the $\rho=1000$ drop, the internal flow is expected to be completely driven by the pressure forces. The internal flow plots shown in \cref{fig:rho_vel} confirm this behavior. The internal flow for the $\rho=10$ drop is highest near the upstream surface and decreases to nearly zero at the downstream pole (see $u_x$), with the corresponding velocity gradient pointing in a direction normal to the upstream surface. The internal flow for $\rho=1000$ (shown in (d)) on the contrary, has the highest x-velocities at the upstream pole, and not at the periphery.All cases also show negligible lateral inertia differences across the drop before $t^*<1$. Thus, for the $\rho=10$ case, the high shear at the periphery leads to a greater local acceleration there than at the center. We also note that the the lower relative velocity of the low-$\rho$ case with the ambient leads to even lower $\Re$ values, resulting in a vortex that is not fully detached from its periphery, increasing the induced stresses. This creates a differential acceleration between the pole and the periphery for the $\rho=10$ case, resulting in a forward-facing pancake. The $\rho=1000$ case on the other hand forms a flat pancake. Thus, the competition between shear stresses and $\delP$ controls the orientation of the pancake.

Beyond the pancake stage (at $t^* \approx 1$), a prominent toroidal rim forms in all cases (\cref{fig:rho_P_lowoho}). This concentrates mass at the periphery, increasing the rim's local inertia relative to the drop's thinning core. The subsequent evolution is dictated by the interplay between this variation in local inertia, aerodynamic forces, and the downstream vortex structure, which varies significantly with $\rho$. However, for all three cases, the rim eventually gains enough mass to decelerate relative to the center of mass, forming a bag-like structure of backward orientation. The rate of evacuation of the core fluid towards the periphery and the lateral stretching of the pancake is significantly different for the three cases, leading to different bag morphologies.

The $\rho=10$ drop experiences the smallest pressure forces driving its internal flow, and hence its core experiences the smallest rates of evacuation. It hence takes longer for a prominent rim to form, coinciding with a delayed flipping of the thinned pancake from forward to backward. The downstream vortex shed from the periphery is also the weakest for this case and cannot induce any significant stretching of the rim. The end result is substantially less deformation and no fragmentation in this case. Conversely, the $\rho=1000$ drop experiences the largest $\delP$ driving its internal flow, and hence the rate of evacuation of the core fluid is very high. The high inertia makes the rim fairly insensitive to downstream vortices and large $\Re$ leads to the vortex detaching from the periphery early and cleanly. It is only the intermediate $\rho=100$ case which shows a sufficiently strong downstream vortex which sheds close and produce larger lateral stretching rates, reflected in the larger transverse internal velocities $u_r$ compared to the other two cases (\cref{fig:rho_vel}). In fact, only an intermediate $\rho$ allows large enough $\Re$ to generate a strong downstream vortex, and yet low enough inertia for the rim to be sensitive to such forces. This stronger stretching coupled with the intermediate rates of evacuation of the core results in a backward bag with a core that has not completely evacuated, leading to a bag-plume morphology. A similar explanation for the formation of a plume is provided by \cite{jackiw_aerodynamic_2021}, where a faster bag inflation rate (due to higher $\We_0$ in the paper) compared to the movement of drop fluid from the center to rim leads to the presence of an undeformed core at the center of the drop. The volume of fluid contained in this undeformed core dictates the specific breakup observed --- bag-plume, multi-bag or sheet thinning.  This deformation process is also shown in \cite{marcotte_density_2019} (in figure 4) for a low $\Oho$ case where a variation in $\rho$ from $10$ to $2000$ is accompanied with a shift in pancake and breakup morphology exactly as observed here.
\begin{figure}
    \centering
    \includegraphics[width=1\textwidth]{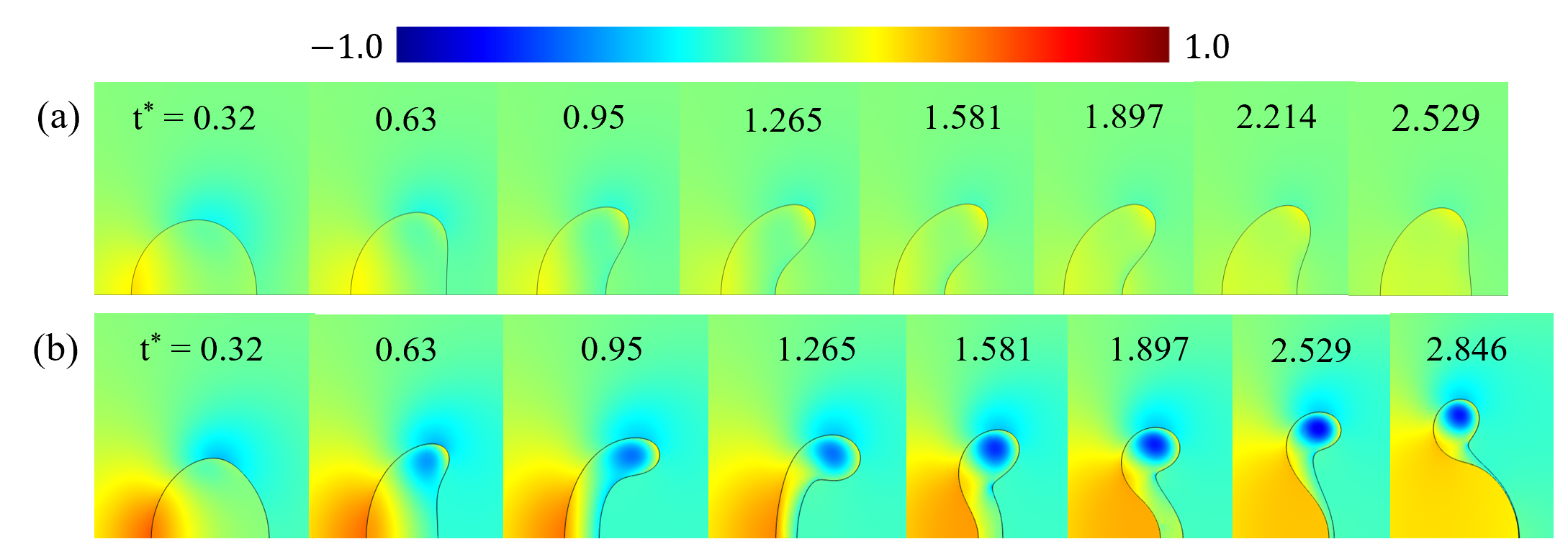}
    \caption{Pressure field plots are plotted for two different $\rho$ values for a high $\Oho$ ambient flow. (a)$\rho=10$ (b)$\rho=1000$.
    All drops shown here have the following common parameters: $\Oho=0.1$, $\Ohd=0.1$, $\We_0=20$. }
    \label{fig:rho_P_highoho}
\end{figure}

Let us now shift our attention to the effect of $\rho$ on high $\Oho$ cases. \Cref{fig:rho_P_highoho} shows the evolution of pressure field around the drops with time. In both cases, $\Oho$ is $0.1$, which corresponds to $\Re_0=44.72$. For such a low $\Re_0$, we expect the shear stresses on the upstream surface to be substantial. Hence, despite observing substantially higher upstream stagnation pressures (and hence higher $\delP$) for (b) ($\rho=1000$) compared to (a) ($\rho=10$), $\delP$ still does not dominate over the shear stresses during pancake formation. As expected, we see a forward pancake at $t^*=0.948$ for both cases. Additionally, because of the low $\Re_0$ of the flow, the ambient flow remains attached to the drop's surface, eliminating the formation of any downstream circulation zones. Subsequently, as the drop core is evacuated and a distinct rim is formed, a backward bag remains the only possible morphology. (b) shows a backward bag breakup, while (a) shows much smaller deformations and does not fragment, the lower deformation is consistent with lower relative velocities, resulting in lower external forces.

In conclusion, the morphology of the pancake depends on the competition between the pressure difference between the poles and the periphery of the drop, with the shear stresses acting on the upstream surface. A flat pancake is observed when $\delP$ is dominant, whereas a forward-facing pancake is observed when shear stresses are dominant. As the drop deforms past the pancake stage, it forms a bag, which can be forward or backward, depending on the local inertia of the rim and the strength of the downstream vortex. Local inertia depends on $\rho$ and the rate of evacuation of fluid from the drop core. The strength and location of downstream vortices, on the other hand, depend on $\Oho$ and the instantaneous velocity of the drop, which again depends on inertia $\rho$. Finally, under conditions where the drop experiences a higher rate of lateral stretching (dependent on $\Re$) compared to the rate of evacuation of the core (dependent on $\delP$), we may also observe a plume.
  \subsection{Ambient Ohnesorge number} \label{subsec:oho}
\begin{figure}
  \centering \includegraphics[width=1\textwidth]{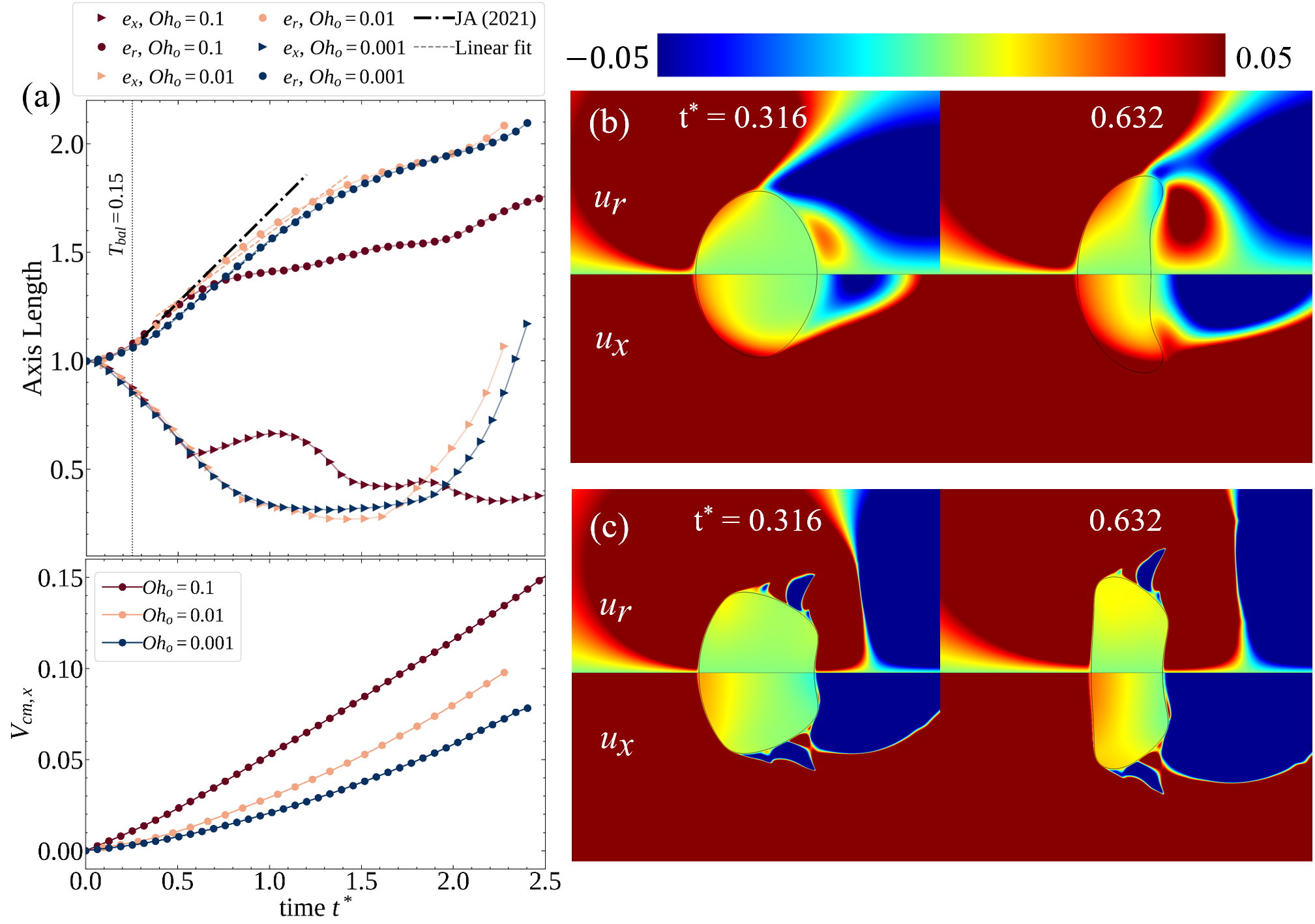}
  \caption{For all plots, $\rho=1000$, $\Ohd=0.1$ and $\We_0=20$. (a) shows the temporal evolution of the (streamwise $e_x$ and transverse $e_r$) axis lengths for different $\Oho$ values.
	The analytical relationship for $\dot e_r$ as given by \cite{jackiw_aerodynamic_2021} is plotted for reference (see line labeled ``JA(2021)'').
	Dashed lines represent linear fit lines of $e_r$ from $t^*=0.3$ to $t^*=1.2$.
  Velocity fields are plotted for $\Oho=0.1$ and $\Oho=0.001$ in (b) and (c) respectively.}
  \label{fig:oho_vel}
\end{figure}

\begin{figure}
  \centering
  \includegraphics[width=1\textwidth]{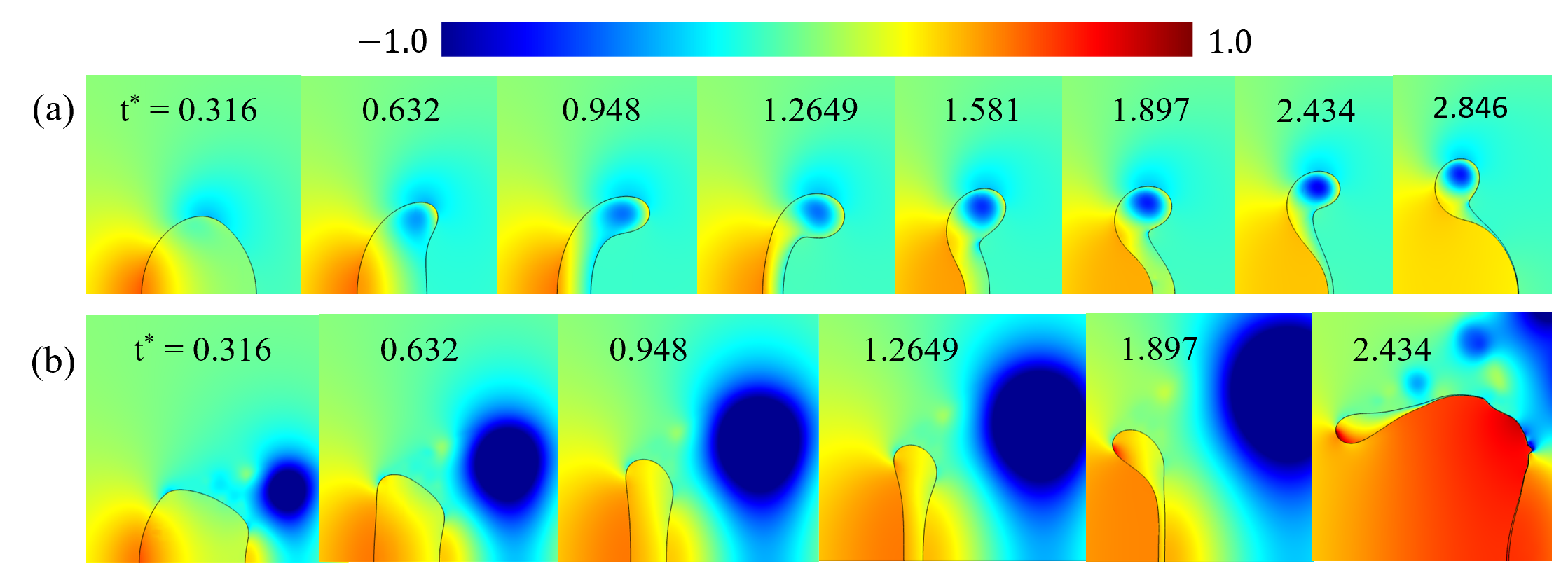}
  \caption{Both the drops shown here have $\rho=1000$, $\Ohd=0.1$ and $\We_0=20$. Pressure fields are shown for drops with (a)$\Oho=0.1$, and (b)$\Oho=0.001$.}
  \label{fig:oho_P}
\end{figure}

\Cref{fig:oho_vel}(a) illustrates the temporal evolution of drop streamwise ($e_x$) and transverse ($e_r$) lengths and x-component of center-of-mass velocities ($V_{cm,x}$), spanning $\Oho$ values from $0.1$ ($\Re_0=44.72$) to $0.001$ ($\Re_0=4472$) for a high density-ratio system. The discussions on the temporal evolution of $e_x$ and $e_r$ made in \cref{subsec:rho} carry ideas that can be utilized to interpret the results shown in \cref{fig:oho_vel}(a). The two lower $\Oho$ cases show a decrease in $e_x$ to idential minimum value, achieved at the initiation time. This decrease in $e_x$ is accompanied by a linear increase in $e_r$ after $T_{bal}$, the rate of increase very similar to the analytical relationship (\ref{eq:dot_er_JA2021}) for $\dot e_r$ as given by \cite{jackiw_aerodynamic_2021}. However, in the case of $\Oho=0.1$, the streamwise length shows two distinct minima. The first minima occurs when the drop starts forming a forward pancake at $t^* \approx 0.6$, as seen in \cref{fig:oho_vel}(b) and \cref{fig:oho_P}(a). At $t^* \approx 1.2$, the forward pancake starts flipping as the drop forms a backward bag, leading to a second minima in $e_x$ right at the onset of bag inflation, which coincides with the initiation time as defined by \cite{jackiw_aerodynamic_2021}. The second minima reached by the drop in the $\Oho=0.1$ case is higher than the minima reached by the other two cases, which hints at the lower rim curvatures achieved by the drop before bag inflation.
It is also observed that this case follows the analytical linear increase in $e_r$ only upto the first minima, after which the rate of increase in $e_r$ decreases.
Thi is expected as the analytical relationship was derived for experimental systems which typically form flat pancakes for high $\Re$ systems.

The drop with the highest value of $\Oho$ experiences the largest drag forces given the same relative velocity, the majority of which is imparted by the shear stresses acting on its surface, owing to the circular symmetry of pressure field around a cylinder in low $\Re$ flows. This case thus shows the largest center-of-mass acceleration, resulting in the highest $V_{cm,x}$ throughout the deformation process. The lower relative velocities of this case result in lower stagnation pressures and thus lower $\delP$ values. This hints at the cause for the lower rim curvatures achieved by the drop before bag inflation.

Pressure fields for two different $\Oho$ values for $\rho=1000$ have been plotted in \cref{fig:oho_P}. All non-dimensional parameters except $\Oho$ are the same for the two cases. For the drop in \cref{fig:oho_P}(a), $\Oho = 0.1$, i.e., $\Re_0$ is very low which corresponds to a large outside viscosity. Thus, the flow does not detach from the drop surface and leads to large viscous stresses on the upstream surface and consequently larger center-of-mass velocities. The drop in \cref{fig:oho_P}(b), on the other hand, has a $\Re_0$ value that is 100 times larger, leading to much smaller shear stresses and consequently smaller center-of-mass velocities. The larger velocities of the drop in case (a) leads to smaller stagnation pressures and, consequently, smaller $\delP$ compared to case (b). It should be noted that the instantaenous Reynolds number $\Re$ (instead of $\Re_0$) would be a more accurate descriptor of the effective shear stresses experienced by a drop. However, this discrepancy between $\Re$ and $\Re_0$ (resulting from non-zero center-of-mass velocities and frontal radius growths) plays a minor role in influencing shear stresses compared to the two orders of magnitude change in $\Re_0$ between the two cases. For the drop shown in \cref{fig:oho_P}(a), owing to lower stagnation pressures, the shear stresses acting on its upstream surface dictate its initial internal flow and resulting deformation. This is clearly demonstrated by the internal flow plot shown in \cref{fig:oho_vel}(b) for the drop, where velocities are the highest at its upstream surface and decreases to zero at its downstream pole, coinciding with the location of largest shear stresses applied by the ambient flow. This dominance of shear stresses results in the formation of a forward facing pancake.

Conversely, for the higher $\Re$ system depicted in \cref{fig:oho_P}(b), shear stresses are significantly lower, which coupled with the larger $\delP$ (compared to (a)) results in a pressure dominated internal flow. Consequently, the drop deforms into a flat pancake and the highest internal velocities occur at its upstream pole rather than at the periphery.

In summary, the orientation of the pancake, determined by the competition between $\delP$ and shear stresses, is influenced by both $\rho$, due to its significant impact on $\delP$, and $\Oho$, due to its significant impact on the $\Re$, and consequently, the shear stresses exerted on the drop.

As the drops deform further, both cases develop a prominent rim, resulting in a large disparity in local inertia between the rim and the center (prospective bag) of the drop. In case (a), the extremely low $\Re$ results in an attached flow downstream of the drop, preventing the formation of a downstream vortex. In contrast, case (b) with its large $\Re$ develops a downstream vortex, but the large local inertia (i.e., $\rho$) allows the vortex to detach early from the drop. Hence for both cases, the rim does not experience any additional forces that can counteract the impact of the large local inertia (smaller local acceleration) of its rim. We observe that the drop in case (a) flips  orientation from a forward pancake to a backward bag, while the drop in case (b) deforms from a flat pancake to a backward bag.
\begin{figure}
  \centering
  \includegraphics[width=1\textwidth]{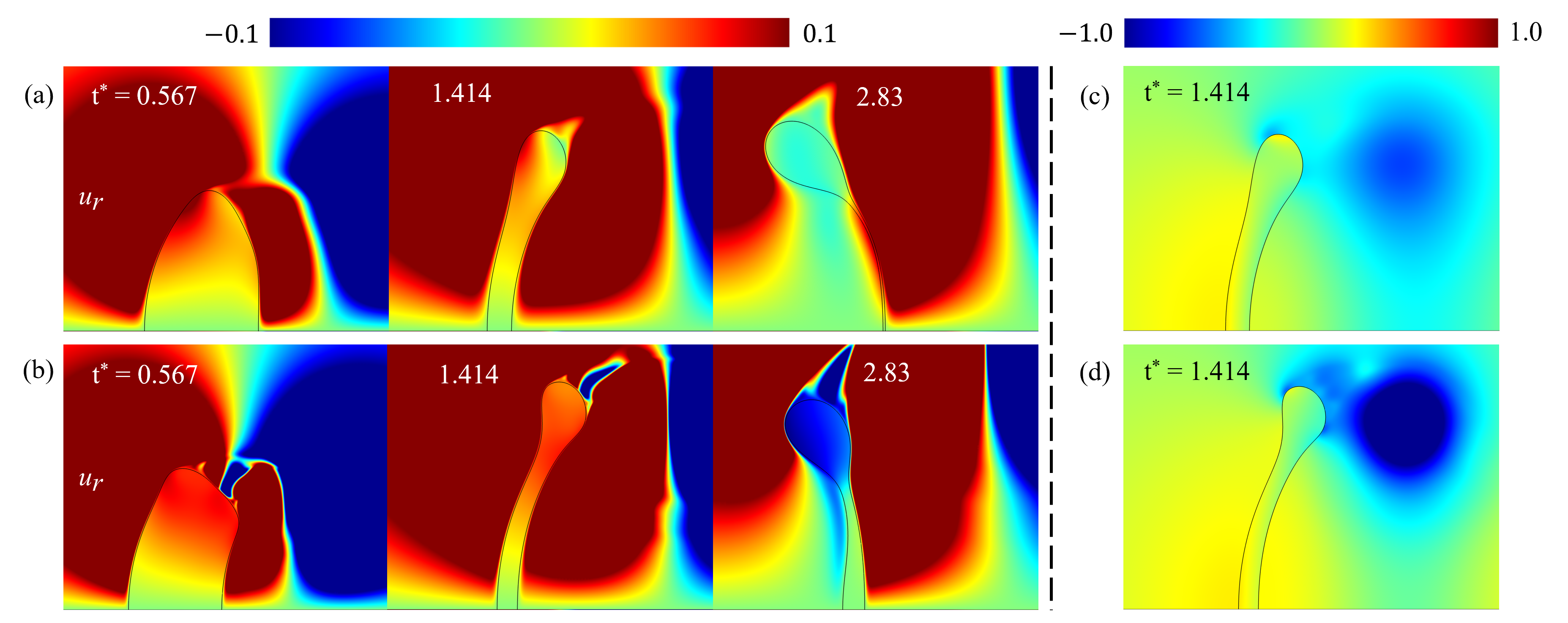}
  \caption{Both cases shown here have $\rho=50$, $\Ohd=0.1$ and $\We_0=20$. (a) and (c) plot y-velocities and pressure field for drop with $\Oho=0.01$; whereas (b) and (d) plot y-velocities and pressure field for drop with $\Oho=0.001$.}
  \label{fig:oho_plume}
\end{figure}

It is worthwhile to note that decreasing $\Oho$ motivates the formation of a plume. The reason for this can be discerned from the velocity field plots shown in \cref{fig:oho_plume}(a) and (b). The lower $\Oho$ case develops a stronger downstream vortex due to its larger $\Re_0$, as is evident from the larger r-velocites at its rim. The common mediocre density-ratio of $\rho=50$ renders the toroidal rims of the drops more susceptible to lateral stretching due to their interaction with the downstream vortices. However, the induced drag on the lower $\Oho$ case is larger leading to comparatively higher rates of stretching. Ultimately, the larger $\Oho$ case fragments with a simple backward bag breakup morphology. The lower $\Oho$ case instead develops a plume which leads to the formation of an annular bag between the center and the periphery, i.e., backward bag-plume morphology.

A case with much lower $\Oho$ ($<0.001$) would have a Reynolds number firmly in the free shear regime \citep{forouzi_feshalami_review_2022}, producing smaller, stronger, and faster shedding vortices that form much closer to the drop periphery. Consequently, the drop in the higher $\Re_0$ case would experience even higher stretching rates from the poles to the periphery, as it is subjected to stronger induced drag forces due to better access to the low pressure zones downstream. This scenario when coupled with a low density ratio drop could prevent the forward pancake from ever flipping, leading to a forward bag formation. This is indeed observed in the threshold fragmentation of the low $\rho$ cases with $\Oho=\{0.001,0.0001\}$ and $\Ohd=0.1$, where the drop forms a forward bag without ever flipping its orientation. The pressure field for these cases is shown in \cref{app:3D_comparison}, and its validity justified through a comparison with analogous 3D simulations.

In short, the rate of stretching, and consequently the size of the resulting plume, increases with increasing proximity and strength of downstream vortices on the rim (due to increase in $\Re_0$). For low $\We_0$ simulations, forward bags are only observed for smaller inertia drops when $\Oho$ values are small, with $\Re$ preferably in the shear layer instability regime ($1000\leq\Re_0\leq10^5$).
  \subsection{Drop Ohnesorge Number} \label{subsec:Ohd}

We start our investigation into the role of drop Ohnesorge number by studying cases with high density ratios and low ambient Ohnesorge numbers. In such drop-ambient systems, drops due to their large inertia maintain high relative velocities with ambient, yielding large stagnation pressures ($\delP$) and large instantaneous Reynolds numbers ($\Re$). The low shear stresses acting on its upstream surface (due to high $\Re$) allow $\delP$ to dictate its internal flow, resulting in the formation of a flat pancake. Given that all other nondimensional parameters are the same across two cases, the case with the larger $\Ohd$ has a larger drop viscosity $\mu_d$, and is expected to have lower internal flow velocities and circulations. Higher $\mu_d$ also results in an exponential decrease in the incidence of surface instabilities \citep{Fuster2009}. While an increase in surface tension decreases the wavelength of the fastest growing surface waves, an increase in $\mu_d$ increases the length and timescales for which a capillary wave generated by an instability remains intact. As a drop accelerates and its relative velocity with ambient decreases, the effective acceleration of the drop relative to the ambient medium decreases. Consequently, a surface wave that might have been unstable at the start of deformation, might be stable at more advanced stages of deformation, under the condition that the lengthscales and timescales of instabilities are large enough for the given $\mu_d$ \citep{goodridge_viscous_1997}.

\begin{figure}
  \centering \includegraphics[width=1\textwidth]{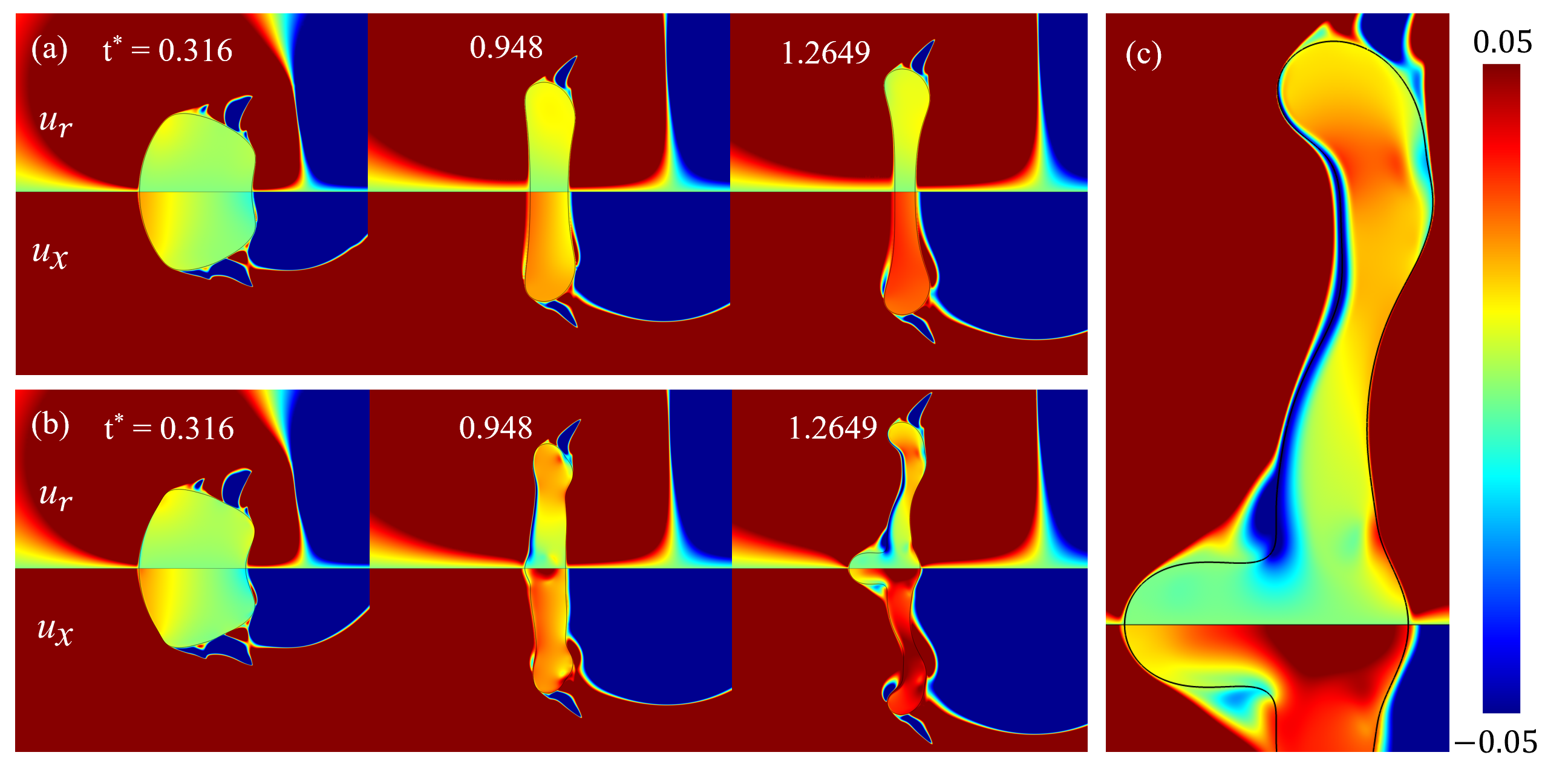}
  \caption{All drops in these plots have $\rho=1000$, $\Oho=0.001$, $\We_0=20$. These Plots show internal flows for two different $\Ohd$ values: (a)$\Ohd=0.1$, and (b)$\Ohd=0.001$. (c) is a zoomed in view of $t^*=1.2649$ for (b)}
  \label{fig:ohd_vel}
  \bigskip \centering \includegraphics[width=1\textwidth]{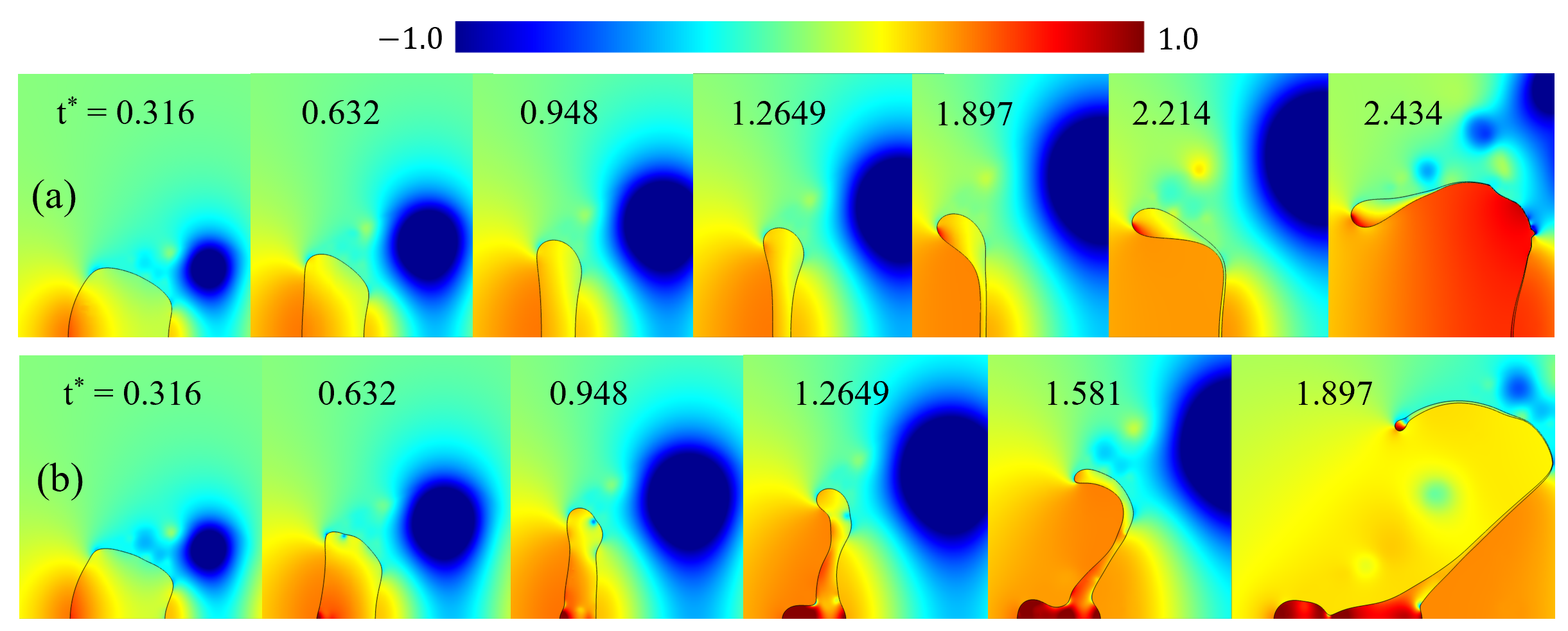}
  \caption{All drops in these plots have $\rho=1000$, $\Oho =0.001$, $\We_0=20$. These plots show the pressure fields around drops of different $\Ohd$ values. (a)$\Ohd=0.1$ (b)$\Ohd=0.001$.}
  \label{fig:ohd_P}
\end{figure}

\Cref{fig:ohd_P} illustrates the pressure fields for two such cases with $\rho = 1000, \Oho = 0.001$, and a varying $\Ohd$. Both cases show very similar (high) stagnation pressures and an $\Re_0$ of $4472$ results in a well-defined downstream vortex detached from their peripheries, indicative of their large local and total inertia. As expected, at $t^*\approx 1$, both cases form a flat pancake and the onset of rim formation. However, for the drop in case (b), we observe a high-pressure zone at its upstream pole that hints at the initiation of a plume. The internal flow field in \cref{fig:ohd_vel}(b) reveals an instability at the upstream pole of the drop, motivating a flow from its periphery to its upstream pole, hugging its upstream surface. The smaller viscosity of the drop in case (b) (100 times smaller) facilitates the development of a prominent capillary instability at its pole and the corresponding pancake-plume shape. Furthermore, the location of the instability (i.e., the upstream pole) is a stagnation point and sees the highest accelerations out of all regions of the drop. This matches with the definition of Rayleigh-Taylor instabilities and might be the primary mechanism behind the development of a plume of this kind. According to \cite{Villermaux2007,Jalaal2014}, an increase in density discontinuity motivates the formation of RT instabilities. Our current simulations also show this behavior, as only the cases with $\rho = 500$ or $1000$ and for the lowest $\Ohd$ values form an unstable plume.

As the drops in the two cases continue to deform and more mass is transferred from their cores to their rims, substantial variations in local inertia (and hence local accelerations) start to develop. Notably, for case (b), the plume has grown further and the drop now has two high local inertia regions --- its core and its rim. The annular region connecting its core and rim has lower local inertia compared to both these regions and hence accelerates downstream relative to both, resulting in the growth of a bag between the plume and the rim. Ultimately, this annular bag fragments, culminating in a backward bag-plume morphology, as seen in figures for case (b).

For the drop in \cref{fig:ohd_P}(b), it should be noted that a reduction in $\We_0$ (while keeping other parameters constant) still results in the development of an instability-driven plume at its upstream pole, albeit of a smaller size. For instance, when $\We_0=16$ for the drop in \cref{fig:ohd_P}(b), it does not deform enough to exhibit fragmentation (of the bag-plume kind). Consequently, solely decreasing $\We_0$ is not sufficient to shift the breakup morphology from a backward bag-plume to simple bag breakup for the specific ($\rho$, $\Ohd$ and $\Oho$) set. This makes a backward bag-plume breakup the critical breakup morphology for this case --- a feature of the physical properties of the system described by ($\rho$, $\Ohd$ and $\Oho$), and not just a function of boundary conditions (i.e., inflow velocity or $\We_0$).

\begin{figure}
  \centering
  \begin{minipage}{0.49\textwidth}
    \includegraphics[width=1.0\textwidth]{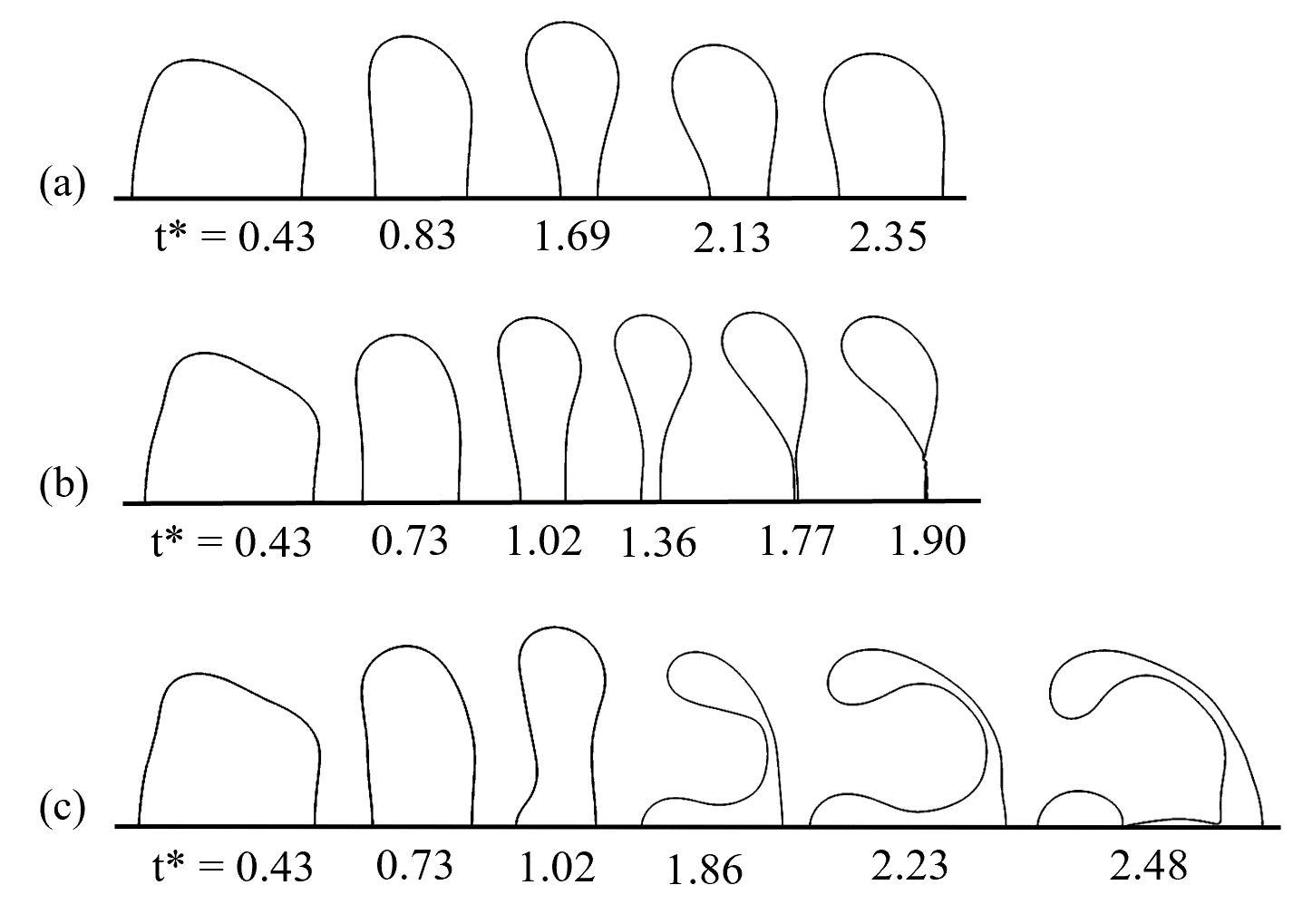}
    \caption{These plots show the fluid interface for three cases with $\rho=1000$, $\Oho =0.001$, $\We_0=15$, and (a) $\Ohd=0.1$ (b) $\Ohd=0.01$ (c) $\Ohd=0.001$.}
    \label{fig:ohd_vof}
  \end{minipage}
  \hfill
  \begin{minipage}{0.45\textwidth}
    \centering \includegraphics[width=1.0\textwidth]{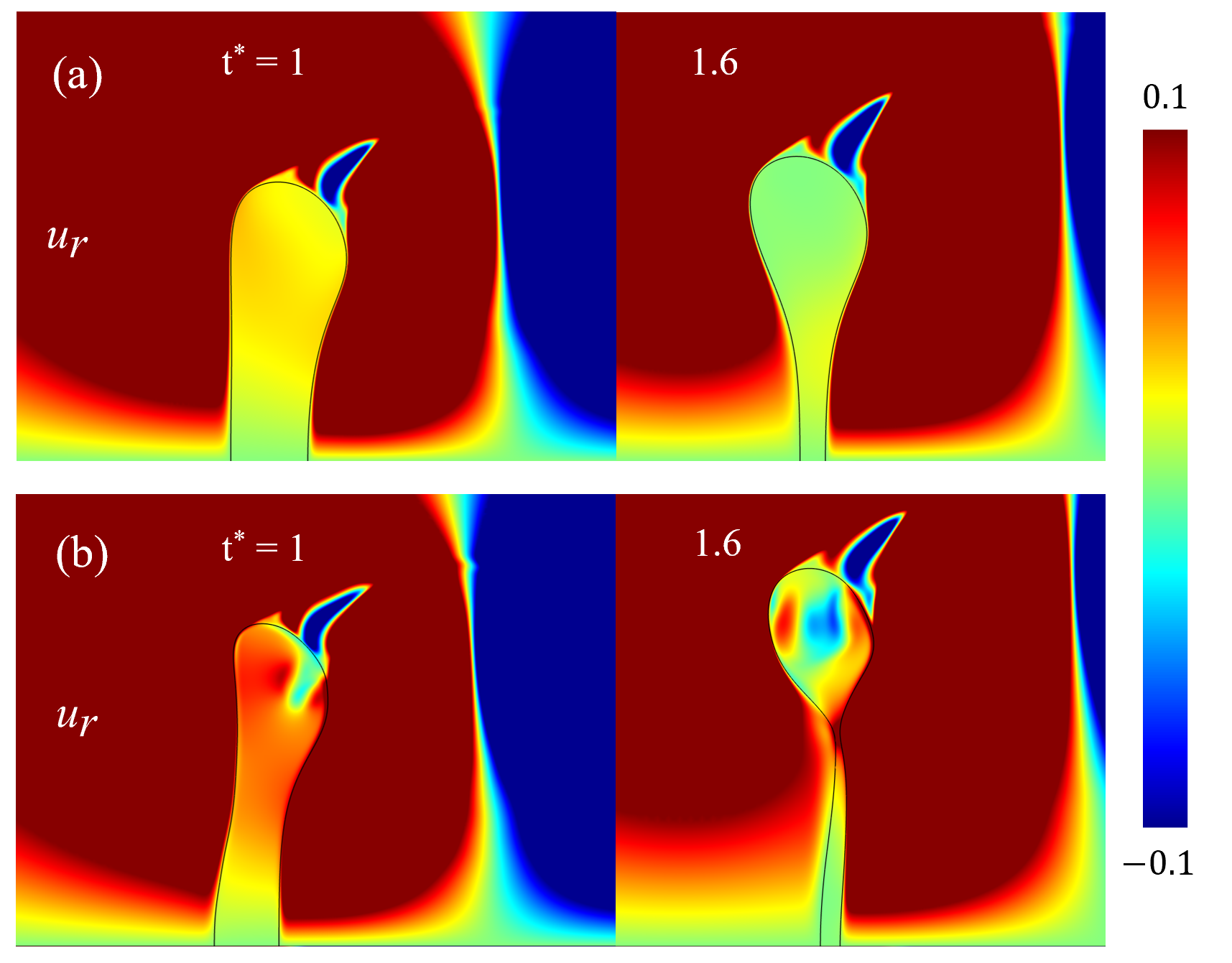}
    \caption{Both cases have $\rho=100$, $\Oho=0.001$, $\We_0=13$. These plots show y-velocities for two different $\Ohd$ values: (a)$\Ohd=0.1$, and (b)$\Ohd=0.001$.}
    \label{fig:ohd_plume}
  \end{minipage}
\end{figure}

In contrast to the plume in \cref{fig:ohd_P}(b) which originates from an instability due to lack of sufficient viscous damping, a decrease in $\Ohd$ can also lead to a plume similar to the one in \cref{fig:oho_plume}(b). One such example is shown in \cref{fig:ohd_plume}.

If we focus our attention on the specific case of drops, $\Ohd$ can be interpreted as the ratio of the capillary timescale $T_\sigma$ to the viscous timescale $T_\mu$, i.e., $\Ohd = T_\sigma/T_\mu$. The capillary timescale $T_\sigma$ ($= \sqrt{\rho_d D^3/\sigma}$) is defined as the duration for a capillary wave of wavelength $D$ to traverse a distance of $D$; while viscous timescale $T_\mu$ ($=\rho_d D^2/\mu_d$) represents the duration for momentum to diffuse across the drop \citep{Popinet2009}. A smaller $\Ohd$ hence implies a relatively small $T_\sigma$ compared to $T_\mu$, indicating that the information about interface deformation travels much faster than the the rate of transfer of momentum to the drop fluid across the diameter. Hence, the downstream vortices could apply some induced drag on the drop rim, causing local acceleration relative to the core and, subsequent deformation. However, this induced drag may not setup equivalent flow throughout the entire drop fluid. This is evident from the y-velocity plots shown in \cref{fig:ohd_plume}, where the lower $\Ohd$ case (b) shows larger y-velocities at the rim, indicating greater rates of stretching compared to the higher $\Ohd$ case (a). Hence, case (b) due to its larger $T_\mu$ results in a plume. It is essential to note that the drop in this case only shows a plume once it has started to form a bag, and the initial pancake at $t^*\approx 1$ is flat. In contrast, the plume in \cref{fig:ohd_P}(b) develops very early in the deformation process, right at the instant of formation of the pancake. Hence, the two types of plumes fundamentally differ in their formation mechanisms.

According to much of the existing literature, if $\Ohd \leq 0.1$, $Oh_d$ tends to have minimal impact on drop breakup mechanism. Therefore, for most studies, the choice of $Oh_d$ is not focused upon, as long as it is ensured to be lower than $0.1$. However, the simulations conducted and analyzed in this study do not corroborate with this understanding.

Another example emphasizing the effect of $\Ohd$ on drop deformation and breakup morphology is shown through drop interface plots in \cref{fig:ohd_vof}. In (a), the drop never achieves large enough deformation to undergo breakup. The drop in (b) on the other hand shows bag breakup for the same parameters except for $Oh_d=0.01$. The lower deformations achieved by the drop (a) can be attributed to higher fluid viscosity, which provides resistance to internal flow and dissipates energy supplied by the ambient flow through surface forces. For $\Ohd=0.001$ in (c), the breakup type shifts from a simple backward bag to a backward bag-plume breakup, with the plume formation driven by RT instabilities. In short, a decrease in $Oh_d$ is expected to reduce the required critical Weber number $\Wecr$ for a backward bag breakup. Hence, for the same $We_0=15$, we observe (c) exhibit a backward bag plume breakup, which is a multimode breakup that we expect to manifest at a $\We_0$ higher than that required for a simple backward bag.

\section{Discussion} \label{sec:discussions}

In this work, a parameter sweep using axisymmetric simulations was performed for multiple values of Weber number for every set of $\{\rho, \Oho, \Ohd \}$ possible in the parameter space defined in \cref{subsec:paramter_space}. From this vast set of simulation data, we set out to achieve two primary objectives:
\textbf{1.} extract the influence of each involved non-dimensional parameter involved --- specifically $\rho$, $\Oho$, and $\Ohd$ --- on drop pancake and breakup morphology; and
\textbf{2.} Determine both Critical Weber number values as well as corresponding critical breakup morphologies for each unique combination of $\{\rho, \Oho, \Ohd \}$ in our parameter space.
The first objective was addressed in detail in \cref{sec:results}. This section focuses on the second objective.

\subsection{The threshold of Impulsive drop breakup}
\begin{figure}
  \centering
  \includegraphics[width=1\textwidth]{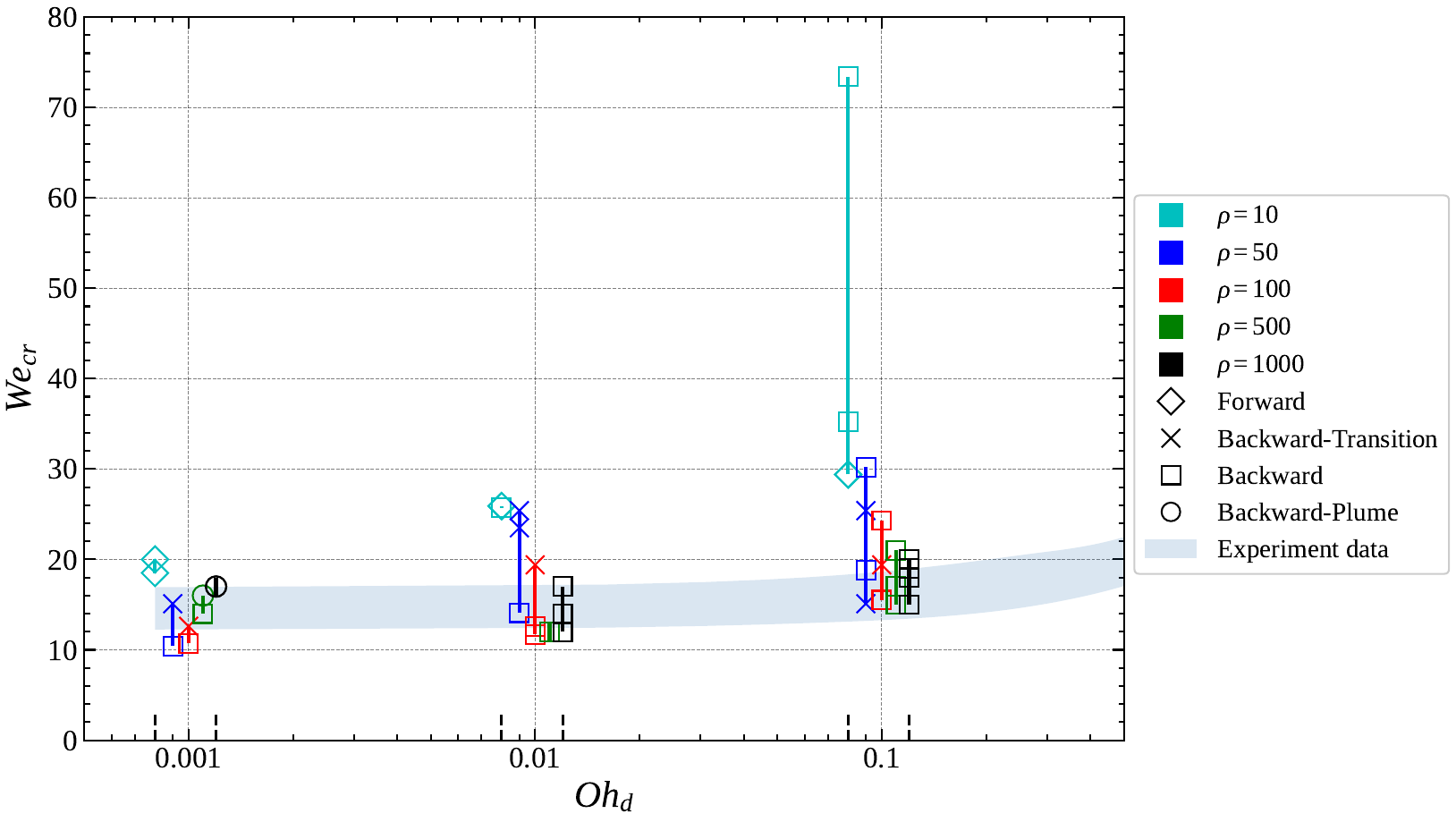}
  \caption{Plot of $\Wecr$ against $\Ohd$.
  Dependence on $\Oho$ is represented using vertical lines, with their vertical extent representing corresponding variation in $\Wecr$. Dependence on $\rho$ is shown through different colored vertical lines (offset from its true x-location to prevent overlaps with other lines) representing each $\rho$ value in the parameter space. Specific markers are used to represent all critical breakup morphologies observed in the simulations. The experimental data for $\Wecr$ from \cref{fig:lit_available_data} is shown as a translucent area in the background of the plot.}
\label{fig:wecr_ohd_sim}
\end{figure}

\Cref{fig:wecr_ohd_sim} shows the variation of critical Weber number ($\Wecr$) against the drop Ohnesorge number ($\Ohd$) for drops of different density ratios ($\rho$) and the outside Ohnesorge numbers ($\Oho$). $\Ohd$ takes three different values in the parameter space: $0.1$, $0.01$ and $0.001$. For every $\Ohd$, a $\rho$ value is represented by a colored vertical line that shows the range of $\Wecr$ values obtained due to variation in $\Oho$. The lower $\Wecr$ values generally correspond to lower $\Oho$ values and vice versa. Therefore, for each $\Ohd$ value in the plot, there exist $5$ colored vertical lines corresponding to the $5$ $\rho$ values explored in the parametric sweep. It should be noted that each colored line has been offset from its $\Ohd$ value by a different amount for preventing overlaps with other $\rho$ lines and hence improve clarity. All the cases are also explicitly marked with a uniquely shaped marker corresponding to each fragmentation morphology. Finally, all the experimental data for $\Wecr$ explored through this work is shown as a translucent area in the background of the plot, also shown as explicit markers in \cref{fig:lit_available_data}.

\begin{figure}
	\centering
	\begin{tikzpicture}
		\node[anchor=south west,inner sep=0] (image) at (0,0) {\includegraphics[width=1\textwidth]{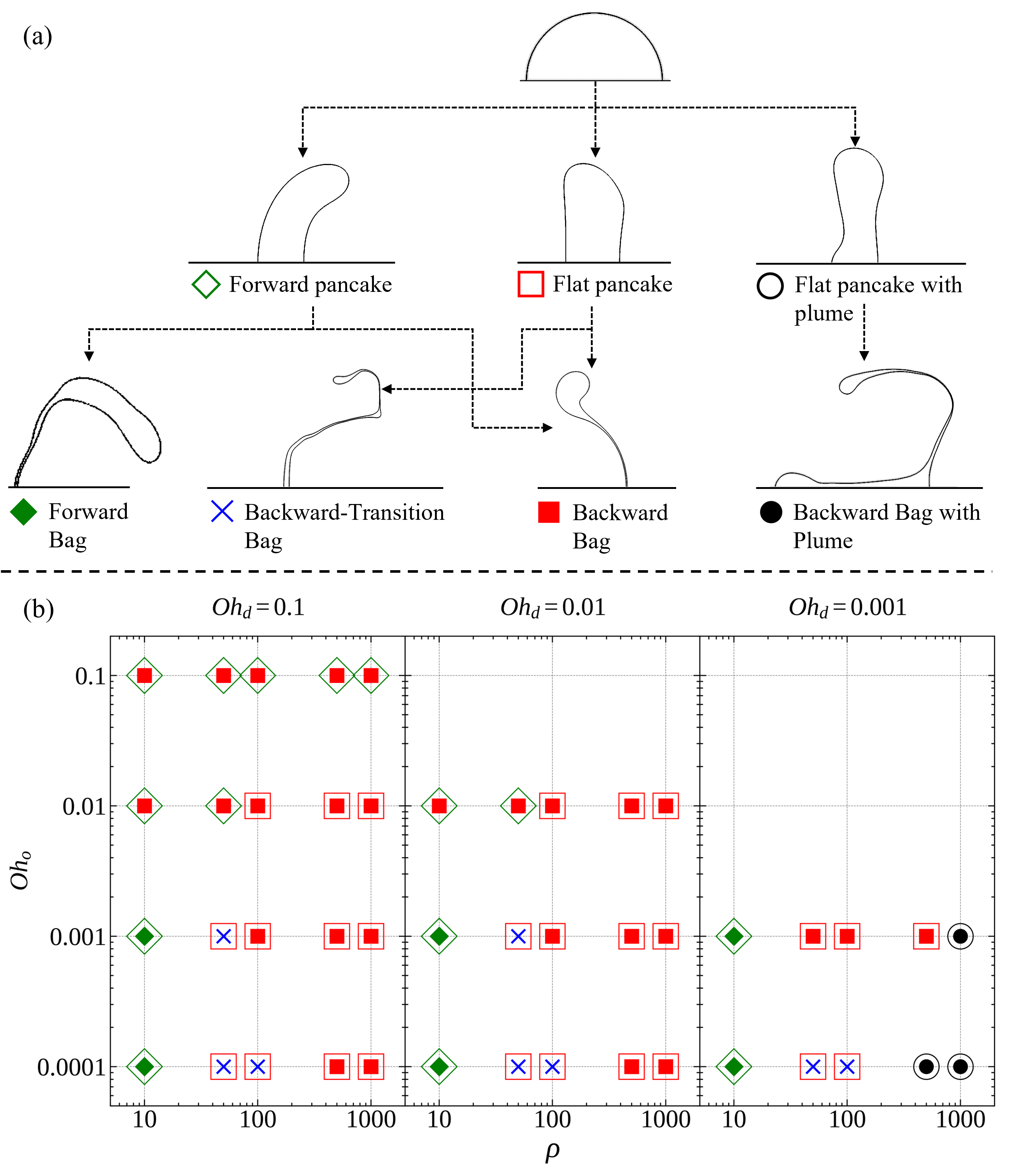}};
		\begin{scope}[x={(image.south east)},y={(image.north west)}]
			\fill[blue, opacity=0.2] (0.70,0.07) rectangle (0.98,0.12);
      \draw[blue, thick] (0.70,0.07) rectangle (0.98,0.12);
		\end{scope}
	\end{tikzpicture}
	\caption{(a) is a path diagram of all deformation paths a spherical drop under impulsive acceleration can take when breaking up critically. A spherical drop can deform into three types of pancakes, each of which can further deform into one of four breakup morphologies, the corresponding ${\rho,\Oho,\Ohd,\We_0}$ parameter space is shown in the phase diagram (b). The \textcolor{blue}{blue-highlighted region} in (b) indicates cases with the lowest $\Oho$ and $\Ohd$ values, for which the axisymmetric simulations may not be representative of reality, as has been discussed in \cref{app:3D_comparison}.
  }
	\label{fig:phase_diagram}
\end{figure}

On the basis of all the simulation results and \cref{fig:wecr_ohd_sim}, the following conclusions emerge:
\begin{enumerate}
    \item \textbf{Consistency with experimental data}: All simulations with high $\rho$ ($\geq 500$) values critically fragment at $\Wecr$ values that very closely match historical experimental works. This is expected given that most experimental studies have historically focused on impulsive fragmentation for water-air analogous systems, which have $\rho \geq 500$. Only cases with $\rho<500$ show any appreciable deviation from the experiments.
 
    % \item \textbf{Sensitivity to $\rho$}: As $\rho$ decreases below 500, we observe greater variations in the corresponding $\Wecr$ values.
    % The lowest $\rho$ cases exhibit the largest variations in $\Wecr$ values with respect to $\Oho$ for the highest $\Ohd$ values.
    % Conversely, for a low $\Ohd$, the influence of $\Oho$ on $\Wecr$ is minimal 
    % The high density ratio cases instead show extremely similar $\Wecr$ values and variations in it corresponding $\Ohd$.
 
    \item \textbf{Sensitivity to $\Oho$}: The lowest $\rho$ cases exhibit the largest variations in critical Weber numbers with respect to changes in $\Oho$ for the high $\Ohd$ cases, as indicated by the length of vertical lines in \cref{fig:wecr_ohd_sim}. A drop with a large $\rho$ ($\geq 500$) experiences larger relative velocities, making $\delP$ the dominant factor driving its deformation in most cases. Even in cases with large external shear stresses on the drop (the largest $\Oho$ cases), once a clear rim has formed in the deformed drop, local inertia variations take over the deformation process. Only the lower inertia ($\rho$) drops show the any appreciable sensitivity to changes in $\Oho$ and the corresponding differences in downstream vortical structures. An exception exists when $\Oho$ is in the free-shear regime ($\Re_0>10000$ (\cref{subsec:oho}), and even the drops with a high $\rho$ respond strongly to the resulting downstream vortices, fragmenting with a forward bag morphology.

    \item \textbf{Variations in fragmentation morphologies}: The critical breakup morphology transitions from a forward bag ($\rho=10$) to a backward-transition ($\rho=50,100$) (see \cref{fig:phase_diagram}) to a backward and backward-plume bag with increase in $\rho\geq 500$. As $\rho$ increases, the drop's rim is expected to start lagging behind the drop core at some point during its deformation, when local inertia of its rim becomes substantially larger than that of its core. The shift in morphology from a forward to a backward bag is observed only for low $\Oho$ values, which produce downstream vortices that are strong enough to compensate for the larger local inertia of the rim. Alternatively, large $\Oho$ cases always exhibit backward bag breakup at critical conditions.
    This is due to the downstream vortices being weak or non-existent for low $\Re_0$ flows.

    \item \textbf{Sensitivity to $\Ohd$}:
    In addition to drop Ohnesorge number's role in controlling the sensitivity of $\Wecr$ to $\Oho$, a decrease in $\Ohd$ also affects the critical fragmentation morphology by motivating the formation of a plume. This is seen for the lowest $\Ohd$ ($0.01, 0.001$) and $\Oho$ ($0.001, 0.0001$) values, and for the largest $\rho$ ($500, 1000$) drops (star shaped marker in \cref{fig:wecr_ohd_sim}). Such drops show an unstable plume at the upstream poles of their flat pancakes, which are also locations of maximum acceleration in the drops and can be attributed to Rayleigh-Taylor instabilities. Both low viscosity and larger density ratios motivate the development of such instabilities \citep{Villermaux2007,guildenbecher_secondary_2009,Jalaal2014}. Since the pancake for this non-dimensional set starts with a plume even for lower non-critical $\We_0$ values, a plain backward bag breakup can never manifest for such systems, and hence a backward bag-plume breakup becomes its critical breakup morphology. $\Ohd$ also influences the rate of evacuation of the drop core, which ultimately results in a forward bag for all high $\Oho$ cases, when shear stresses drive pancake formation.
\end{enumerate}

The conclusions drawn from \cref{fig:wecr_ohd_sim} show that the accepted idea of critical Weber number being almost independent of drop Ohnesorge number for $\Ohd<0.1$, might not always hold, especially for systems that stray too far from properties analogous to Water-Air. Furthermore, the critical breakup morphology need not necessarily be a backward bag breakup. Backward bag-plume and forward bag morphologies can be the critical morphologies for certain low $\rho$ and low $\Oho$ cases.

A path diagram, as illustrated in \cref{fig:phase_diagram}(a), can be drawn that summarizes all the deformation paths a spherical drop might follow as it deforms to critical fragmentation under impulsive acceleration. A companion phase diagram (\cref{fig:phase_diagram}(b)) provides the non-dimensional parameter space that results in one of the four observed fragmentation morphologies. It is worth highlighting that all the breakup paths provided in the diagram are for their respective critical conditions, and consequently the $\Wecr$ values corresponding to different paths need not be the same.

For small $\rho$ or high $\Oho$ values or both, shear stresses drive the internal flow, resulting in a forward pancake. The fate of this forward-facing orientation is then contingent on the balance between local inertia differences and the strength and proximity of downstream vortices to the rim. For systems with large $\Oho$, irrespective of density ratio, downstream vortices either do not form or shed further downstream. The drop is not subjected to large lateral stretching rates, allowing the formation of a prominent toroidal rim. The resulting lateral inertia differences lead to the flipping of the bag from forward to backward orientation, eventually fragmenting with a backward bag morphology. On the other hand, for small $\Oho$ (and hence small $\rho$ for a forward pancake) coupled with large $\Ohd$, the strong, fast, and proximal downstream vortices generate large lateral stretching rates, preventing the formation of a prominent rim. As a result, the drop continues to hold its forward facing orientation and breaks up with a forward bag morphology.

When $\rho$ is large and $\Oho$ is small, the drop under critical conditions either deforms into a flat pancake or a flat pancake with a plume depending on whether $\rho$ and $\Oho$ are at the extreme ends of the parameter space. The largest values of $\rho$ and the smallest values of $\Oho$ and $\Ohd$ lead to the appearance of a plume at the upstream pole of the flat pancake. It can be hypothesized that the low viscosities of both outside and drop fluids do not provide sufficient viscous dissipation to stabilize the jet ejected at the upstream pole (due to Rayleigh-Taylor instability) of the drop. From this pancake shape, the only possible critical breakup morphology is a backward bag-plume breakup. All other intermediate cases form a flat pancake which can form either a backward-transition breakup (for intermediate $\rho$ values and low $\Oho$ values) or a backward bag (for all the remaining cases). A backward-transition breakup is a forward bag with a flipped rim, i.e. a drop which at its final moments gains enough inertia in its rim to start the bag flipping process. As expected, this is observed for intermediate $\rho$ values where neither local inertia differences nor downstream vortices outright dominate the dynamics of the drop's periphery.

The path diagram, together with the phase diagram, only informs of the types of pancakes and corresponding general breakup morphologies observed for a specific set of $\{\rho, \Oho, \Ohd\}$ under threshold conditions. For information on the lowest $\We$ required to achieve the corresponding non-vibrational breakup (i.e. $\Wecr$), one may refer to \cref{fig:wecr_ohd_sim}.

\subsection{Bag Inflation Characteristics}
\label{subsec:bag_inflation}

\begin{figure}
	\centering
	\includegraphics[width=1\textwidth]{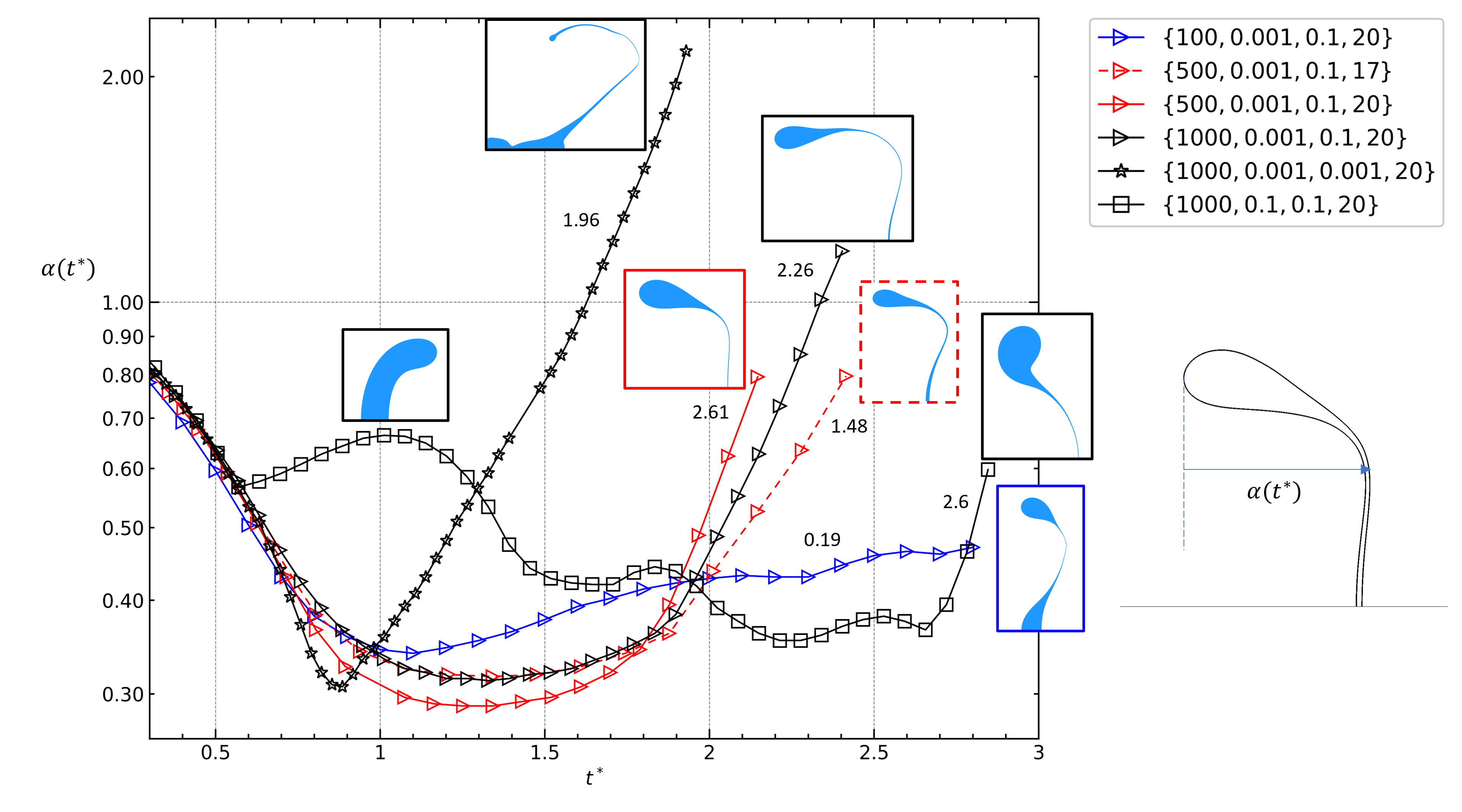}
	\caption{Bag inflation $\alpha(t^*)$ with time $t^*$ is shown for some unique backward bag breakup cases. The non-dimensional parameter set for each case is of the form $\{ \rho, \Oho, \Ohd, \We_0 \}$. The cases plotted here include simple backward bags ($\{500, 0.001,0.1,17\}$, $\{1000, 0.1,0.1,20\}$ $\{1000, 0.001,0.1,20\}$) and backward plume-bags ($\{100, 0.001,0.1,20\}$, $\{500, 0.001,0.1,20\}$, $\{1000, 0.001,0.001,20\}$). In addition, $\{1000, 0.1,0.1,20\}$ initially forms a forward pancake shape which then flips to backward bag. Through this plot, the effect of $\We_0$, $\Oho$ and $\rho$ on bag inflation rates is highlighted.}
	\label{fig:bag_inflation}
\end{figure}

An understanding of the timescales involved for the inflation of bags during bag breakups is essential in correct estimation of bag burst timescales and effective center-of-mass velocities for bag breakups. \cite{Villermaux2009} were the first to give an analytical description of bag inflation rates for backward bag morphology. They found that the amplitude of bag inflation increased exponentially with time with an exponent factor of 2 for an inviscid drop. \cite{Kulkarni2014} extended the work of \cite{Villermaux2009} to include drop viscosity and numerically obtained a similar exponential relationship for bag inflation, now also dependent on $\Ohd$. However, $\Ohd$ was found to have a very small impact on bag inflation rates. $\We_0$ on the other hand has a more dramatic effect on the exponent factors governing inflation, with higher $\We_0$ showing faster inflation rates. The large variety of simulations in the current work across a large parameter space present an opportunity to explore the role of the relevant non-dimensional parameters on bag inflation. Before we proceed, it is essential to emphasize that the axisymmetric simulations conducted in this work cannot capture the bag inflation too far beyond the initiation time \citep{lingDetailedNumericalInvestigation2023,tangBagFilmBreakup2023}.
Hence the bag inflation rates presented here are not useful in their absolute values, and only representative of the early stages of bag inflation.
However, the differences in bag inflation rates for different cases, solely in context of other axisymmetric simulations, still provides useful insights into the functional dependence of early bag inflation rates on these parameters.

With this caveat in mind, we highlight some key observations related to the bag inflation rates for a few simulations where backward bag breakup is observed. The evolution of bag inflation $\alpha(t^*)$ with time is illustrated in \cref{fig:bag_inflation}, where $\alpha(t^*)$ is equal to the horizontal extent of the drop $e_x$ (as shown in \cref{fig:drop_params}). For every plot, the drop shows a decrease in $\alpha(t^*)$ up until it reaches the end of pancake stage at $t^* \approx 1$, beyond which the drop begins to inflate. The only exception is the $\{1000, 0.1,0.1,20\}$ case, which initially shows an increase in $\alpha(t^*)$ due to deforming to a forward pancake. Bag inflation stage only starts at $t^* \approx 2.5$ for this case. Once firmly in the bag inflation stage, all cases plotted here show an exponential growth $\alpha(t^*)$ with time with a specific exponential growth factor, marked in \cref{fig:bag_inflation} as numbers alongside each plot.

From the plots, a number of key observations can be made. The $\{1000, 0.001, 0.001, 20\}$ case, which has properties closest to a water-air drop-ambient system, shows an inflation growth factor of 1.96, which is very close to the value that was analytically found by \cite{Villermaux2009}, and was matched against a bag breakup experiment of a water drop. Even though this case is a Backward-plume breakup, its inflation growth rate matches that of a simple bag and is not affected by the presence of the plume.

For $\{500, 0.001, 0.1\}$ case, out of the two Weber numbers shown in the plot $(\We_0 = 17,\, 20)$, the higher $\We_0$ drop shows a faster bag inflation rate, indicated by the larger exponential growth of its $\alpha(t^*)$. This is an expected observation, since the overall deformation rate is also higher for a higher $\We_0$ case, owing to the lower surface tension forces relative to the dynamic pressure forces driving the drop's deformation. Hence, the same amount of energy (supplied to the drop by the external forces) should lead to a larger change in the interface area as a result of a lower surface tension.

We also observe that the $\rho=100$ case shows a dramatically lower inflation rate compared to an analogous $\rho=500$ or $\rho=1000$ cases. This may be attributed to the lower relative velocities with ambient observed for the low $\rho$ case, which results in a decrease in its effective Weber number, and hence lower bag inflation rates.

Drop and ambient viscosities appear to have an inverse effect on bag inflation rates. Among the three $\rho=1000$ cases shown in the plot, the higher drop and ambient Ohnesorge number cases show the larger exponential growth rates of $\alpha(t^*)$.

More generally, \cref{fig:bag_inflation} shows that $t \approx \tau$ (or $t^* \approx 1$) is a good representative timescale for the start of the bag inflation process. This observation is consistent with the aspect ratio plots shown in \cref{fig:rho_vel,fig:oho_vel}, where the pancake formation stage almost always ended at $t^*\approx 1$.

If the aerodynamic forces acting on a drop are large enough to initiate bag inflation in its pancake, it is expected that the drop will go on to or be very close to fragmentation. Therefore, we hypothesize that the balance of all forces acting on a drop at this stage (aerodynamic forces driving its deformation, and capillary and viscous forces resisting it), i.e., at the end of its pancake formation and at the start of its bag inflation, is representative of the overall stability of the drop fragmentation process. We will use this understanding to obtain a better estimate of the effective aerodynamic forces on the drop in the next section.

\subsection{ A parameter for Prediction of Breakup Threshold} 
\label{subsec:parameter_for_threshold}

Most previous studies have characterized the threshold for impulsive breakup of spherical drops using a Weber number based on the initial relative velocity with ambient ($\We_0=\rho_o V_0^2D/\sigma$). For all cases with properties analogous to water-air system, i.e., $\rho>500$, $\Oho<0.01$ and $\Ohd<0.1$, critical breakup occurs consistently at a critical Weber number of $\Wecr \approx 14$. However, as has been discovered and exhaustively described through the simulation results in the previous sections (summarized succinctly in \cref{fig:wecr_ohd_sim,fig:phase_diagram}), different cases which stray away from the non-dimensional space described by water-air systems do not show the same threshold Weber number value. Substantially higher $\Wecr$ values are observed for cases with low density ratios and high Ohnesorge numbers. Weber number $\We_0$ represents the ratio of the pressure forces applied on a drop surface by the ambient medium, based on its initial relative velocity to the surface tension acting against any change in surface energy. However, drop deformation also depends on the viscous forces applied by the surrounding flow, the inertia and hence the acceleration, and the viscous dissipation against internal fluid flow. These effects have been explained in detail in \cref{sec:results} through simulations for varying $\Oho$, $\Ohd$ and $\rho$ values respectively. $\We_0$ hence does not capture the role of all the factors relevant in the drop deformation process. We aim to construct a new non-dimensional group which aggregates the effect of all the parameters, namely $\We_0$, $\Oho$, $\Ohd$, and $\rho$, and shows a consistent critical value demarcating the threshold of breakup under impulsive acceleration for the complete parameter space explored in this study.

Let us assume that this new non-dimensional number, denoted by $\Cbr$, is a function of all the dimensional variables involved in the breakup process (\ref{eq:c_breakup_dimensional_dependance}). There are $3$ independent dimensions in the problem. Hence, through Buckingham-Pi analysis, we can obtain at most $4$ independent non-dimensional numbers. The four non-dimensional numbers relevant to this problem have already been defined in \cref{subsec:problem_description}, namely $\rho$, $\Oho$, $\Ohd$, and $\We_0$. It is expected for $\Cbr$ (\ref{eq:c_breakup_non_dimensional_dependance}) to be described as
\begin{equation}
  \label{eq:c_breakup_dimensional_dependance}
  \Cbr = f (V_0, D, \rho_o, \rho_d, \mu_o,\mu_d, \sigma),
\end{equation}
which when non-dimensionalized gives us
\begin{equation}
  \label{eq:c_breakup_non_dimensional_dependance}
  \Cbr = f (\We_0, \rho, \Oho, \Ohd).
\end{equation}

We derive $\Cbr$ using a method analogous to that of \cite{blackwell_sticking_2015} (section IV). In their work, Blackwell et al. developed a non-dimensional group to describe the stick-splash behavior of a drop impacting a surface. This group was based on the ratio of forces promoting splashing (inertial forces) to those promoting sticking (dissipative forces). Similarly, we define $\Cbr$ based on the competition between forces driving fluid radially outward from the drop's core and forces resisting this outward flow. We hypothesize that this competition determines the threshold for secondary fragmentation (the breakup of the primary drop into smaller droplets), where $\Cbr$ is defined as:
\begin{equation}
  \Cbr = \frac{\text{Forces driving drop deformation}}{\text{Forces resisting drop deformation}}.
  \label{eq:breakupfactor1}
\end{equation}

The forces will be evaluated at $t^*\approx 1$, which represents the end of the pancake stage and the start of the bag inflation process. The choice of $t^*\approx 1$ as the critical time controlling the bag inflation process has been justified in \cref{subsec:bag_inflation}.

\begin{figure}
  \centering
   \includegraphics[width=1.0\textwidth]{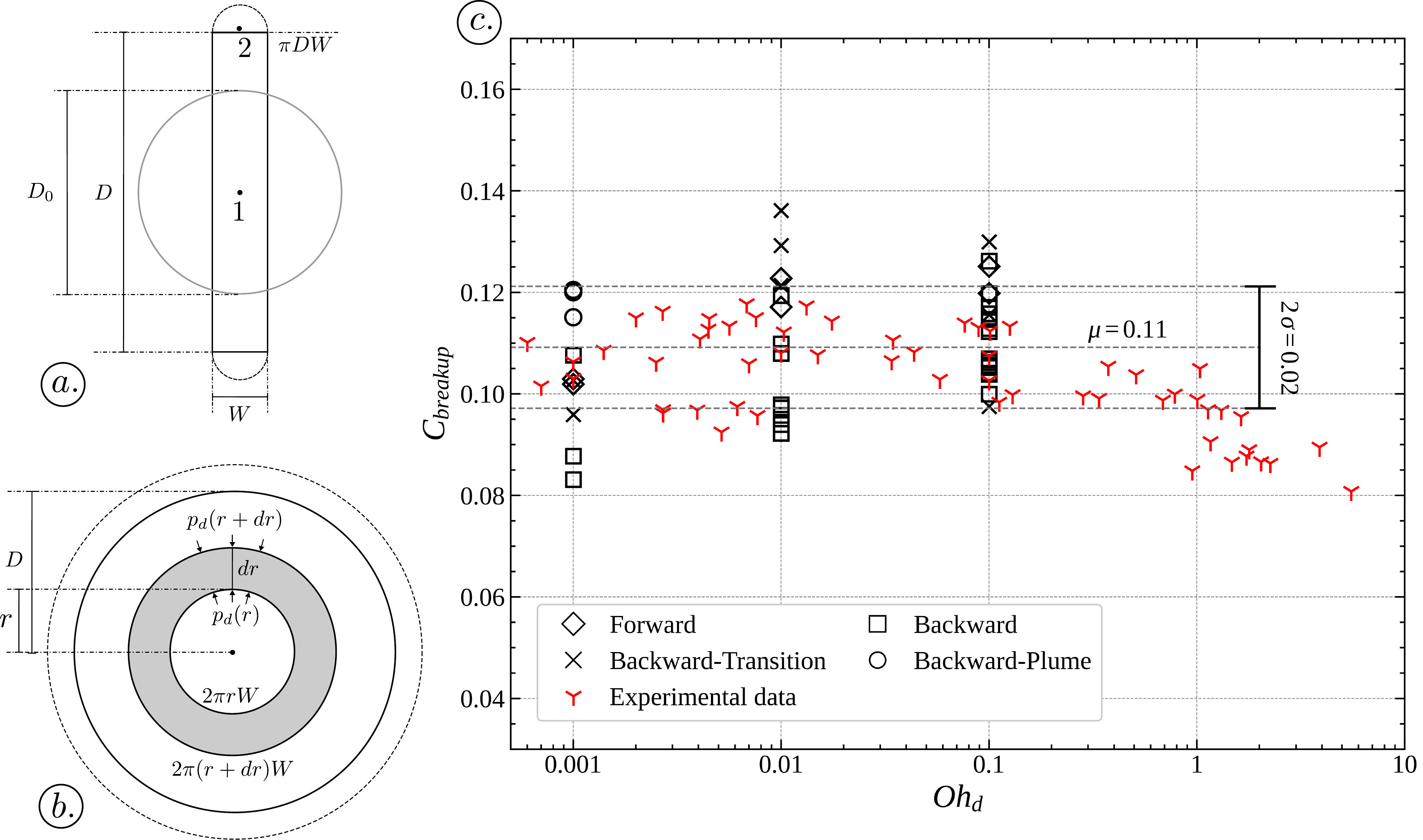}
   \caption{This figure shows an alternate version of \cref{fig:wecr_ohd_sim} where instead of $\Wecr$, the variation of $\Cbr$ with respect to $\rho$, $\Oho$, and $\Ohd$ is plotted. In addition to simulation data, all available experimental data in the relevant non-dimensional space for critical bag breakup (plotted in \cref{fig:lit_available_data}) is also plotted here for reference.}
   \label{fig:dropforces_cbreakup}
\end{figure}

Let us assume that a spherical drop of diameter $D_0$ at $t^*=0$ deforms in to a disk of diameter $D$, bounded radially by a semi-circular ring of diameter $W$, and has an aspect ratio $A_{xr}$ at $t^*=1$ (\cref{fig:dropforces_cbreakup}(a)). Mass conservation when applied to the drop gives us $D \approx (2/A_{xr})^{(1/3)} D_0$ and $W \approx (2/3) (D_0^3/D^2)$, if we ignore the mass that the curved periphery holds.

All regions of the pancake within the flat disk region ($r<D$) have a very large local radius of curvature, and hence contribute negligibly to surface tension forces locally. The periphery is the only region with any appreciable curvature equal to the sum of local longitudinal and azimuthal curvatures, i.e., $(2/W + 2/D) = 2(1+A_{xr})/W$. Therefore, crossing the drop interface at its periphery produces a pressure jump of $2\sigma (1+A_{xr})/W$ according to the Young-Laplace equation, that acts on the narrow cylinder between the semi-circular peripheral ring and the flat disk of area $\pi D W$ at a radial distance $D/2$ (point $2$ in \cref{fig:dropforces_cbreakup}(a)). The total contribution of surface tension can be estimated by calculating the force applied by the excess pressure on this area, given as
\begin{equation}
    F_\sigma = 2 \pi \sigma D\, (1+A_{xr}).
    \label{eq:surfacetension_force}
\end{equation}

The surface tension force is directed against the movement of the drop fluid from the core to the periphery.

If the flow around the pancake is assumed to be a potential flow with a stagnation point at the upstream pole, the pressure at the upstream surface of the pancake at radial distance $r$ from the longitudinal (axisymmetric) axis can be estimated \citep{Villermaux2009,jackiw_aerodynamic_2021} as:
\begin{equation}
    p_o\left( r \right) = p_o(0) - \dfrac{\rho_o a^2 \Vrel^2}{8 D_0^2} r^2.
    \label{eq:drop_pr}
\end{equation}

Here, $p_o(0) = 0.5 \rho_o \Vrel^2$ is the stagnation pressure at the upstream pole. Thus, pressure is the maximum at the upstream pole and drops as we move radially away from the upstream stagnation point towards the periphery. The potential flow around a rigid body has a stretching factor $a$ of $6$ for a sphere and $4/\pi$ for a flat disk, the latter being applicable in this work for the flow around a flat pancake. The true pressure inside the drop is a superposition of aerodynamic pressures and excess pressure due to surface tension. Since the local curvature of the drop interface is negligible for $0<r<(D-0.5W)$, the pressure inside the drop $p_d(r)$ near its upstream surface in the disk region is almost equal to $p_o(r)$. We can thus calculate the independent contribution of the aerodynamic pressure forces in driving the evacuation of the drop core --- by integrating the infinitesimal force due to $p_o(r)$ acting on a cylindrical surface of radius $r$ and width $W$ across the radius of the pancake.
\begin{subequations}
  \begin{align}
    F_p &= \int_0^{0.5D} \left[ p_o(r)\,(2\pi r W) - p_o(r+dr)\,(2\pi (r+dr)W) \right],
    \\
        &= - 2\pi W \int_0^{0.5D} \left[ p_o(r) + \dfrac{\partial p_o(r)}{\partial r}\,r \right] dr,
        \label{eq:pressure_force1}
  \end{align}
\end{subequations}
\begin{equation}
    F_p = \dfrac{a^2}{48} \pi \rho_o \Vrel^2 D_0\, D - \dfrac{1}{3} \pi \rho_o \Vrel^2 \dfrac{D_0^3}{D} = F_{p,d} - F_{p,s}.
    \label{eq:pressure_force2}
\end{equation}

$F_{p,d}$ is the contribution of the radial pressure drop in the ambient flow in support of evacuation of the core, whereas $F_{ps}$ is the contribution of stagnation pressure against it.

In addition to surface tension and external dynamic pressures, we must estimate the contribution of drop fluid viscosity in dissipating (part of) the kinetic energy of the internal flow. An estimate for the differential viscous dissipation power can be obtained by using the relation \citep{batchelor1967introduction,deville2002high,rimbert_spheroidal_2020}:
\begin{equation}
    \Phi = 2\mu_d\, \mathbf{d}:\mathbf{d} = 2\mu_d\; d_{ij}d_{ij},
    \label{eq:differential_dissipation_power}
\end{equation}

where $\mathbf{d} = 0.5 \left[ \nabla \mathbf{u} + (\nabla \mathbf{u})^T \right]$ is the rate of deformation tensor, and $\mathbf{d}:\mathbf{d}$ is its dyadic product. $\mathbf{u}$ is the velocity vector field. Using mass conservation, \cite{Kulkarni2014} derived a relation $u_r = r\dot{D}/D$ for the radial velocity of the drop fluid in terms of its radial expansion rate, which in Cartesian coordinates can be written as $\mathbf{u} = (\dot{D}/D)(x\, \mathbf{\hat{j}} + y\, \mathbf{\hat{k}})$, given that the transverse plane is synonymous with the yz-plane. Substituting this velocity function in \ref{eq:differential_dissipation_power} and dividing by the local velocity scale, we can obtain a scale for the small local viscous force, which is always oriented in a direction opposite to the internal flow velocity (i.e., negative radial direction) as
\begin{equation}
    dF_\mu \propto 4\mu_d\, \dfrac{\dot{D}}{D}\, \dfrac{1}{r}.
    \label{eq:differential_dissipation_force}
\end{equation}

Integrating \ref{eq:differential_dissipation_force} over the entire volume of the pancake, we get an estimate for the total viscous dissipation force working against the internal flow as
\begin{equation}
    F_\mu \propto \dfrac{4}{3} \pi \mu_d D_0^3 \dfrac{\dot{D}}{D^2}.
    \label{eq:total_dissipation_force}
\end{equation}

A drop under impulsive acceleration starts at zero velocity and then asymptotically accelerates towards a maximum velocity equal to that of the ambient flow (given that the drop is still intact). This acceleration is driven by the pressure and viscous stresses applied by the outside flow on the drop interface, the corresponding magnitude of which is dictated by the instantaneous Reynolds number and relative velocity with the ambient medium $\Vrel$. As the drop continues to accelerate, its relative velocity with respect to the ambient medium continues to decrease, which reduces both the pressure and the viscous stresses. For water-air systems where $\Oho$ is low and $\rho$ is high, the drop does not show significant accelerations and $\Vrel$ remains close to the initial relative velocity of $V_0$ for almost the entire duration of the breakup process. For such cases, setting $\Vrel$ equal to $V_0$ would be a valid assumption. Conversely, if the drop shows significant accelerations and gains velocities that are a substantial fraction of $V_0$ (which is the case for low $\rho$ or high $\Oho$ systems), $\Vrel$ cannot be assumed to be equal to $V_0$ anymore. It then becomes essential to derive a scaling for $\Vrel$ that is valid for the whole parameter space. For this purpose, we utilize the drag equation for a sphere:
\begin{equation}
  F_\mathit{drag} = \dfrac{1}{2} C_D \rho_o V_\mathit{rel}^2 A.
  \label{eq:drag_equation}
\end{equation}
to evaluate the drag forces acting on the drop at time $t=0$ (i.e., when the drop is spherical and at rest):
\begin{equation}
  \rho_d \frac{\pi}{6} D^3 a_0 = F_\mathit{drag}(t=0) = \frac{\pi}{8} C_{D0} \rho_o V_0^2 D^2.
  \label{eq:drop_initial_drag}
\end{equation}
\ref{eq:drop_initial_drag} gives us a scale for the acceleration experienced by the drop at $t=0$:
\begin{equation}
  a_0 = \frac{3}{4} \frac{1}{\rho D} C_{D0} V_0^2.
  \label{eq:drop_initial_acc}
\end{equation}
A scale describing the velocity of the drop relative to the ambient medium, after a time of the order of the deformation timescale $\tau$ (\cref{subsec:bag_inflation}) can be obtained by evaluating $V_0 - \tau a_0$, which when simplified gives
%\begin{equation}
%  V_\mathit{rel}
%  \propto V_0 - \left( \frac{D}{V_0} \sqrt{\rho} \right) \;
%  \left( \frac{3}{4} \frac{1}{\rho D} C_{D0} V_0^2 \right)
%\end{equation}
\begin{equation}
  V_\mathit{rel}
  \propto K_v V_0 \, ,
  \text{ where }
  K_v = \left( 1- \frac{3}{4} \frac{C_{D0}}{\sqrt{\rho}} \right).
  \label{eq:relative_vel_scale}
\end{equation}

In \cref{fig:rho_vel,fig:oho_P}, we observe that the drop's velocity decreases with an increase in $\rho$ and a decrease in $\Oho$. These observations are in line with the relative velocity scale obtained in \ref{eq:relative_vel_scale}, where $C_{D0}$ is a function of $\Re_0 = \sqrt{\We_0}/\Oho$. Any increase in the ratio between $C_{D0}$ and $\sqrt{\rho}$ leads to a decrease in $V_\mathit{rel}$. It is important to note that the use of a drag coefficient for a sphere in the derivation of $V_\mathit{rel}$ is a simplification that allows us to obtain an explicit estimate for $V_\mathit{rel}$. However, this choice results in an underestimation of the average acceleration experienced by the drop, since a pancake shape is less aerodynamic and has a larger frontal area compared to a sphere. This results in an overestimation of $V_{rel}$, and hence an overestimation of pressure forces as well as viscous dissipation. In addition, as the drop accelerates and deforms, its instantaneous Reynolds number also changes, which would affect the drag coefficient. However, the threshold is expected to be overestimated as a result of this simplification, since the aerodynamic forces are proportional to the square of the relative velocity.

$\dot{D}$ is estimated to be equal to $2\Vrel/(\sqrt{\rho}+1)$, described as Dimotakis velocity \citep{dimotakis1986two} and also used by \cite{marcotte_density_2019} in their work. Finally, in all simulations conducted in the current work across the parameter space, it was observed that the aspect ratio $A_{xr}$ at $t^*\approx 1$ was almost always $0.175$, and we will use this value and the corresponding $D$ for the calculation of the forcing scales.

At this stage, we have derived all the relevant forcing scales and related parameters for deriving a new parameter describing fragmentation threshold. However, this derivation hinges upon an important assumption that the pancake is a flat disk for all the cases. This is however not true for all the high $\Oho$ cases which form a forward facing pancake (concave shape pointing downstream). This difference in geometry affects the potential flow and the corresponding pressure field around the drop, the pressure jumps due to surface curvature, and the corresponding internal flow vectors.

It is also important to note that the use of the deformation time scale $\tau$ to estimate the velocity scale implicitly inherits the assumption that the drop flattens into a pancake in time $\tau$, and the balance of forces acting on the drop control whether the pancake invariably develops a bag. Hence, we expect $\Cbr$ to correctly capture the threshold only for drops that go through a critical pancake shape during their deformation process.

\ref{eq:breakupfactor1} can now be rewritten incorporating the pressure deficit and stagnation pressure forces (\ref{eq:pressure_force2}), surface tension force (\ref{eq:surfacetension_force}), and viscous resistance in drop fluid (\ref{eq:total_dissipation_force}):
\begin{equation}
  \Cbr = \frac{ F_{p,d} }{ F_{p,s} + F_\sigma + F_\mu },
  \label{eq:breakupfactor2}
\end{equation}
which finally results in:
\begin{equation}
  \Cbr = \dfrac{ (a^2/48)\,\We_0 K_v^2 D^2 }{ 2D^2(1+A_{xr}) + (4/3)\Ohd\, (\rho \We_0)^{0.5}\dot{D} + (1/3)\We_0 K_v^2 }.
  \label{eq:C_breakup_final}
\end{equation}

For a water-air analogous systems, $\Ohd<<1$ and $\rho>500$, which implies that $K_v \approx 1$ ($C_{D0}\approx 1$ for large $\Re_0$) and $\dot{D} \approx \Vrel/\sqrt{\rho}$. $\Cbr$ can then be simplified to:
\begin{equation}
  \Cbr = \dfrac{ (a^2/48)\,\We_0 D^2 }{ 2D^2(1+A_{xr}) + (1/3)\We_0 }.
  \label{eq:C_breakup_simplified}
\end{equation}

For obtaining $\Cbr$ using \ref{eq:C_breakup_final}, an explicit equation for drag coefficient is required. We use the relationship provided by \cite{turton_short_1986}:
\begin{equation}
  C_{D0} = \frac{24}{\Re_0} \left( 1 + 0.173 \Re_0^{0.657} \right) + \frac{0.413}{1+16300 \Re_0^{-1.09}}.
  \label{eq:C_D_turton_levenspeil}
\end{equation}
Note that $C_{D0}$ is a function of $\Re_0$ but can also be expressed in terms of $\Oho$ and $\We_0$ since $\Re_0 = \sqrt{\We_0}/\Oho$.

The $\Wecr$ vs $\Ohd$ plot in \cref{fig:wecr_ohd_sim} is recreated using $\Cbr$ (\ref{eq:C_breakup_final}) in the y-axis in \cref{fig:dropforces_cbreakup}(c). When scaled according to this new non-dimensional parameter, almost all simulation points move to a narrow range of $0.083<\Cbr<0.137$ with a standard deviation of $0.02$ about a mean of $0.11$, when compared to $10<\Wecr<74$ in the former plot. The highest $\Wecr$ cases corresponding to $\Ohd$ and $\Oho=0.1$ and $\rho=10,50$ in \cref{fig:wecr_ohd_sim} when plotted according to \ref{eq:C_breakup_final} achieve substantially lower threshold values, much more in line with the $\Cbr$ values obtained for other simulation cases. It is also observed that the non-trivial fragmentation morphologies such as backward bag-plume and backward-transition breakup cases show relatively the largest $\Cbr$ values among all other cases with the same $\Ohd$.

We also plot $\Cbr$ values of all the available experimental data for critical backward bag breakup as a set of reference data for the plot. As expected, the experimental data remains bound within a narrow extent of $\Cbr$ values, similar to its $\Wecr$ analog. The $\Cbr$ values do start to drop off as we approach larger and larger $\Ohd$ values, i.e., $\Cbr$ appears to overestimate the effort required to achieve critical breakup in very high $\Ohd$ drops compared to experiments. This is in line with the observations in previous experimental works \citep{Hinze1955,Hsiang1995} where it was observed that bag breakup becomes progressively difficult for higher drop Ohnesorge numbers and ultimately stops manifesting for $\Ohd>2$. Other breakup modes such as multimode and sheet-thinning become the critical breakup modes for very high $\Ohd$ systems. The inherent assumption in the derivation of $\Cbr$ was to assume that critical breakup morphology is a bag breakup, i.e., a drop first flattens into a pancake which then blows in to a bag. Any major deviation from this general breakup mechanism is expected to drastically change the deformation and breakup physics of such drops. Hence, as $\Ohd$ increases past $1$, physics related to multimode and sheet thinning breaks starts to become significant and hence must be considered in the derivation of any parameter attempting to define the critical breakup criteria.

Another factor worth considering is the initial Reynolds number required (i.e., $\Re_0 = \sqrt{\We_0}/\Oho$) for very large $\We_0$ values to show a breakup for high $\Ohd$ drops. For very large $\We_0$ values, very large $\Re_0$ values are expected, which would make the external flow more chaotic, which can have a major impact on the pressure forces experienced by the drop, its interaction with downstream vortices, and intensity of surface waves (\cref{subsec:oho}). These effects have not been taken into account in our derivation of $\Cbr$.

For the non-dimensional parameter space considered in the current work, $\Cbr$ captures the dominant physics very well and succeeds in compressing the rather large variance in $\Wecr$ values observed due to very low $\rho$ and high $\Ohd$ and $\Oho$ values. Therefore, for the purposes of the work, $\Cbr$ fulfills our requirements.

\section{Conclusions} \label{sec:conclusions}

The current work aimed to clarify two major questions:
\textbf{1.} The effect of  $\We_0$, $\rho$, $\Oho$ and $\Ohd$ on drop deformation and breakup characteristics, at or near the $\Wecr$ and for physical properties that are vastly different from commonly available water-air systems; and 
\textbf{2.} The effect of these non-dimensional parameters on drop critical breakup morphology.

As has been explained extensively in \cref{section:introduction}, most of the currently accepted ideas on threshold secondary atomization, such as the independence of $\Wecr$ with respect to $\Ohd$ given it is below $0.1$, or the critical breakup morphology being solely a backward bag, etc., originate from experimental works done for a small parametric space occupied by water-air analogous systems. The current work aimed to shed some light on these accepted ideas and provide a more complete picture of the process of secondary atomization of Newtonian drops. For this purpose, a parametric sweep across all the involved non-dimensional parameters, i.e. $\{\rho, \Oho, \Ohd\}$ (\cref{subsec:paramter_space}) was performed using axisymmetric simulations on Basilisk, with $\We_0$ values increased until a non-vibrational (critical) breakup was achieved for a given $\{\rho, \Oho, \Ohd\}$ set. The large amount of simulation data obtained from the comprehensive parameter exploration resulted in the following key findings:
\begin{enumerate}
  \item The internal flow inside a drop away from its core can be motivated by two forces: the pressure difference between its poles ($\delP$) and its periphery, and the shear stresses acting on its upstream surface. The competition between the two and the relative strengths of the two dictates whether the the drop pancake is flat or forward-facing. An internal flow predominantly driven by pressure difference results in a flat pancake, whereas a shear-stress driven internal flow results in a forward-facing pancake. (\cref{sec:results})
  \item Density-ratio controls $\delP$, whereas $\Oho$ controls the shear-stresses on the upstream surface. Thus, both these non-dimensional parameters are important when predicting the orientation of the pancake.
  \item  Given an external forcing, the sensitivity (i.e., local accelerations) of the different parts of the drop when experiencing external forces is directly controlled by the local inertia of that region. The morphology of the drop (forward or backward bag) as it continues to deform past the pancake stage is then controlled by the relative accelerations of different parts, which depends on the dominant mechanism driving the internal flow (pressure vs. shear stresses), the rate of evacuation of the core, the strength of downstream vortices, and the density-ratio of the drop fluid.
  \item $\Oho$ controls the lengthscales and timescales of the vortices produced in the wake of the drop. If the vortices do not detach from the periphery of the drop, higher local accelerations at the periphery are observed, which results in the formation of a forward facing bag. This is generally only the case for low $\rho$ drops.
  \item Given $\rho$ and $\Oho$ are the same, drop Ohnesorge number $\Ohd$ controls the response of the drop fluid to the external forcing. A low $\Ohd$ may allow the development of a surface instability at the upstream pole, resulting in the formation of a plume, and thus making a backward bag-plume morphology as the threshold morphology. $\Ohd$ also controls the rate of evacuation of the core, with lower $\Ohd$ cases showing slower evacuation and thus motivate the formation of a plume.
  \item  The critical Weber number $\Wecr$ over the explored parameter space is obtained and plotted against $\Ohd$ (\cref{fig:wecr_ohd_sim}) to recreate the plot presented in \cite{Hsiang1995}. It is found that $\Wecr$ varies significantly in the space $\Ohd<0.1$, achieving values as high as $70$ for the lowest density-ratio and ambient Ohnesorge number systems. Furthermore, backward bag-plume and forward bag morphologies are observed for threshold fragmentation in addition to the trivial backward bag morphology.
  \item  Based on all the insights gained from the simulation results, a phase diagram (\cref{fig:phase_diagram}) is constructed describing the various deformation pathways a spherical drop may undertake under impulsive acceleration depending on the properties. The drop deformation path shows its first split at the shape of pancake, which then leads to the possible threshold fragmentation morphologies.
  \item  The simulations also allow us to explore bag inflation characteristics for backward bag breakups for a greater parameter space, and extract some general conclusions on the influence of each parameter in modulating the growth rate and the associated timescales. Additionally, $t^* \approx 1$ is found to be a good measure for the timescale required for the initiation of bag inflation.
  \item Finally, a non-dimensional parameter ($\Cbr$) is derived by finding the ratio of forces supporting the fragmentation of the drop to the forces opposing it. Surface tension forces, pressure forces, and viscous dissipation is considered. The change in velocity of the drop relative to the ambient medium (which can be significant for low $\rho$ drops) is accounted for when calculating the forces. $\Cbr$ thus obtained provides a more complete alternative to Weber number as an indicator of drop threshold criteria in the dimensional space of the current study, by capturing the effects of all three studied non-dimensional parameters, i.e., $\rho$, $\Oho$, and $\Ohd$, on drop deformation and breakup.
\end{enumerate}

\section{Acknowledgment} \label{sec:acknowledgement}
Computational resources provided by the Covid-19 HPC Consortium through time on Blue Waters (NCSA) were used for the simulations.
S.D. and A.P.'s participation was supported by the DOE Office of Science through the National Virtual Biotechnology Laboratory (NVBL), a consortium of DOE national laboratories focused on response to COVID-19, with funding provided by the Coronavirus CARES Act. We would also like to thank Texas Advanced Computing Center for providing compute time on the NSF funded Frontera supercomputer, through the Pathways grant program. The compute time allowed us to simulate the 3D cases that were used to validate the accuracy of the axisymmetric simulations. 

\vspace{10pt}
\textbf{Declaration of interests}. The authors report no conflict of interest.

\appendix
\section{Numerical Scheme} \label{app:numerical_scheme}
Basilisk solves incompressible Navier Stokes multiphase flow equations (\ref{eq:NS} \ref{eq:vof} \& \ref{eq:c_adv}) on a quad/octree discretized grid, which allows variable mesh densities at the interface \citep{Popinet2003} and therefore accurately capture capillary scale phenomena.
\begin{subequations}
  \label{eq:NS}
  \begin{align}
    \rho(\partial_t \mathbf{u} + \pmb{u}.\pmb{\nabla u}) &= -\pmb{\nabla} p + \pmb{\nabla}.(2 \mu \pmb{D}) + \sigma \kappa \delta_s \pmb{n},
    \\
    \partial_t \rho + \pmb{\nabla}.(\rho\,\pmb{u}) &= 0,
    \\
    \pmb{\nabla.u} &= 0
  \end{align}
\end{subequations}

where $\pmb{u} = (u, v, w)$ is the fluid velocity, $\rho \equiv \rho(\pmb{x}, t)$ is the fluid density, $\mu \equiv \mu(\pmb{x}, t)$ is the dynamic viscosity, and $\pmb{D}$ is the deformation tensor defined as $D_{ij} = (\partial_i u_j + \partial_j u_i)/2$.
The Dirac distribution function $\delta_s$ allows inclusion of surface tension forces in the momentum equation by switching on the surface tension term only at the interface between the fluids; $\sigma$ is the surface tension coefficient, $\kappa$ and $\pmb{n}$ the curvature and the normal to the interface, respectively.
$\kappa$ is computed using Height Function (HF) formulation as described by \cite{torrey_nasa-vof2d_1985}, with attention given to address under-resolved interfaces.
The surface tension term is calculated using Continuum Surface Force (CSF) approach first described in \cite{brackbill_continuum_1992}, with special care taken to ensure that the conditions described in \cite{francois_balanced-force_2006} are satisfied to prevent parasitic currents.

To maintain the single equation formulation of the momentum equation, the two fluids are represented using a volume fraction $c(\pmb{x}, t)$ according to which $\rho$ and $\mu$ are defined as:
\begin{subequations}
  \label{eq:vof}
  \begin{align}
    \rho &= c\,\rho_1 + (1-c)\,\rho_2,
    \\
    \mu &= c\,\mu_1 + (1-c)\,\mu_2
  \end{align}
\end{subequations}

$\rho_1, \rho_2$ and $\mu_1, \mu_2$ are the densities of the first and second fluid in the domain, respectively.
In this formulation, the density advection equation is replaced with a volume fraction advection equation:
\begin{align}
  \label{eq:c_adv}
  \partial_t c + \pmb{\nabla}.(c\,\pmb{u}) = 0
\end{align}

The entire computational domain is discretized using squares for 2D (quadtree) and cubes for 3D (octree) and then organized in a hierarchy of cells.
The mesh resolution is adaptive in nature, and hence the two-fluid interface is resolved at a much higher resolution than other computationally less interesting regions of the domain.
This allows for large savings in the computational costs for two-phase simulations.
Any cell serving as a parent computational element can undergo further refinement into four or eight equal children cells for 2D and 3D computations, respectively.
Each of the children cells, in turn, can act as a parent cell if further refinement is warranted.
This successive refinement continues until a (user-defined) threshold criterion for error is satisfied or a maximum refinement level is reached.
A wavelet-based error estimation is used to estimate errors associated with the specified fields \citep{popinet_quadtree-adaptive_2015,vanhooftAdaptiveGridsAtmospheric2018}.
The maximum allowed refinement, corresponding to the smallest allowed cell size, is constrained by a user-specified minimum allowed cell dimension, which is defined by a parameter called ``Maximum Level'', a maximum level of $N$ corresponding to a minimum cell size of $L/2^N$.

\section{Choice of numerical parameters} \label{app:convergence}
Before using Basilisk for the production runs, it is essential to test for independence of the solutions with respect to both the maximum mesh resolution (normally achieved at the interface), wavelet-error thresholds for relevant field variables, and the tolerance of the Poisson solver.
For drop simulations, the accuracy of the calculated interface and the velocity fields must be ensured for correct retrieval of surface stresses, and correspondingly the temporal development of drop deformation process.
Hence in the current simulations, the extent of adaptive mesh refinement performed at a computational cell is restricted by the allowable maximum wavelet errors for velocity ($\chi_u$) and volume fraction ($\chi_c$) fields.
Additionally, a maximum allowed refinement level ($N$) is specified which enforces a strict allowed minimum cell size of $L/2^N$ across all the computational cells.
This helps prevent unbounded mesh refinement if the error threshold criteria converges slowly with decrease in cell size in a computational cell, which in turn eases the restriction on simulation timestep (dependent on the cell with the highest CFL number in the domain).
Finally, the tolerance of the Poisson solver ($\epsilon$) is also a critical parameter that can affect the accuracy of the velocity field and hence the interface.

\begin{figure}
  \centering \includegraphics[width=1.0\textwidth]{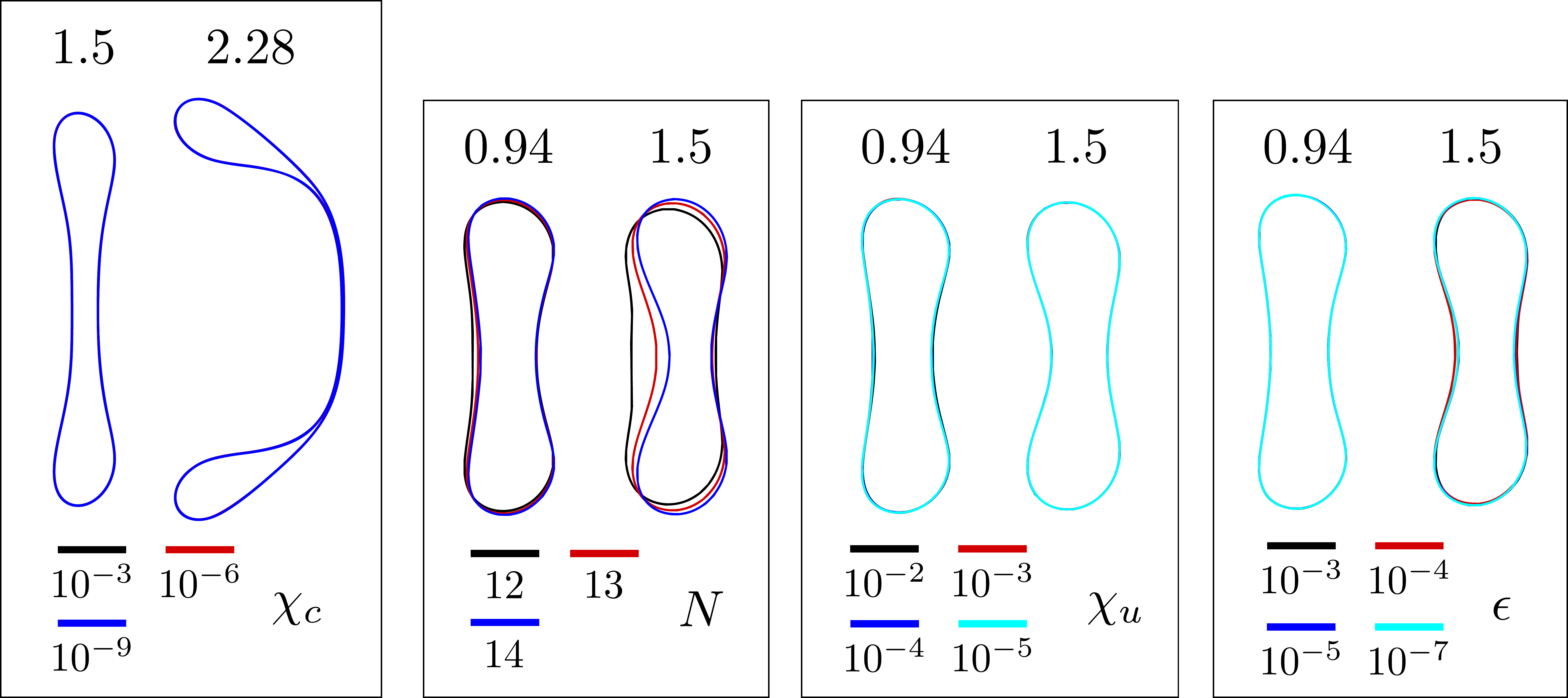}
  \caption{
  VOF plots at specific times for different error thresholds for volume fraction field $c$ ($\chi_c$) and velocity fields ($\chi_u$) is shown.
  The role of the tolerance of the Poisson solver ($\epsilon$) is also shown through its effect on the interface.
  }
  \label{fig:convergence_vof}
\end{figure}
\begin{figure}
\centering
\begin{subfigure}{0.32\textwidth}
    \includegraphics[width=\textwidth]{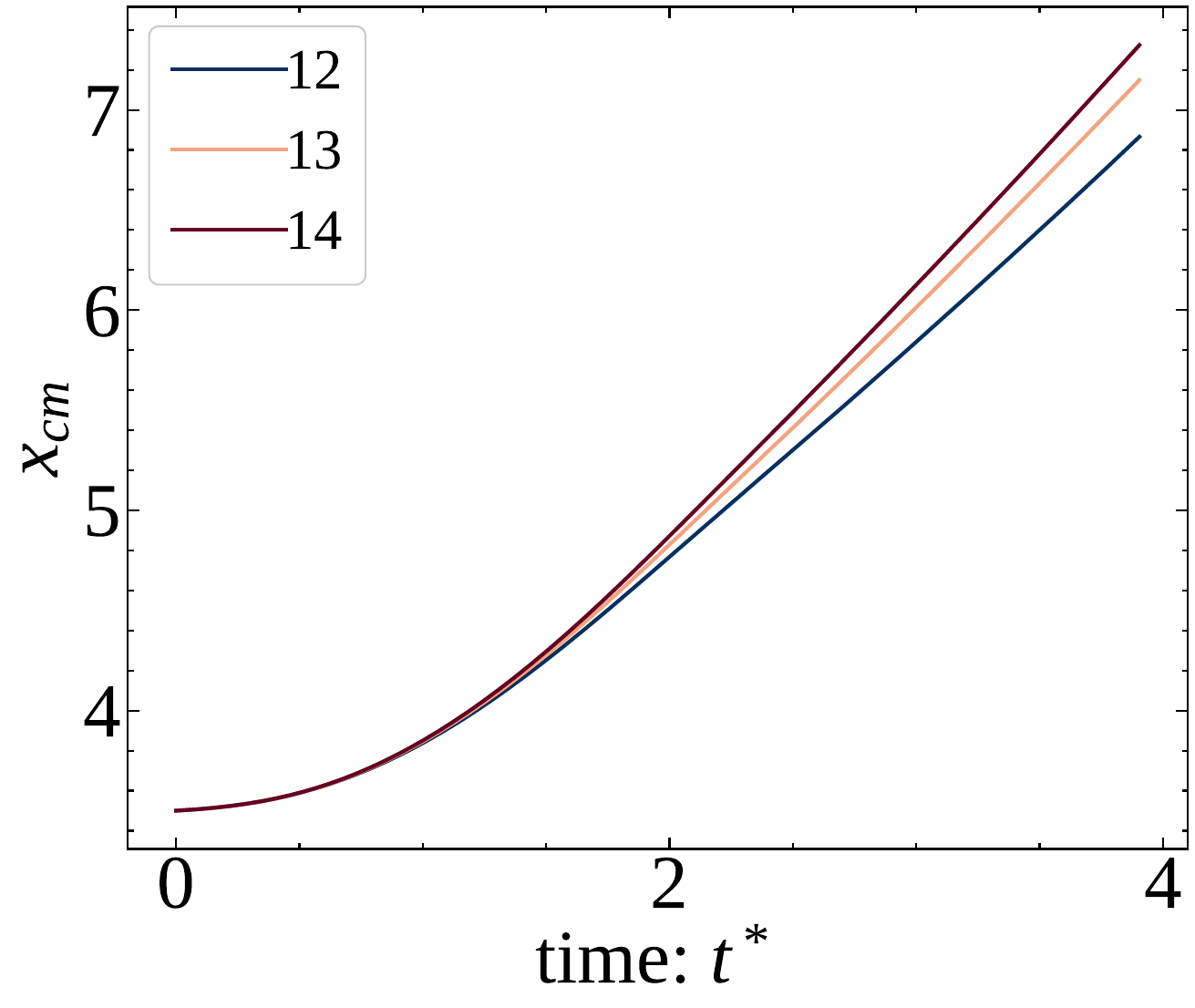}
    \caption{}
    \label{fig:con_lev_xpos}
\end{subfigure}
\hfill
\begin{subfigure}{0.32\textwidth}
    \includegraphics[width=\textwidth]{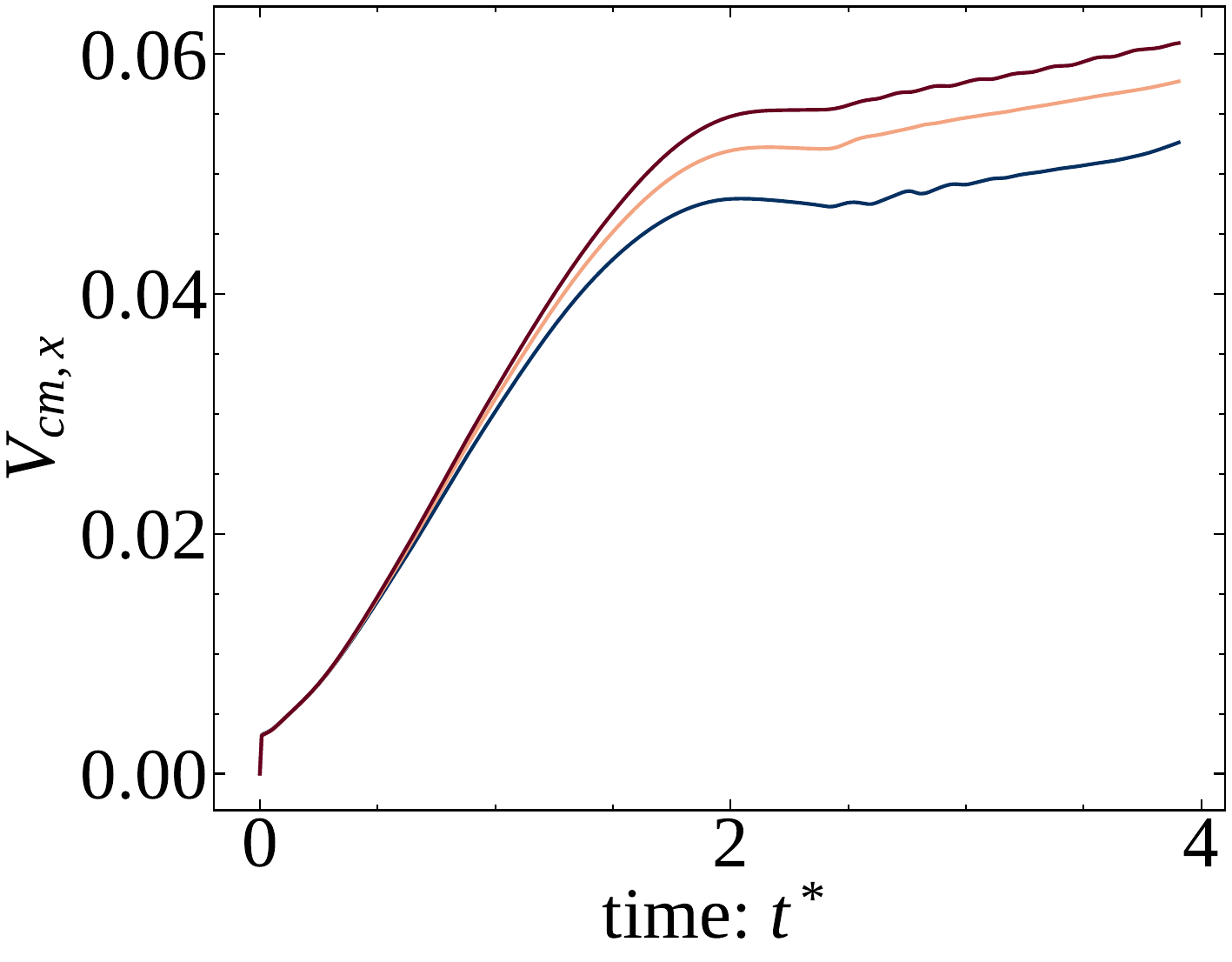}
    \caption{}
    \label{fig:con_lev_xvel}
\end{subfigure}
\hfill
\begin{subfigure}{0.32\textwidth}
    \includegraphics[width=\textwidth]{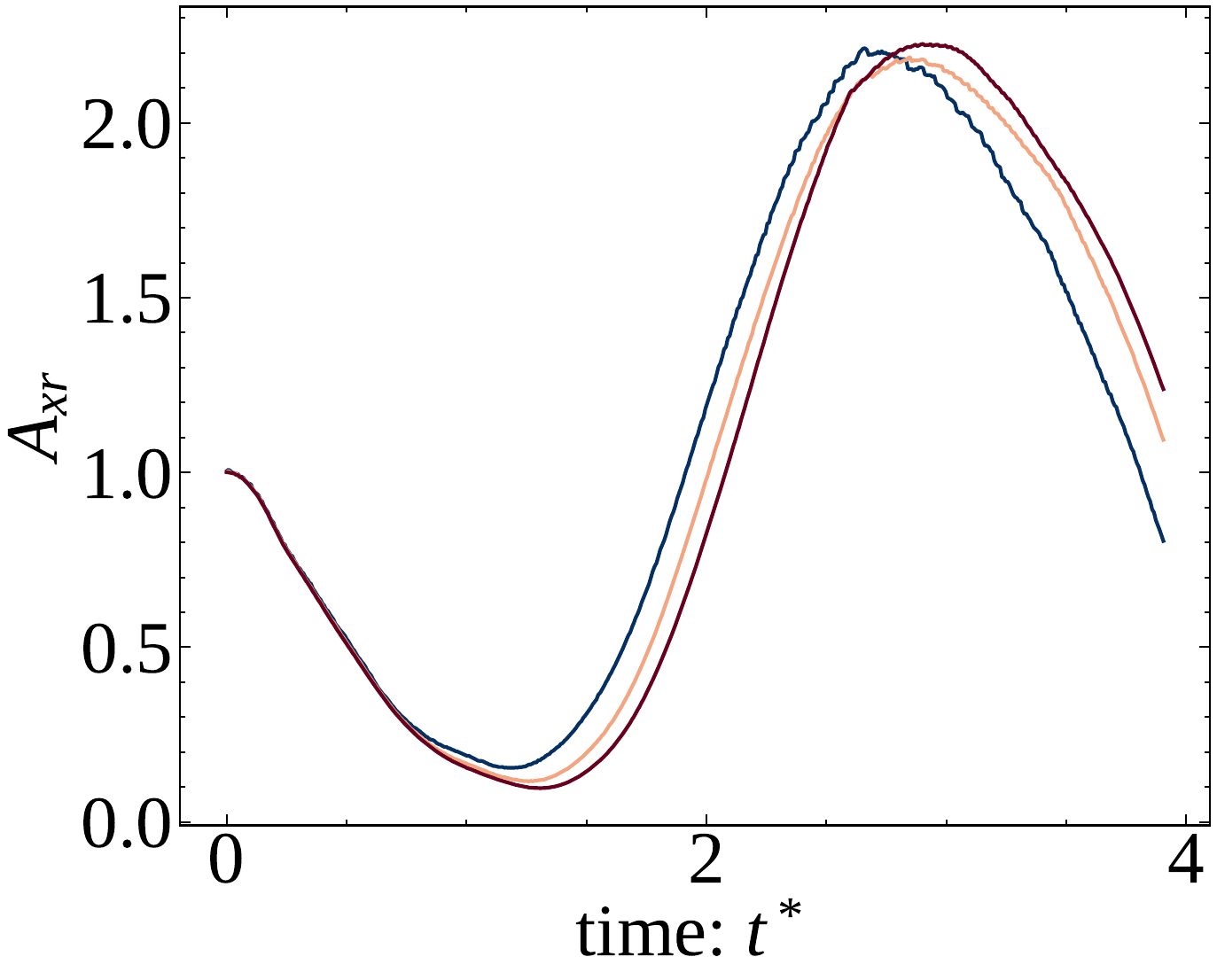}
    \caption{}
    \label{fig:con_lev_ar}
\end{subfigure}

\begin{subfigure}{0.32\textwidth}
    \includegraphics[width=\textwidth]{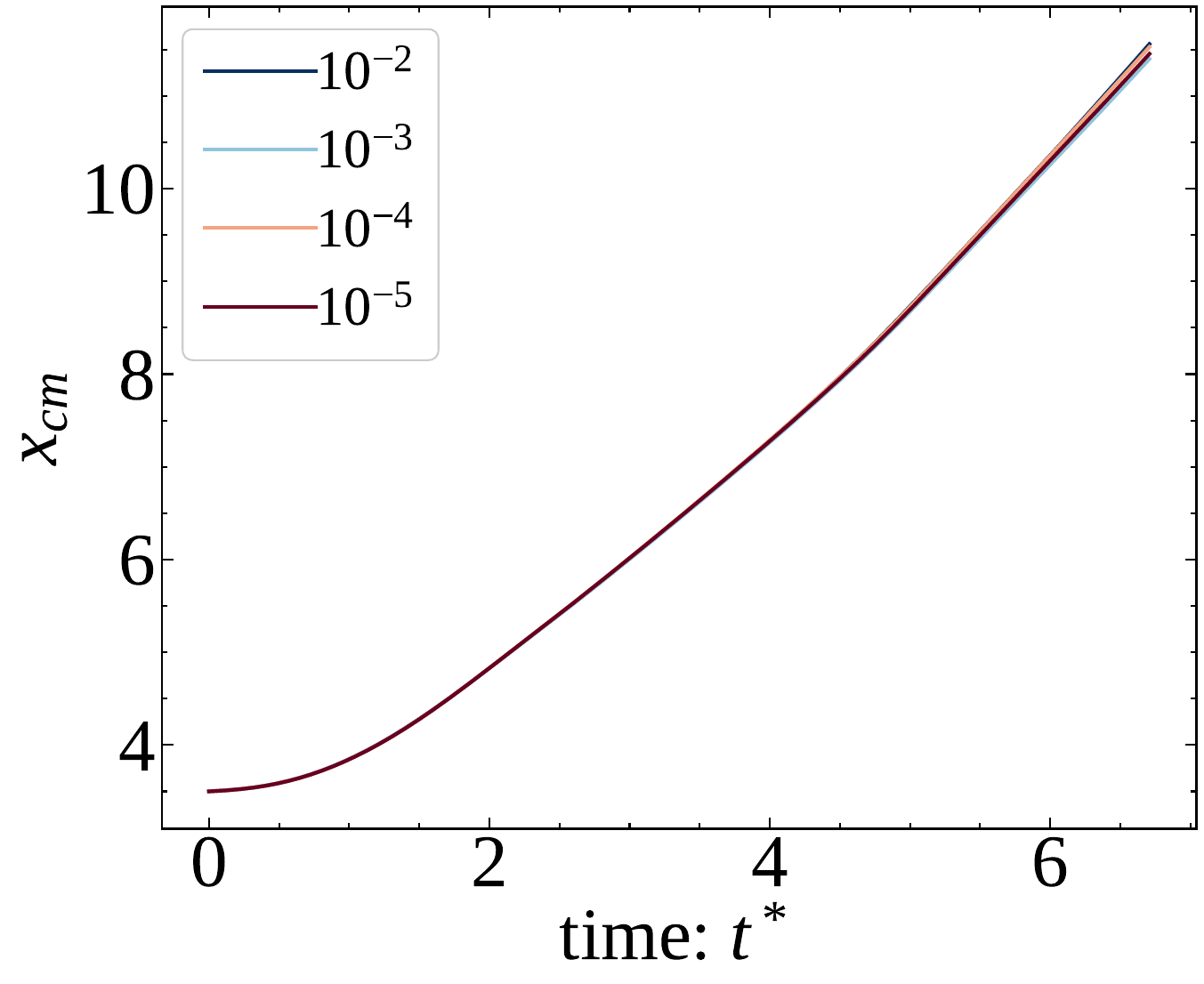}
    \caption{}
    \label{fig:con_ue_xpos}
\end{subfigure}
\hfill
\begin{subfigure}{0.32\textwidth}
    \includegraphics[width=\textwidth]{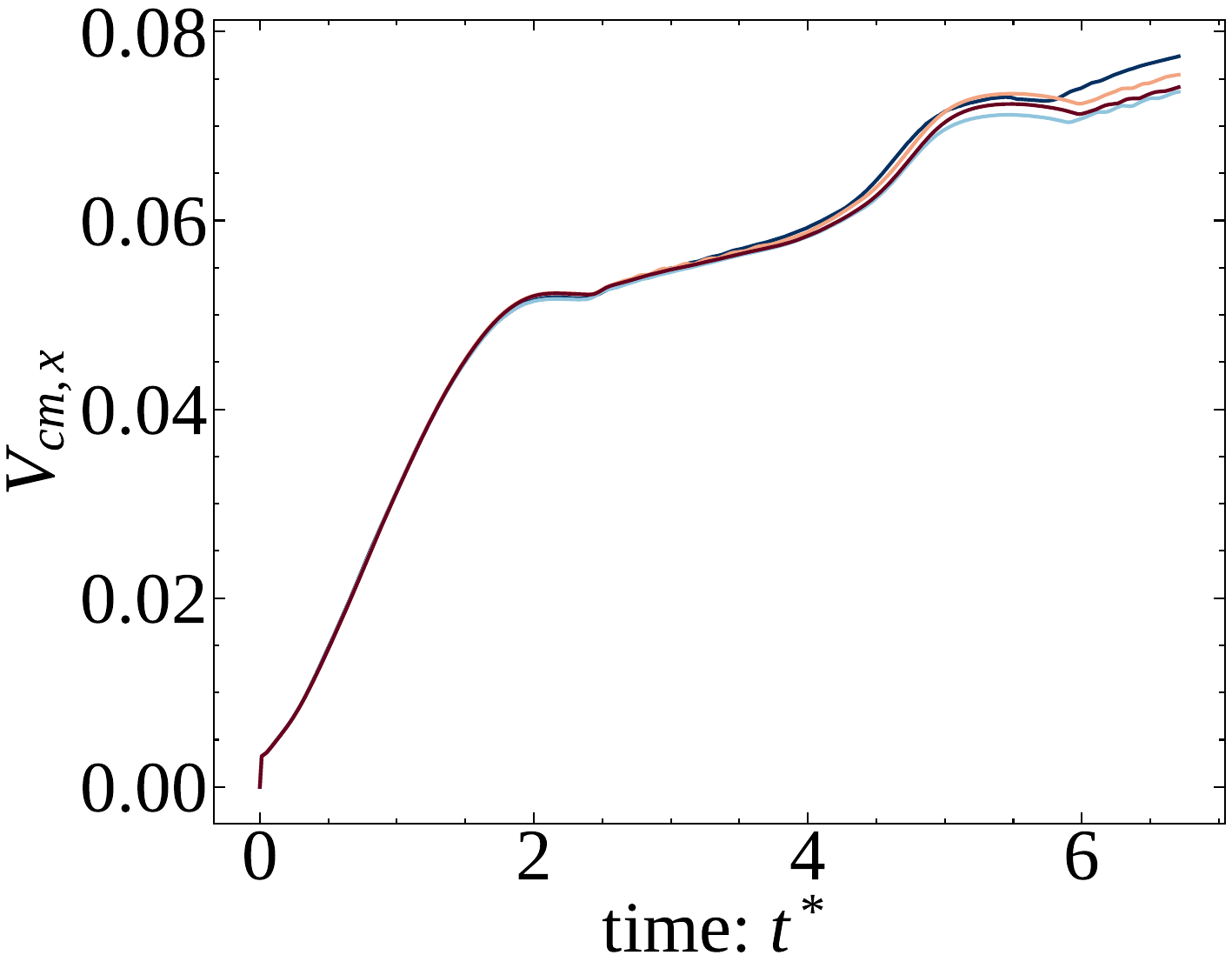}
    \caption{}
    \label{fig:con_ue_xvel}
\end{subfigure}
\hfill
\begin{subfigure}{0.32\textwidth}
    \includegraphics[width=\textwidth]{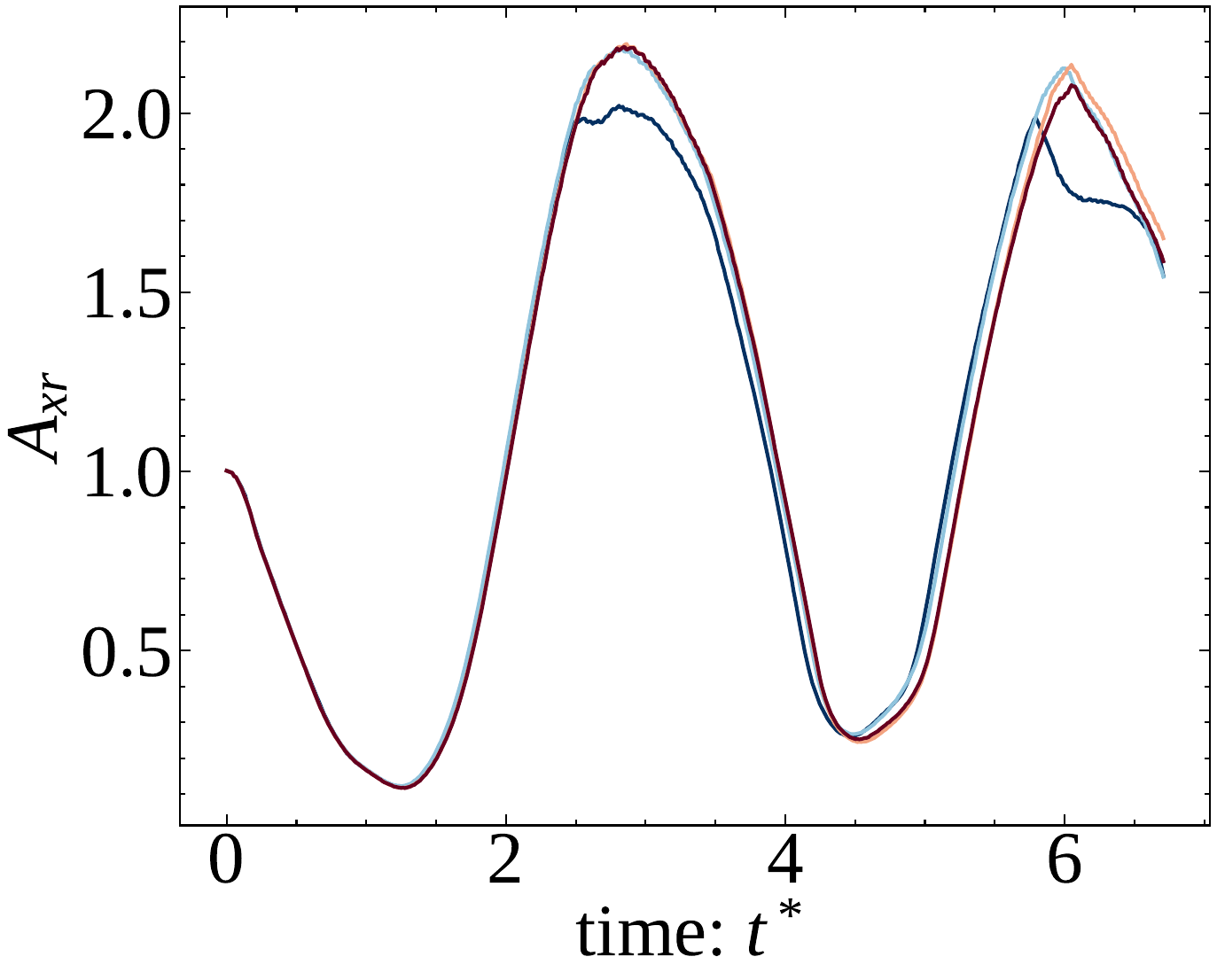}
    \caption{}
    \label{fig:con_ue_ar}
\end{subfigure}

\begin{subfigure}{0.32\textwidth}
    \includegraphics[width=\textwidth]{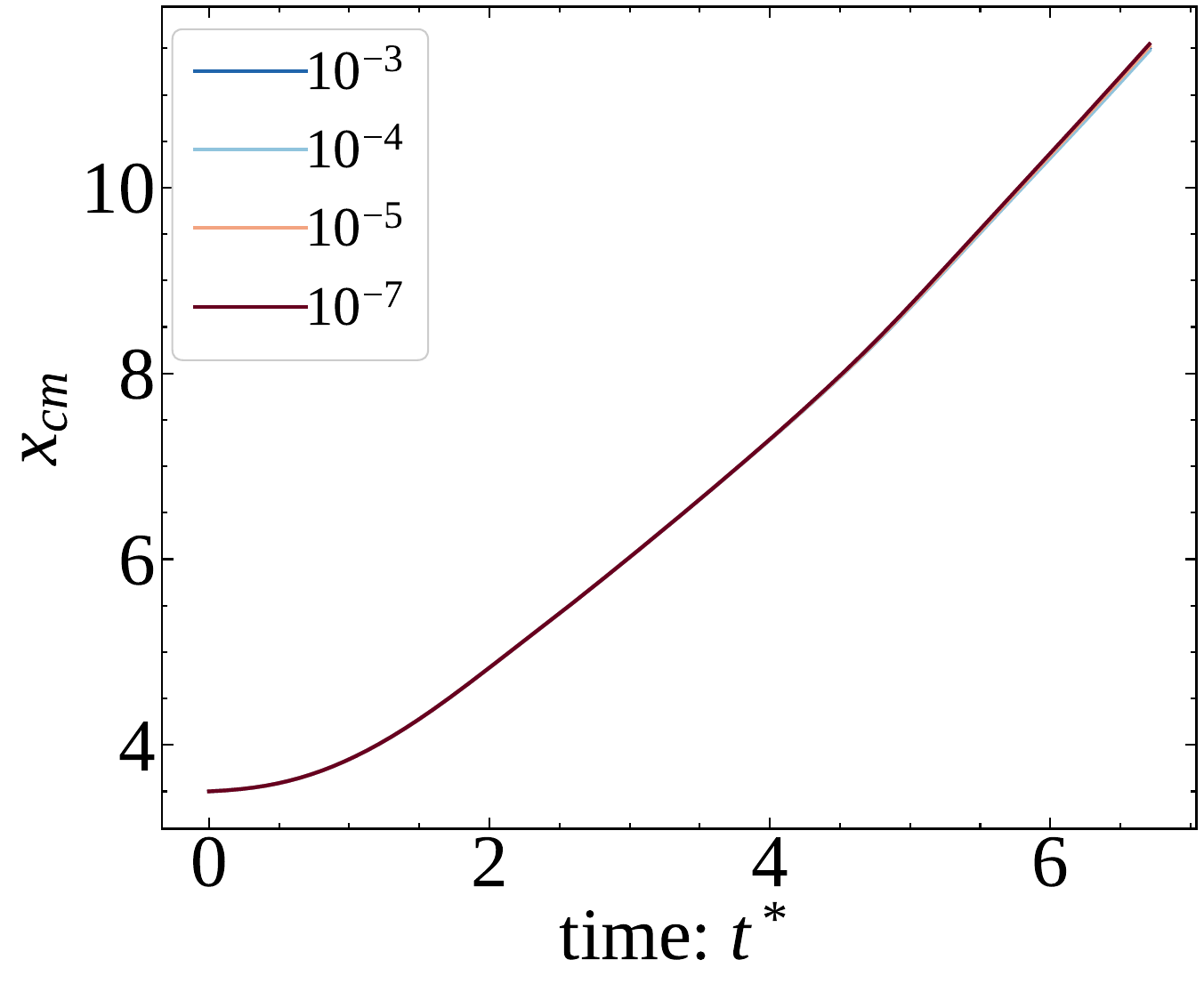}
    \caption{}
    \label{fig:con_tol_xpos}
\end{subfigure}
\hfill
\begin{subfigure}{0.32\textwidth}
    \includegraphics[width=\textwidth]{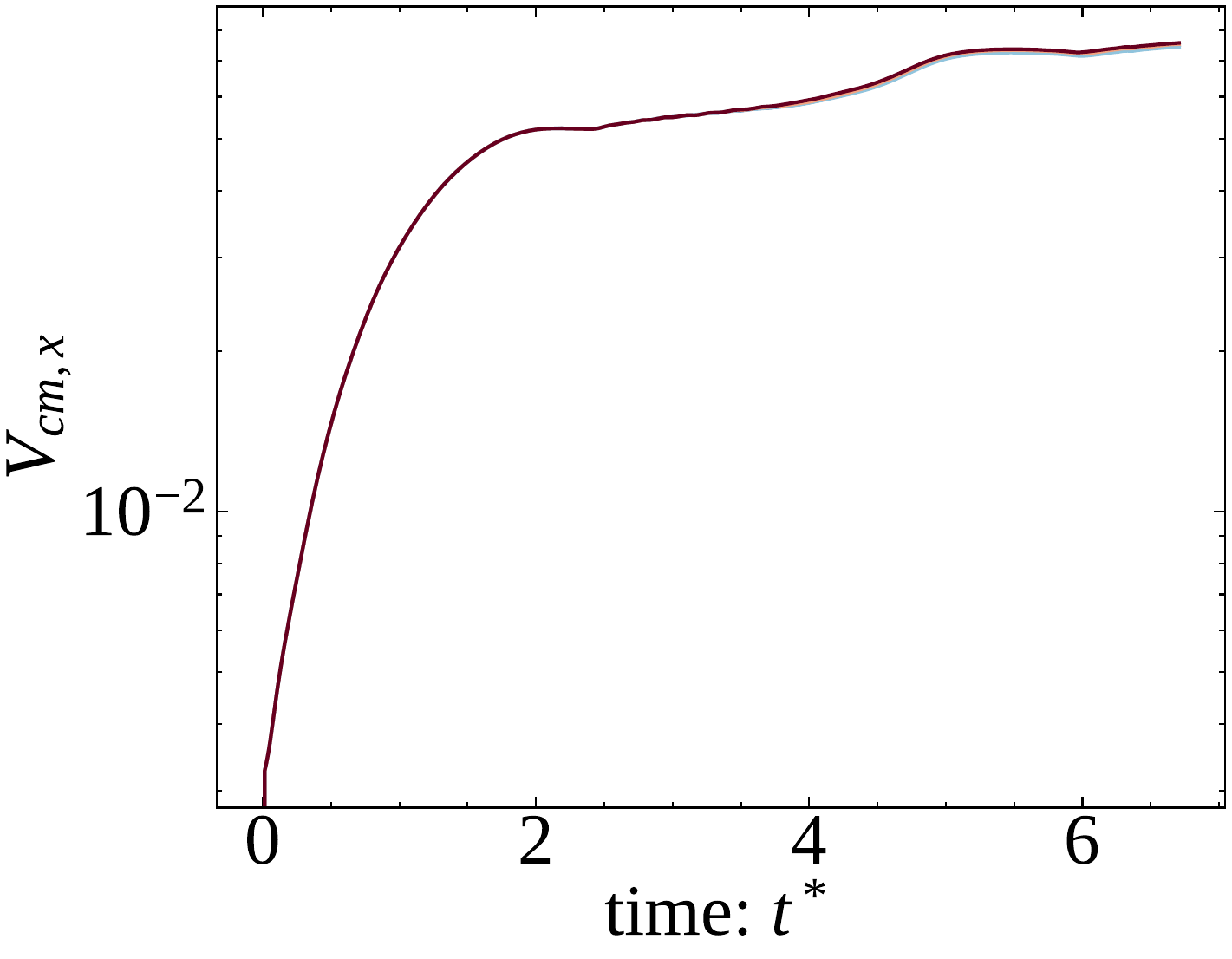}
    \caption{}
    \label{fig:con_tol_xvel}
\end{subfigure}
\hfill
\begin{subfigure}{0.32\textwidth}
    \includegraphics[width=\textwidth]{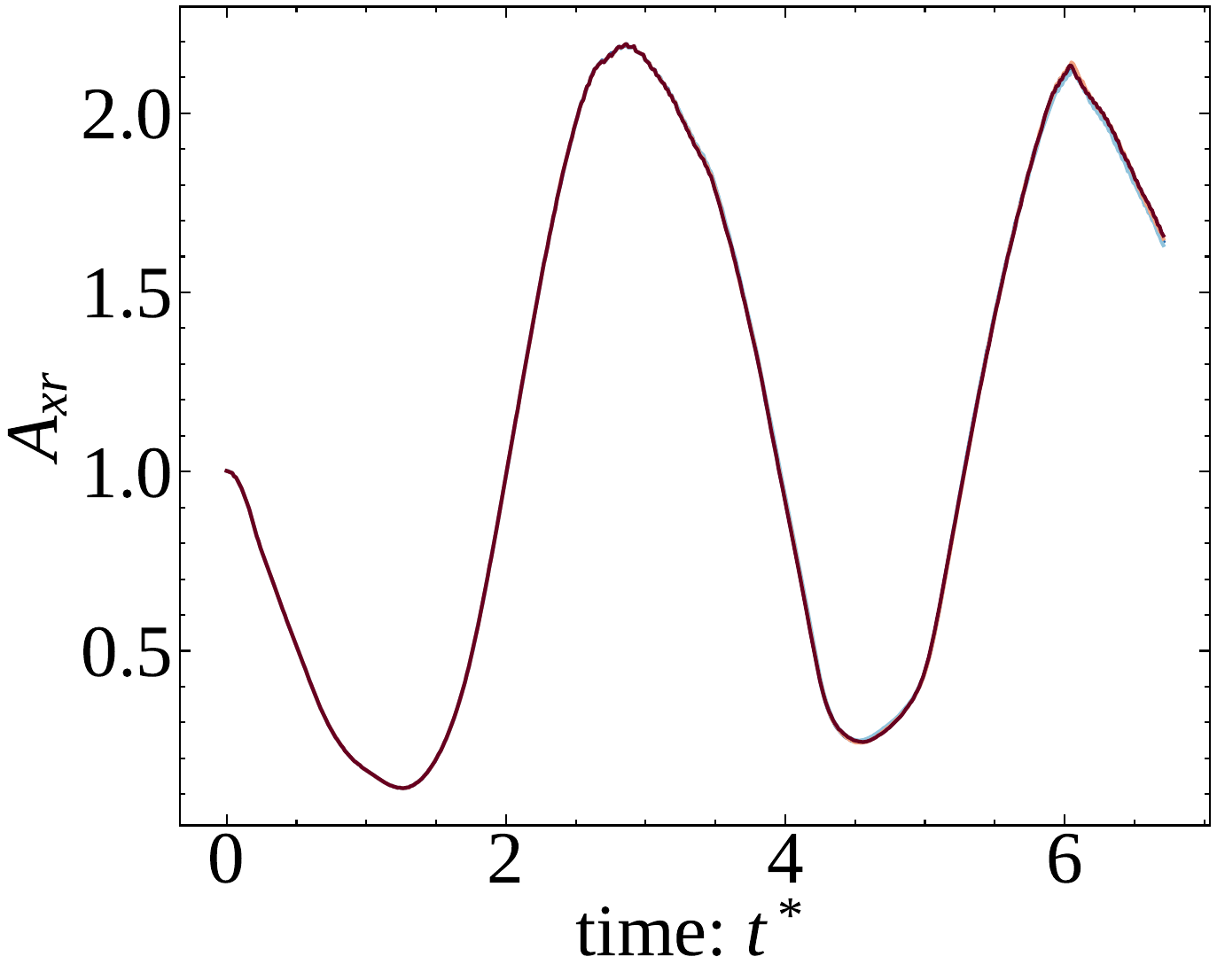}
    \caption{}
    \label{fig:con_tol_ar}
\end{subfigure}
\caption{
x-coordinates of center-of-mass, x-velocity of center-of-mass, and axis ratio are shown for different thresholds for wavelet errors of the volume fraction fields ($\chi_c$) in (a) to (c), different threshold for wavelet error of the velocity field ($\chi_u$) in (d) to (f), and different tolerances of the Poisson solver in (g) to (i), respectively.
In (a) to (c), the maximum allowed refinement level $N$ refers to a minimum cell size of $L/2^N$.
Thus, 256, 512 and 1024 cells per diameter correspond to $N=12$, $N=13$, and $N=14$ respectively, given $L=16$ and $D=1$.
}
\label{fig:convergence}
\end{figure}

We test the independence of the axisymmetric simulations with respect to $\chi_u$, $N$, and $\epsilon$ using a test case with $\{\rho, \Oho, \Ohd, \We_0\} = \{500, 0.001, 0.01, 10\}$ on an axisymmetric domain (\cref{fig:axi_domain}) with $L=16$, $D=1$ and $V_0=1$.
A drop-ambient system with these physical properties is expected to deform close to fragmentation, but instead of fragmenting, it retracts towards its neutral spherical shape.
Thus, it provides a good representative case for identifying any spurious diffusion accumulated in the system.
For testing the influence of $\chi_c$, we simulate a case with $\rho, \Oho, \Ohd, \We_0 = 500, 0.01, 0.1, 16$, which is expected to inflate into a bag morphology and fragment, and hence is a good representative case for the accuracy of the interface calculation.
The default values of $\chi_c$, $\chi_u$, $\epsilon$, and $N$ for all cases is set to $10^{-6}$, $10^{-3}$, $10^{-4}$, and $13$ respectively, and the parameter of concern is varied to test the independence of the solution with respect to it.

In \cref{fig:convergence_vof}, the drop interface at specific times is shown for different values of $\chi_c$, $N$, $\chi_u$, and $\epsilon$.
$\chi_c$ is varied across $10^{-3}$, $10^{-6}$, and $10^{-9}$, and we observe that the interface is identical for all the three values.
This indicates that the interface is converged with respect to $\chi_c$.
The effect of $N$ however is significant, with the drop showing far more numerical diffusion for $N=12$ compared to $N=13$ and $N=14$.
This results in the drop achieving less deformation and thus starts retracting earlier for $N=12$.
The effect of $\chi_u$ and $\epsilon$ is negligible for the two time instances shown, with the interface being identical for all the explored values of $\chi_u$ and $\epsilon$.

\Cref{fig:convergence} shows the effect of $N$, $\chi_u$, and $\epsilon$ on the x-coordinate of the center-of-mass, x-velocity of the center-of-mass, and the aspect ratio of the drop, resulting in three sets of plots for each parameter.
Plots for $\chi_c$ are not shown since we have already established that the interface is converged with respect to $\chi_c$.
It is also found that the computational costs associated with the three $\chi_c$ values are within 1\% of each other.
This makes the choice of $\chi_c$ less critical, and hence the default value of $\chi_c=10^{-6}$ is chosen for the production runs.

\Cref{fig:convergence}(a), (b), and (c) show the effect of $N$ on the x-coordinates of the center-of-mass, x-velocities of the center-of-mass, and aspect ratios of the drop, respectively.
It is observed that the three $N$ values start to diverge as the drop reaches its minimum aspect ratio, with $N=12$ showing the highest values corresponding to the least deformation.
As the maximum allowed refinement increases, the drop starts to deform more, with $N=14$ showing the highest deformation.
This points to the larger numerical diffusion in the $N=12$ case.
$N=13$ and $N=14$ show much closer results for all the three plots.
$N$ also has a dramatic effect on computational costs, with $N=14$ requiring approximately 3 times the computational time as required for $N=13$.
$N=13$, which is equivalent to $512$ cells per diameter for $L=16$, provides a good balance between accuracy and computational costs.
For all simulations in this study, $N=13$ is chosen as the default value for $L=16$.
For the highest $Re_0$ cases ($\Oho=0.0001$), $N=14$ is chosen to ensure that the low viscosity ambient flow is resolved accurately.

\Cref{fig:convergence}(d), (e), and (f) plot $x_{cm}$, $V_{cm,x}$, and $A_{xr}$ for four different values of $\chi_u$.
We observe that apart from the lowest value of $\chi_u=10^{-2}$, the three properties are almost identical for all values of $\chi_u$.
The $\chi_u=10^{-2}$ case shows a lower $A_{xr}$, resulting from higher dissipation of the energy contained in the drop, compared to the other three cases, resulting in a slightly higher $x_{cm}$ and $V_{cm,x}$.
The results for $\chi_u=10^{-4}$ and $\chi_u=10^{-5}$ are almost identical, with the computational cost for $\chi_u=10^{-4}$ being approximately 2.5 times less than that for $\chi_u=10^{-5}$.
Hence, $\chi_u = 10^{-4}$ is chosen for the production runs.

Finally from the convergence plot for the tolerance of the Poisson solver in \cref{fig:convergence}(g), (h), and (i), we observe that the x-coordinates of the center-of-mass, x-velocities of the center-of-mass, and axis ratios are almost identical for all the three values of $\epsilon$.
We hence use the lowest value of $\epsilon=10^{-3}$ for the production runs, as it provides a converged result with the least computational cost.

\section{Comparison to 3D simulations}
\label{app:3D_comparison}
\begin{figure}
  \centering \includegraphics[width=1.00\textwidth]{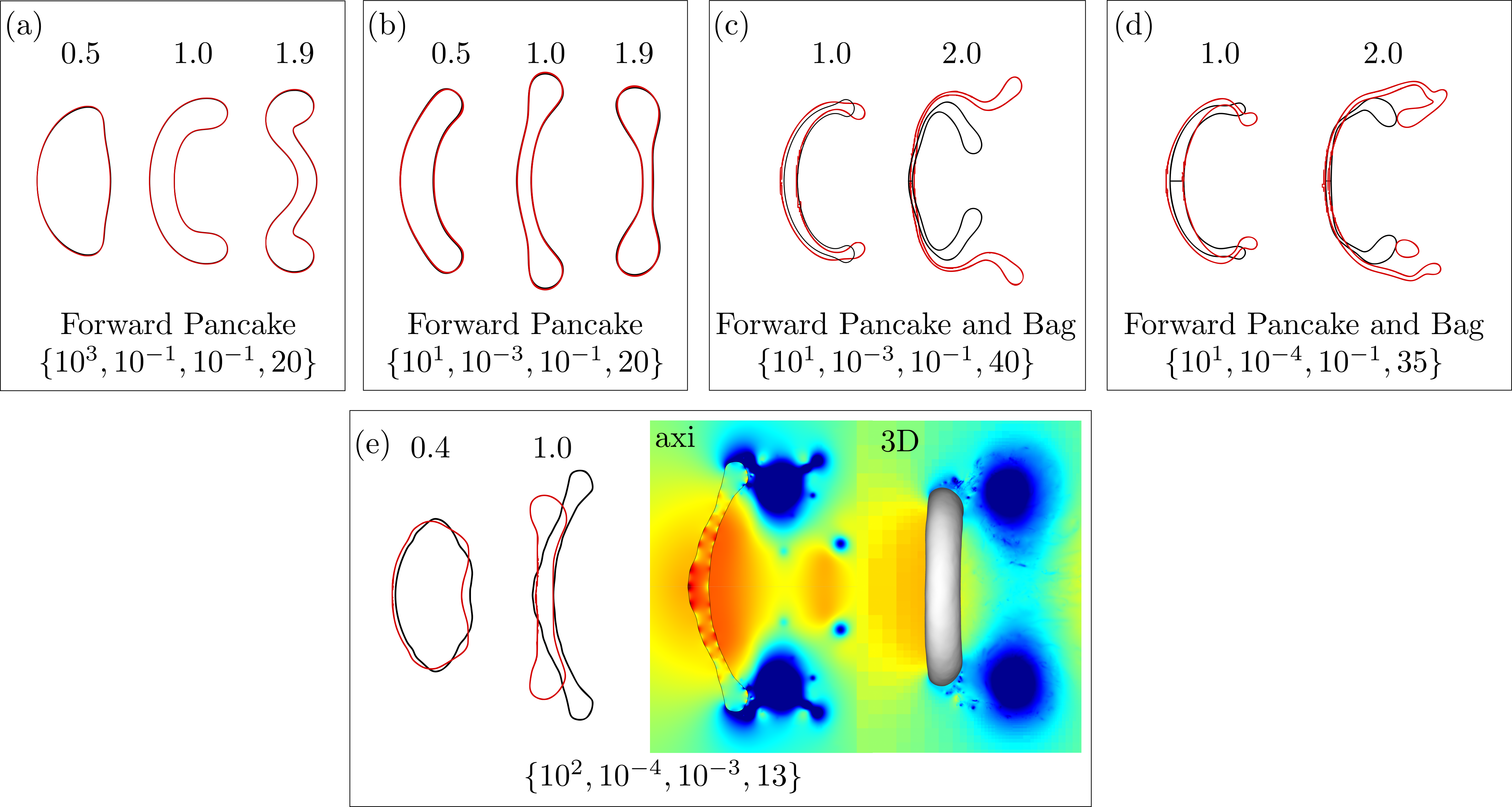}
  \caption{
  The figure shows the fluid interface for different time instances for axisymmetric (shown in black) and 3D (shown in red) simulations for five different cases, with the drop-ambient system properties mentioned in the figure as $\{\rho, \Oho, \Ohd, \We_0\}$.
  Two of these cases, shown in (a) and (b), are chosen to verify if the axisymmetric simulations can capture the formation of a forward pancake in two different contexts: a large density-ratio drop ($\rho=500$) and a low density-ratio drop ($\rho=10$).
  The cases shown in (c) and (d) help us verify the physical validity of forward bags observed during threshold fragmentation of low $\rho$ drops and small $\Oho$ values.
  The final case (e) serves to highlight the differences between axisymmetric and 3D simulations for cases where both the drop and ambient Ohnesorge numbers are the smallest. The pressure fields at $t^*\approx 1$ is also shown.
  }
  \label{fig:3daxi_comp}
\end{figure}

To justify the use of axisymmetric simulations for the wide parameter space of density ratios and viscosities under consideration, we conduct 3D simulations for some benchmark cases in order to verify if 3D simulations produce similar non-trivial morphologies as the axisymmetric simulations. A qualitative match with axisymmetric simulations would justify their use for prediction in the respective pancake and bag orientations for the parameter space. \cite{jain_secondary_2019} states that ``For the drops with high $\rho$, flow around the drop has relatively low effect on the drop deformation, morphology and the breakup''. If this is the case, the ideal benchmark drop-ambient systems should have either low $\rho$ drops in low $\Oho$ flows, or high $\rho$ drops in high $\Oho$ flows in order to magnify the role of ambient flow on the drop. This is a computationally favorable scenario for us, since both low $\rho$ and low $\Re_0$ systems have significantly reduced computational requirements. Keeping this in mind, we specify six benchmark cases in 3D:
\begin{enumerate}
  \item Two of the cases are designed to verify the formation of forward pancakes in axisymmetric simulations, one is a large density-ratio drop ($\rho=1000$) drop in a shear stress dominated (large $\Oho$) system, another is a low density-ratio drop ($\rho=10$) in a pressure dominated system. These simulations are run at the same refinement level as the axisymmetric simulations.
  \item Two of the cases test the ability of axisymmetric simulations to correctly predict the transition from a forward pancake to forward bag for low density-ratio drops in high $\Re_0$ systems. Both the simulations are run at the same refinement level as the axisymmetric simulations.
  \item An additional case at an intermediate density ratio of $\rho=100$ are used to highlight the differences between the two simulation types for large ambient Reynolds number cases ($\Oho=0.0001$) and a low viscosity drop ($\Ohd=0.001$). This choice renders the ambient flow non-axisymmetric since the formation of turbulent vortices is a purely 3D phenomenon, making such a flow system impossible to perfectly reproduce with only axisymmetric simulations. The high $\Re_0$ of the flow coupled with an intermediate density-ratio makes the simulations computationally extremely expensive making a 3D simulation until fragmentation unfeasible. These simulations have been run only until $t^* \approx 1$, i.e., until the formation of the pancake.
\end{enumerate}
Backward bag formation has not been tested here, since it has already been considered in \cref{subsec:validation}.
The fluid interface for axisymmetric and 3D simulations for the cases described in (i) and (ii) are shown in \cref{fig:3daxi_comp}(a)-(d), with red color representing the 3D simulations. All four cases show a good qualitative match between the two simulation types, with the formation of a forward pancake observed in all cases. The cases with $\Oho \le 0.001$ (shown in (a) and (b)) show an excellent match in the interface shape all throughout the deformation process. For the $\Oho=0.0001$ cases (shown in (c) and (d)), the interface shape differs more, with the 3D simulations showing a more pronounced forward pancake and a larger forward bag compared to the axisymmetric simulations. This can be attributed to the differences in viscous dissipation between axisymmetric and 3D simulations. However, most importantly, the 3D simulations transition from a forward pancake to a forward bag in a manner similar to axisymmetric simulations, thus supporting the physical validity of forward bag morphology observed in the axisymmetric simulations.

\Cref{fig:3daxi_comp}(e) shows the pressure field at $t^* \approx 1$ for the $\Oho=0.0001$ case for $\rho=100$, as described in (iii). It is observed that the pressure fields even for such high $\Re_0$ cases is close to toroidal and interacts similarly with the periphery of the drop. However, the downstream stagnation pressures are much higher for the axisymmetric simulations. This is consistent with the observations of \cite{lingDetailedNumericalInvestigation2023}. Thus, for all very large $\Re_0$ simulations with low $\Ohd$ performed in this work, both the pancake shape and subsequent fragmentation morphology is not well justified. All simulation results for such cases are to be considered with this caveat in mind. It is hypothesized that highly turbulent ambient flow in low $\Oho$ systems produces non-axisymmetric interfacial perturbations which are not reproduced in the axisymmetric simulations. These perturbations however have little significance for the high $\Ohd$ cases, since they are rapidly dissipated at timescales smaller than the deformation timescale. However for low $\Ohd$ drops, these surface perturbations persist for longer timescales and hence can significantly influence the subsequent deformation. Thus, the axisymmetric simulations are not expected to accurately predict the fragmentation morphology for cases with high ambient Reynolds numbers coupled with very low drop viscosities.

\bibliographystyle{jfm}
\bibliography{draft.bib}

\end{document}